\documentclass[11pt]{article}

\pdfoutput=1
\usepackage[utf8]{inputenc}
\usepackage[linktocpage,bookmarks,plainpages=false,pdfborder={0 0 0},pdfborderstyle={}]{hyperref}
\usepackage{amsmath}
\usepackage{amsthm}
\usepackage{amsfonts}          % if you want the fonts
\usepackage{amssymb}           % if you want extra symbols
\usepackage{color}
\usepackage{graphicx}
\usepackage{cite}
\usepackage{multirow,setspace,microtype}
\usepackage{tikz}
\usepackage{booktabs}
\usepackage[hang,flushmargin]{footmisc} 
\usepackage[framemethod=tikz]{mdframed}
\AtBeginEnvironment{mdframed}{%
	\tikzset{every picture/.style={}}%
}
\mdfsetup{roundcorner=.5ex}
{\begin{mdframed}[backgroundcolor=black!5!white]}%
	{\end{mdframed}}

\PassOptionsToPackage{linecolor=blue,backgroundcolor=blue!25,bordercolor=blue,textsize=scriptsize}{todonotes}
\usepackage{todonotes}

\numberwithin{equation}{section}

\usepackage[bf]{caption}
\let\originalleft\left
\let\originalright\right
\renewcommand{\left}{\mathopen{}\mathclose\bgroup\originalleft}
\renewcommand{\right}{\aftergroup\egroup\originalright}

\makeatletter
\g@addto@macro\bfseries{\boldmath}
\makeatother

\newlength{\xtrawidth}
\newlength{\xtraheight}
\newcommand{\Z}{\ensuremath{\mathbb{Z}}}

\newcommand{\R}{\ensuremath{{\mathbb{R}}}}
\newcommand{\C}{\ensuremath{{\mathbb{C}}}}
\newcommand{\Rep}[1]{\ensuremath{\boldsymbol{\underline{#1}}}}

\newcommand{\Kcone}{\ensuremath{\mathcal{K}}}
\newcommand{\tmod}{~\mathrm{mod}~}
\newcommand{\Ocal}{\ensuremath{{\cal O}}}
\newcommand{\Vvis}{\ensuremath{V^{(1)}}}
\newcommand{\Vhid}{\ensuremath{V^{(2)}}}
\newcommand{\Rhat}{\ensuremath{{\widehat R}}}
\newcommand{\ah}{\hat{\alpha}_{\text{GUT}}}

\global\long\def\dd{\text{d}}%
\global\long\def\ii{\text{i}}%
\global\long\def\SO#1{{SO}(#1)}%
\global\long\def\Uni#1{{U}(#1)}%
\global\long\def\SU#1{{SU}(#1)}%
\global\long\def\Ex#1{{E}_{#1}}%
\global\long\def\eqspace{\mathrel{\phantom{{=}}{}}}%
\global\long\def\op#1{\operatorname{#1}}%

\DeclareMathOperator{\tr}{tr}
\DeclareMathOperator{\Span}{span}
\DeclareMathOperator{\rank}{rank}
\DeclareMathOperator{\re}{Re}

\renewcommand{\thefootnote}{\fnsymbol{footnote}}

\usepackage{subcaption}
\usepackage{float}
\usepackage{caption}

\begin{document}

\begin{titlepage}
%\vspace{-4cm}
\title{\LARGE \bf{Line Bundle Hidden Sectors for Strongly{\\[.15cm]  } Coupled Heterotic Standard Models}\\[.3cm]}
                       
\author{{
   Anthony Ashmore,
   Sebastian Dumitru
   and Burt A.\,Ovrut}\\[0.8cm]
   {\it Department of Physics, University of Pennsylvania} \\[.1cm]
   {\it Philadelphia, PA 19104--6396}}         

\date{}

\maketitle

\begin{abstract}
\noindent
The compactification from the eleven-dimensional Hořava--Witten orbifold to five-dimensional heterotic M-theory on a Schoen Calabi--Yau threefold is reviewed, as is the specific $SU(4)$ vector bundle leading to the ``heterotic standard model'' in the observable sector. Within the context of strongly coupled heterotic M-theory, a formalism for consistent hidden-sector bundles associated with a single line bundle is presented, and a specific line bundle is introduced as a concrete example. Anomaly cancellation and the associated bulk space five-branes are discussed in this context, as is the constraint that the hidden sector bundle be compatible with the slope-stability requirements of the observable sector $SU(4)$ gauge bundle.
The further compactification to a four-dimensional effective theory on a linearized BPS double domain wall is then presented to order $\kappa_{11}^{4/3}$. Specifically, the generic constraints required for anomaly cancellation and the restrictions imposed by positive squared gauge couplings to order $\kappa_{11}^{4/3}$ are presented in detail.  Three additional constraints are imposed, one guaranteeing that the $S^{1}/{\mathbb{Z}}_{2}$  orbifold length is sufficiently larger than the average Calabi--Yau radius, and two enforcing that the hidden sector be compatible with both the unification mass scale and unified gauge coupling of the $SO(10)$ group in the observable sector. Finally, the expression for the Fayet--Iliopoulos term associated with an anomalous $U(1)$ symmetry is presented and its role in $N=1$ supersymmetry in the low-energy effective theory is discussed. It is shown that $N=1$ supersymmetry can be preserved by cancelling the tree-level and genus-one contributions against each another. As a check on our results, we calculate several quantities to order $\kappa_{11}^{6/3}$ and show that they remain physically acceptable, and even improve, when computed to higher order.

\noindent

\let\thefootnote\relax\footnotetext{aashmore@sas.upenn.edu\\sdumitru@sas.upenn.edu\\ovrut@elcapitan.hep.upenn.edu}

%\vspace{.3in}
%\noindent
\end{abstract}

\thispagestyle{empty}
\end{titlepage}

\tableofcontents

\section{Introduction}

Within the context of heterotic M-theory \cite{Horava:1995qa,Horava:1996ma,Lukas:1997fg, Lukas:1998yy, Lukas:1998tt}, there have been a number of $N=1$ supersymmetric theories introduced that have a phenomenologically realistic observable sector \cite{Braun:2005nv,Braun:2005bw,Braun:2005ux,Bouchard:2005ag,Anderson:2009mh,Braun:2011ni,Anderson:2011ns,Anderson:2012yf,Anderson:2013xka,Nibbelink:2015ixa,Nibbelink:2015vha,Braun:2017feb}. Various aspects of such theories, such as the spontaneous breaking of their supersymmetry \cite{Lukas:1999kt,Antoniadis:1997xk,Dudas:1997cd,Lukas:1997rb,Nilles:1998sx,Choi:1997cm}, the role of five-branes in the orbifold interval \cite{Lukas:1998hk,Lehners:2006ir,Carlevaro:2005bk,Gray:2003vw,Brandle:2001ts,Lima:2001nh,Grassi:2000fk}, moduli and their stabilization \cite{Anderson:2010mh, Anderson:2011cza, Anderson:2011ty,Correia:2007sv}, their low-energy phenomenology \cite{Deen:2016vyh,Dumitru:2018jyb,Dumitru:2018nct,Dumitru:2019cgf,Ambroso:2009jd,Ovrut:2014rba,Ovrut:2015uea}, non-perturbative superpotentials~\cite{Witten:1999eg,Buchbinder:2002ic,Beasley:2003fx,Basu:2003bq,Braun:2007xh,Braun:2007tp,Braun:2007vy,Bertolini:2014dna,Buchbinder:2016rmw,Buchbinder:2019eal,Buchbinder:2019hyb,Buchbinder:2017azb,Buchbinder:2018hns,Buchbinder:2002pr,Anderson:2015yzz} and so on have been discussed in the literature.
Within the context of heterotic M-theory, it was shown that compactifying the observable sector on a specific Schoen Calabi--Yau threefold equipped with a particular holomorphic gauge bundle with structure group $SU(4)$ produces a low-energy $N=1$ supersymmetric theory with precisely the spectrum of the MSSM \cite{Braun:2005bw,Braun:2005zv,Braun:2005nv}; that is, three families of quarks and leptons with three right-handed neutrino chiral supermultiplets, one per family, and a Higgs-Higgs conjugate pair of chiral superfields. There are no vector-like pairs and no exotic fields. However, in addition to the gauge group $SU(3)_{C} \times SU(2)_{L} \times U(1)_{Y}$ of the MSSM, there is an extra gauged $U(1)_{B-L}$ group. In a series of papers \cite{Ovrut:2014rba,Ovrut:2015uea,Deen:2016vyh}, it was shown that if $N=1$ supersymmetry is softly broken at the unification scale, there exists an extensive set of initial soft breaking parameters -- dubbed ``viable black points'' -- such that, when the theory is scaled down to low energy, all phenomenological requirements are satisfied. More precisely, the $B-L$ symmetry is broken at a sufficiently high scale, the electroweak symmetry is spontaneously broken at the correct scale with the measured values for the $W^{\pm}$ and $Z^{0}$ gauge bosons, the Higgs boson mass is within three sigma of the experimentally measured value, and all sparticle masses exceed their present experimental lower bounds. Remarkably, the initial viable soft supersymmetry breaking parameters appear to be uncorrelated and require no fine-tuning. This realistic theory is referred to as the $B-L$ MSSM or the strongly coupled heterotic standard model.

However, in order to be a completely viable vacuum state, it is essential that there exists a holomorphic gauge bundle on the Schoen threefold in the hidden sector. That such a hidden sector gauge bundle can exist and be compatible with the observable sector $SU(4)$ bundle and the Bogomolov inequality was shown in \cite{Braun:2006ae}. In conjunction with the $SU(4)$ observable sector bundle, this hidden sector gauge bundle must be consistent with a number of constraints. These are~\cite{Ovrut:2018qog}: 1) the $SU(4)$ holomorphic vector bundle must be slope-stable so that its gauge connection satisfies the Hermitian Yang--Mills equations, 2) allowing for five-branes in the $S^{1}/{\mathbb{Z}}_{2}$ orbifold interval, the entire theory must be {\it anomaly free}, and 3) the squares of the unified gauge coupling parameters in both the observable and hidden sectors of the theory must be {\it positive definite}. Furthermore, the hidden sector gauge bundle should be chosen so that it is 4) slope-stable and, hence, its gauge connection satisfies the Hermitian Yang--Mills equations, and 5) it does not spontaneously break $N=1$ supersymmetry. In a previous paper \cite{Braun:2013wr}, several such hidden sector bundles, composed of both a single line bundle and a direct sum of line bundles,  were presented and proven to satisfy all of these constraints. Unfortunately, it was shown that the effective four-dimensional theory corresponded to the {\it weakly coupled} heterotic string. Hence, amongst other problems, the correct value for the observable sector unification scale and gauge coupling could not be obtained.

It is the purpose of this paper to rectify this problem. We will provide a hidden sector gauge bundle -- characterised by a line bundle $L$ that gives an induced rank-two bundle $L\oplus L^{-1}$ so as to embed properly into $E_{8}$ -- which not only satisfies all of the above ``vacuum'' constraints, but, in addition, corresponds to the {\it strongly coupled} heterotic string. We will show that there is a substantial region of K\"ahler moduli space for which a) the  $S^{1}/{\mathbb{Z}}_{2}$ orbifold length is sufficiently larger than the average Calabi--Yau radius, and b) that the effective strong coupling parameter is large enough to obtain the correct value for the observable sector $SO(10)$ unification mass and gauge coupling. We refer to these additional criteria as the ``dimensional reduction'' and ``physical'' constraints respectively. Finally, we will show, via an effective field theory analysis, that this hidden sector line bundle preserves $N=1$ supersymmetry in the effective $D=4$ field theory.

The fact that, by necessity, our results are only computed to first order, $\kappa_{11}^{4/3}$, within the context of the strongly coupled heterotic string, leads one to ask what the effect of higher-order corrections might be. To partially address this, we calculate a number of important quantities to next order, that is, order $\kappa_{11}^{6/3}$, and compare the results against our first-order computations. We find that, in all cases, by going to higher order the physical behavior of the quantities analyzed actually improves over the lower-order results. These higher-order results are computed and analyzed in detail in Appendix D.

\section{The \texorpdfstring{$B-L$}{B-L} MSSM Heterotic Standard Model}

The $B-L$ MSSM vacuum of heterotic M-theory was introduced in \cite{Braun:2005bw,Braun:2005zv,Braun:2005nv} and various aspects of the theory were discussed in detail in \cite{Marshall:2014kea,Marshall:2014cwa,Dumitru:2018jyb,Dumitru:2018nct,Dumitru:2019cgf}. This phenomenologically realistic theory is obtained as follows. First, eleven-dimensional Hořava--Witten theory \cite{Horava:1995qa,Horava:1996ma} -- which is valid to order $\kappa_{11}^{2/3}$, where $\kappa_{11}$ is the eleven-dimensional Planck constant -- is compactified on a specific Calabi--Yau threefold $X$ down to a five-dimensional $M_{4} \times S^{1}/\mathbb{{Z}}_{2}$ effective theory, with $N=1, D=5$ supersymmetry in the bulk space and $N=1, D=4$ supersymmetry on the orbifold boundaries \cite{Lukas:1998yy, Lukas:1998tt}. By construction, this five-dimensional theory is also only valid to order $\kappa_{11}^{2/3}$. A BPS double domain wall vacuum solution of this theory was then presented \cite{Lukas:1998tt}. This BPS vacuum of the five-dimensional theory -- which will be discussed in detail in Appendix D -- can, in principle, be computed to all orders as an expansion in $\kappa_{11}^{2/3}$ and used to dimensionally reduce to a four-dimensional, $N=1$ supersymmetric theory on $M_4$. However, since the five-dimensional effective theory is only defined to order $\kappa_{11}^{2/3}$, and since solving the BPS vacuum equations to higher order for the Calabi--Yau threefold associated with the $B-L$ MSSM is very difficult, it is reasonable to truncate the BPS vacuum at order $\kappa_{11}^{2/3}$ as well. Dimensionally reducing with respect to this ``linearized'' solution to the BPS equations then leads to the four-dimensional $N=1$ supersymmetric effective Lagrangian for the $B-L$ MSSM vacuum of heterotic M-theory. By construction, this four-dimensional theory is also only valid to order $\kappa_{11}^{2/3}$ -- except for several quantities, specifically the dilaton, the gauge couplings of both the observable and hidden sectors and the Fayet--Iliopoulos term associated with any $U(1)$ gauge symmetry of the hidden sector, which are well-defined to order $\kappa_{11}^{4/3}$ \cite{Lukas:1997fg, Lukas:1998tt, Lukas:1998hk, Ovrut:2018qog}. All geometric moduli are obtained by averaging the associated five-dimensional fields over the fifth dimension. 

Having discussed the generic construction of the four-dimensional effective theory, we will, in this section, simply present the basic mathematical formalism and notation required for the analysis in this paper. 

\subsection{The Calabi--Yau Threefold}

The Calabi--Yau manifold $X$ is chosen to be a torus-fibered threefold
with fundamental group $\pi_1(X)=\Z_3 \times \Z_3$. More specifically, the Calabi--Yau threefold $X$ is the fiber product of two rationally elliptic $\dd\mathbb{P}_{9}$ surfaces, that is, a self-mirror Schoen threefold \cite{MR923487,Braun:2004xv}, quotiented with respect to a freely acting $\mathbb{Z}_{3} \times \mathbb{Z}_{3}$ isometry. Its Hodge data is $h^{1,1}=h^{1,2}=3$, so there are three K\"ahler and three complex structure
moduli. The complex structure moduli will play no role in the present
paper. Relevant here is the degree-two Dolbeault cohomology group
\begin{equation}
  H^{1,1}\big(X,\C\big)=
  \Span_\C \{ \omega_1,\omega_2,\omega_3 \}  \ ,
  \label{1}
\end{equation}
where $\omega_i=\omega_{ia {\bar{b}}}$ are harmonic $(1,1)$-forms on
$X$ with the properties
\begin{equation}
  \omega_3\wedge\omega_3=0 \ ,\quad
  \omega_1\wedge\omega_3=3\,\omega_1\wedge\omega_1 \ ,\quad
  \omega_2\wedge\omega_3=3\,\omega_2\wedge\omega_2 \ .
  \label{2}
\end{equation}
Defining the intersection numbers as
\begin{equation}
  d_{ijk} = \frac{1}{v}
  \int_X \omega_i \wedge \omega_j \wedge \omega_k
   \quad i,j,k=1,2,3\ ,
  \label{3}
\end{equation}
where $v$ is a reference volume of dimension (length)$^6$,
it follows that
\begin{equation}\label{4}
  (d_{ijk}) = 
  \left(
    \begin{array}{ccc}
      (0,\tfrac13,0) & (\tfrac13,\tfrac13,1) & (0,1,0) \\
      (\tfrac13,\tfrac13,1) & (\tfrac13,0,0) & (1,0,0) \\
      (0,1,0) & (1,0,0) & (0,0,0)
    \end{array} 
  \right) \ .
\end{equation}
The $(i,j)$-th entry in the matrix corresponds to the triplet
$(d_{ijk})_{k=1,2,3}$.
The K\"ahler cone is the positive octant
\begin{equation}
  \Kcone = H^2_{+}(X,\R)
  \subset H^2(X,\R)\ .
\label{7}
\end{equation}
The K\"ahler form, defined to be $\omega_{a {\bar{b}}}=ig_{a
  {\bar{b}}}$, where $g_{a {\bar{b}}}$ is the Ricci-flat metric on $X$, can be
any element of $\Kcone$. That is, suppressing the Calabi--Yau indices,
the K\"ahler form can be expanded as
\begin{equation}
  \omega = a^i\omega_i 
  , \quad \text{where } a^i >0 \ .
\label{8}
\end{equation}
The real, positive coefficients $a^i$ are the three $(1,1)$ K\"ahler
moduli of the Calabi--Yau threefold. Here, and throughout this paper,
upper and lower $H^{1,1}$ indices are summed unless otherwise
stated. The dimensionless volume modulus is defined by
\begin{equation}
  V=\frac{1}{v} \int_X \sqrt{g}
  \label{9}
\end{equation}
and, hence, the dimensionful Calabi--Yau volume is ${\bf{V}}=vV$. Using the
definition of the K\"ahler form and the intersection numbers \eqref{3}, $V$ can be written as
\begin{equation}
  V=\frac{1}{6v}\int_X
  \omega \wedge \omega \wedge \omega=
  \frac{1}{6} d_{ijk} a^i a^j a^k \ .
  \label{10}
\end{equation}
It is sometimes useful to express the three $(1,1)$ moduli in terms of $V$ and
two additional independent moduli. This can be accomplished by
defining the scaled shape moduli
\begin{equation}
  b^i=V^{-1/3}a^i \ , \qquad i=1,2,3 \ .
  \label{11}
\end{equation}
It follows from \eqref{10} that they satisfy the constraint
\begin{equation}
d_{ijk}b^ib^jb^k=6
\label{12}
\end{equation}
and, hence, represent only two degrees of freedom.

\subsection{The Observable Sector Bundle}

On the observable orbifold plane, the vector bundle $\Vvis$ on $X$
is chosen to be a specific holomorphic bundle with structure group $SU(4)\subset E_8$. This bundle was discussed in detail in \cite{Braun:2005nv,Braun:2005bw,Gomez:2005ii,Braun:2006ae}. Here we will present only those properties of this bundle relevant to the present paper. First of all, in order to preserve $N=1$ supersymmetry in the low-energy four-dimensional effective theory on $M_{4}$, this bundle must be both slope-stable and have vanishing slope~\cite{Braun:2005zv,Braun:2006ae}. Recall that the slope of any
bundle or sub-bundle $\cal{F}$ is defined as
\begin{equation}
  \mu({\cal{F}})=
  \frac{1}{\rank({\cal{F}})v^{2/3}} 
  \int_X{c_1(\cal{F})\wedge \omega \wedge \omega} \ ,
  \label{50}
\end{equation}
where $\omega$ is the K\"ahler form in \eqref{8}. Since the first Chern class $c_1$ of any $SU(N)$ bundle must vanish, it follows immediately that  $\mu(\Vvis)=0$, as required.
However, demonstrating that our chosen bundle is slope-stable is non-trivial and was proven in detail in several papers \cite{Braun:2005nv,Braun:2005bw,Gomez:2005ii}. The $SU(4)$ vector bundle will indeed be slope-stable in a restricted, but large, region of the positive K\"ahler cone. As proven in detail in~\cite{Braun:2006ae}, this
will be the case in a subspace of the K\"ahler cone defined by seven
inequalities. In this region, all sub-bundles of $V^{(1)}$ will have negative
slope. These can be slightly simplified into the statement that the moduli $a^{i}$, $i=1,2,3$, must satisfy at least one of the two inequalities
\begin{equation}\label{51}
  \begin{gathered}
    \left(
      a^1
      < 
      a^2
      \leq 
      \sqrt{\tfrac{5}{2}} a^1
      \quad\text{and}\quad
      a^3
      <
      \frac{
        -(a^1)^2-3 a^1 a^2+ (a^2)^2
      }{
        6 a^1-6 a^2
      } 
    \right)
    \quad\text{or}\\
    \left(
      \sqrt{\tfrac{5}{2}} a^1
      <
      a^2
      <
      2 a^1
      \quad\text{and}\quad
      \frac{
        2(a^2)^2-5 (a^1)^2
      }{
        30 a^1-12 a^2
      }
      <
      a^3
      <
      \frac{
        -(a^1)^2-3 a^1 a^2+ (a^2)^2
      }{
        6 a^1-6 a^2
      }
    \right) \ .
  \end{gathered}
\end{equation}
The subspace $\Kcone^s$ satisfying \eqref{51} is a full-dimensional
subcone of the K\"ahler cone $\Kcone$ defined in \eqref{7}. 
It is a cone because the inequalities are homogeneous. In other words, only the angular part of the K\"ahler moduli (the $b^i$) are constrained, but not the overall volume. 
\begin{figure}[htbp]
  \centering
  \includegraphics[width=0.9\textwidth]{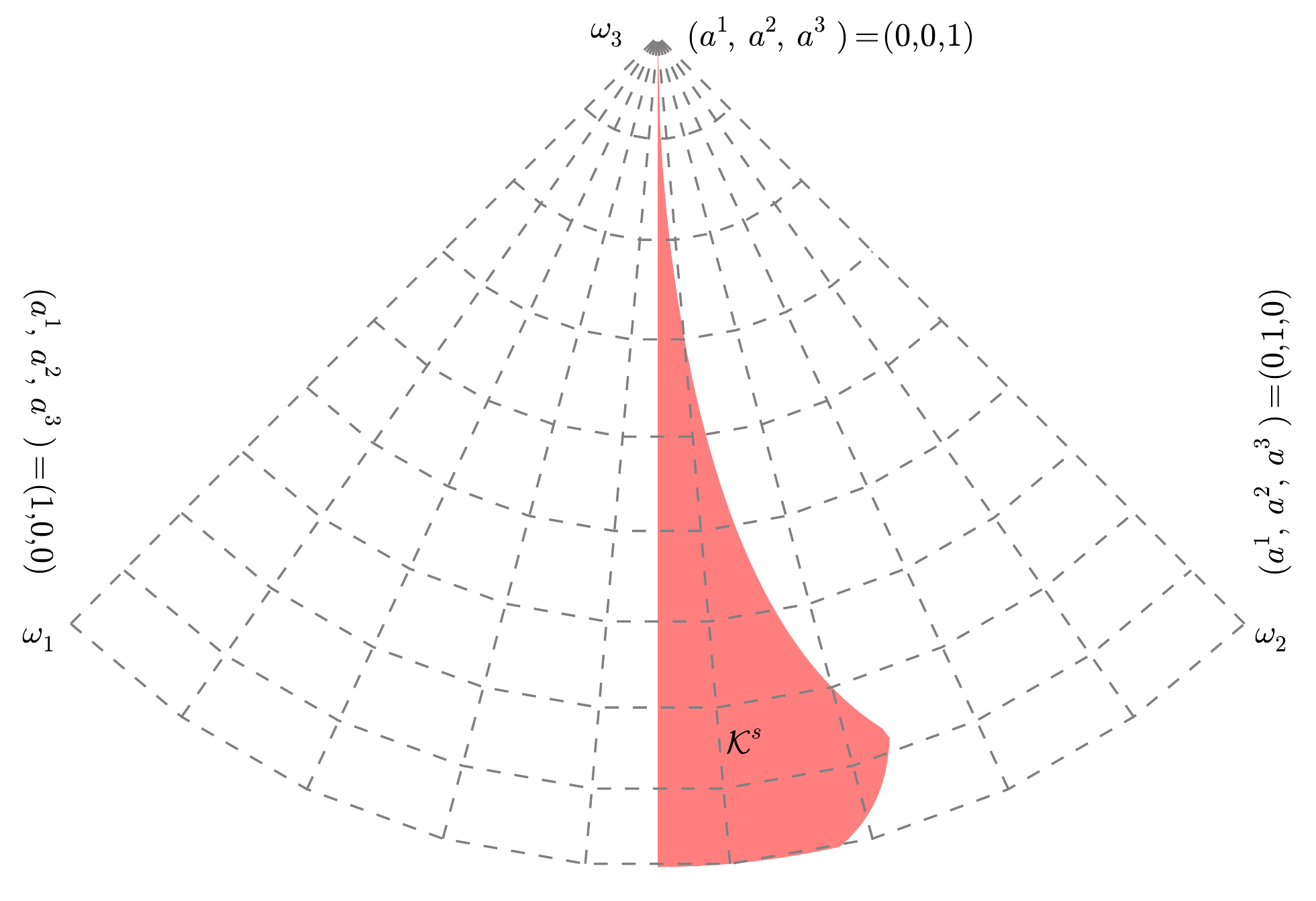}
  \caption{The observable sector stability region in the K\"ahler cone.}
  \label{fig:starmap}
\end{figure}
Hence, it is best displayed as a two-dimensional ``star map'' as seen by an
observer at the origin. This is shown in Figure 1. For
K\"ahler moduli restricted to this subcone, the four-dimensional low-energy theory in the observable sector is $N=1$ supersymmetric.

Having discussed that our specific $SU(4)$ holomorphic vector bundle preserves four-dimensional $N=1$ supersymmetry, let us examine the physical content of the effective theory on $M_{4}$. To begin with, $SU(4) \times  Spin(10)$ is a maximal-rank subgroup of $E_{8}$.
Hence, $SU(4)$ breaks the $E_{8}$ group to
\begin{equation}
  E_8 \to Spin(10) \ .
  \label{13}
\end{equation}
However, to proceed further, one must break this $Spin(10)$ ``grand unified'' group down to the gauge group of the MSSM. This is accomplished by turning on
two \emph{flat} Wilson lines, each associated with a
different $\Z_3$ factor of the $\Z_3 \times \Z_3$ holonomy of $X$. Doing this preserves the $N=1$ supersymmetry of the effective theory, but breaks the 
observable gauge group down to
\begin{equation}
  Spin(10) 
  \to 
  SU(3)_C \times SU(2)_L \times U(1)_Y \times U(1)_{B-L} \ .
  \label{17}
\end{equation}
As discussed in Section 5 below, the mass scale associated with the Wilson lines can be approximately the same, or separated by up to one order of magnitude. Be that as it may, for energies below the lightest Wilson line mass, the particle spectrum of the $B-L$ MSSM is exactly that of the MSSM; that is, three families of quarks and leptons, including three right-handed neutrino chiral supermultiplets -- one per family -- and exactly one pair of Higgs-Higgs conjugate chiral superfields. There are no vector-like pairs of particles and no exotics of any kind. It follows from \eqref{17} however, that the gauge group is that of the MSSM plus an additional gauged $U(1)$ associated with the $B-L$ quantum numbers. The physics of this additional gauge symmetry -- which is broken far above the electroweak scale -- is discussed in detail in a number of papers \cite{Deen:2016vyh,Ambroso:2009jd,Ovrut:2012wg,Ovrut:2014rba,Ovrut:2015uea,Barger:2008wn,FileviezPerez:2009gr,FileviezPerez:2012mj} and is phenomenologically acceptable.

\subsection{The Hidden Sector Bundle}

In \cite{Ovrut:2018qog}, the hidden-sector vector bundle was chosen to have the generic form of a Whitney sum
\begin{equation}
V^{(2)}={\cal{V}}_{N} \oplus {\cal{L}}\ , \qquad {\cal{L}}=\bigoplus_{r=1}^R L_r \ ,
\label{dude1}
\end{equation}
where ${\cal{V}}_{N}$ is a slope-stable, non-abelian bundle and 
each $L_{r}$, $r=1,\dots,R$, is a holomorphic line bundle with  structure group $U(1)$. In Appendix A below, a careful analysis is given of more restrictive vector bundles, consisting of the Whitney sum of line bundles only; that is
\begin{equation}
\qquad {V^{(2)}}={\cal{L}}=\bigoplus_{r=1}^R L_r \ .
\label{dude1aA}
\end{equation}
This is presented to set the context for future work involving hidden sectors with several line bundles.
However, in this present paper, we will further simplify the hidden sector vector bundle by requiring it to be defined by a {\it single} holomorphic line bundle $L$, in such a way that its $U(1)$ structure group embeds into $E_{8}$. 
%That is, we will take
%
%\begin{equation}
%\qquad {V^{(2)}}=L \ .
%\label{dude1cA}
%\end{equation}
%
It follows from the discussion in Appendix A that a line bundle $L$ is associated with a divisor of $X$ and is conventionally expressed as 
\begin{equation}
  L=\Ocal_X(l^1, l^2, l^3)  \ ,
\end{equation}
where the $l^i$ are integers satisfying the condition
\begin{equation}
  (l^1+l^2) \op{mod} 3 = 0 \ .
  \label{22}
\end{equation}
This additional constraint is imposed in order for these bundles to arise from $\Z_3 \times
\Z_3$ equivariant line bundles on the covering space of $X$. The structure group of  $L$ is $U(1)$. However, there are many distinct ways in which this $U(1)$ subgroup can be embedded into the hidden-sector $E_{8}$ group. The choice of embedding determines two important properties of the effective low-energy theory. First, a specific embedding will define a commutant subgroup of $\Ex8$, which appears as the symmetry group for the four-dimensional effective theory.  Second, the explicit choice of embedding will determine a real numerical constant
\begin{equation}
  a=\tfrac{1}{4} \tr _{E_{8}}Q^2 \ ,
  \label{26A}
\end{equation}
where $Q$ is the generator of the $U(1)$ factor embedded in the $\Rep{248}$ adjoint representation of the hidden sector $E_{8}$, and the trace $\tr$ includes a factor of $1/30$.
This coefficient will enter several of the consistency conditions, such as the anomaly cancellation equation, required for an acceptable vacuum solution.

\subsection{Bulk Space Five-Branes}

In strongly coupled heterotic M-theory, there is a one-dimensional interval $S^{1}/{\mathbb{Z}}_{2}$ separating the observable and hidden orbifold planes. Denoting by $\rho$ an arbitrary reference radius of $S^{1}$, the reference length of this one-dimensional interval is given by $\pi \rho$. A real coordinate on this interval is written as $x^{11} \in [0, \pi \rho]$. As discussed in Appendix A, arbitrary dimensionless functions on $M_{4} \times S^{1}/{\mathbb{Z}}_{2}$ can be averaged over this interval, leading to moduli that are purely functions on $M_{4}$. Averaging the $b$ function in the five-dimensional metric,
$\dd s_{5}^{2} = \dots +b^{2}(\dd x^{11})^{2}$, defines a four-dimensional modulus
\begin{equation}
\frac{\Rhat}{2}=\langle   b \rangle _{11} \ .
\label{case1}
\end{equation}
The physical length of this orbifold interval is then given by $\pi \rho \Rhat$. It is convenient to define  a new coordinate $z$ by $z=\frac{x^{11}}{\pi \rho}$, which runs over the interval $z \in [0,1]$.

In addition to the holomorphic vector bundles on the observable and
hidden orbifold planes, the bulk space between these planes can
contain five-branes wrapped on two-cycles ${\cal{C}}_2^{(n)}$,
$n=1,\dots,N$ in $X$. Cohomologically, each such five-brane is
described by the $(2,2)$-form Poincar\'e dual to ${\cal C}_2^{(n)}$,
which we denote by $W^{(n)}$. Note that to preserve $N=1$
supersymmetry in the four-dimensional theory, these curves must be
holomorphic and, hence, each $W^{(n)}$ is an effective class. In Appendix A, we present the formalism associated with an arbitrary number $N$ of such five-branes.
However, in the main text of this paper, we will consider only a {\it single} five-brane. We denote its location in the bulk space by $z_{1}$, where 
$z_{1} \in [0,1]$. When convenient, we will re-express this five-brane location in terms of the parameter $\lambda=z_{1}-\frac{1}{2}$, where $\lambda \in [-\frac{1}{2},\frac{1}{2}]$.

\section{The Vacuum Constraints}

There are three fundamental constraints that any consistent vacuum state of the $B-L$ MSSM must satisfy. These are the following.

\subsection{The SU(4) Slope Stability Constraint}
The $SU(4)$ holomorphic vector bundle discussed in subsection 2.2 must be slope-stable so that its gauge connection satisfies the Hermitian Yang--Mills equations. As presented in \eqref{51}, this constrains the allowed region of K\"ahler moduli space to be
\begin{equation}\label{51A}
  \begin{gathered}
    \left(
      a^1
      < 
      a^2
      \leq 
      \sqrt{\tfrac{5}{2}} a^1
      \quad\text{and}\quad
      a^3
      <
      \frac{
        -(a^1)^2-3 a^1 a^2+ (a^2)^2
      }{
        6 a^1-6 a^2
      } 
    \right)
    \quad\text{or}\\
    \left(
      \sqrt{\tfrac{5}{2}} a^1
      <
      a^2
      <
      2 a^1
      \quad\text{and}\quad
      \frac{
        2(a^2)^2-5 (a^1)^2
      }{
        30 a^1-12 a^2
      }
      <
      a^3
      <
      \frac{
        -(a^1)^2-3 a^1 a^2+ (a^2)^2
      }{
        6 a^1-6 a^2
      }
    \right) \ .
  \end{gathered}
\end{equation}
This constraint depends entirely on the phenomenologically acceptable non-abelian vector bundle in the observable sector. 

However, there are two remaining fundamental constraints that strongly depend on the choice of the hidden sector bundle and on the number of bulk space five branes.
These two constraints, which are required for any consistent $B-L$ MSSM vacuum, were discussed in general in \cite{Ovrut:2018qog}, and are presented in Appendix A for heterotic vacua for any number of bulk space five-branes and in which the hidden-sector gauge bundle consists of a Whitney sum of line bundles. In the text of this paper, however, we will limit our analysis to hidden-sector vacua constructed from a {\it single} line bundle $L$ only and to the case of a {\it single} five-brane. Under these restrictions, the fundamental vacuum constraints given in \cite{Ovrut:2018qog} and Appendix A simplify to the following conditions.

\subsection{Anomaly Cancellation Constraint}

In \eqref{29} of Appendix A, the condition for anomaly cancellation between the observable sector, a hidden sector composed of the Whitney sum of line bundles and an arbitrary number of bulk space five-branes is presented. Restricting this to a single hidden-sector line bundle $L$, a single bulk-space five-brane and using the formalism presented in that Appendix, the anomaly cancellation equation can be simplified and then rewritten in the form
\begin{equation}
  W_i= \bigl( \tfrac{4}{3},\tfrac{7}{3},-4\big)\big|_i
  + a \, d_{ijk} l^j l^k \geq 0 \  \qquad i=1,2,3  \ ,
\label{33A}
\end{equation}
where the coefficient $a$ is defined in \eqref{26A}. The positivity constraint on $W$ follows from the requirement that the five-brane wraps an effective class to preserve $N=1$ supersymmetry.

\subsection{Gauge Coupling Constraints}

The general expressions for the square of the unified gauge couplings in both the observable and hidden sectors -- that is, ${4\pi}/{(g^{(1)})^{2}}$ and ${4\pi}/{(g^{(2)})^{2}}$ respectively -- were presented in \cite{Ovrut:2018qog}. In Appendix A, these are discussed within the context of a hidden-sector bundle \eqref{dude1aA} consisting of the Whitney sum of line bundles, as well as an arbitrary number five-branes in the bulk-space interval. Here, we restrict those results to the case of a hidden-sector bundle constructed from of a single line bundle $L$ and a single five-brane located at $\lambda= z_{1}-\frac{1}{2} \in \left[-\tfrac{1}{2},\tfrac{1}{2}\right]$. The charges $\beta_i^{(0)}$ and $\beta_i^{(1)}$, and the constant coefficient $\epsilon'_S$ are discussed in Appendix A and given by 
\begin{equation}
\beta_i^{(0)}=\left(\tfrac{2}{3},-\tfrac{1}{3},4\right)_{i} \ , \qquad \beta_i^{(1)}=W_{i}  \ ,
\label{ruler1}
\end{equation}
and 
\begin{equation}
  \epsilon'_S = \pi \epsilon_{S} \ , \qquad
  \epsilon_{S}= \left(\frac{\kappa_{11}}{4\pi} \right)^{2/3}\frac{2\pi\rho}{v^{2/3}} \ .
  \label{40AA}
\end{equation}
The parameters $v$ and $\rho$ are defined above and $\kappa_{11}$ is the eleven-dimensional  Planck constant. Written in terms of the K\"ahler moduli $a^{i}$ using \eqref{11}, 
the constraints that $(g^{(1)})^{2}$ and $(g^{(2)})^{2}$ be positive definite are then given by
\begin{equation}
  \label{68AA}
  \begin{split}
    d_{ijk} a^i a^j a^k- 3 \epsilon_S' \frac{\Rhat}{V^{1/3}} \bigl(
    -(\tfrac{8}{3} a^1 + \tfrac{5}{3} a^2 + 4 a^3)
    \qquad &\\
    + 2(a^1+a^2) -(\tfrac{1}{2}-\lambda)^2 a^i \,{W}_i
    \bigr) &> 0 \ ,
  \end{split}
\end{equation}
and
\begin{equation}
  \label{69AA}
  \begin{split}
    d_{ijk} a^i a^j a^k- 3 \epsilon_S' \frac{\Rhat}{V^{1/3}}
    \bigl(a\,d_{ijk}a^i l^j l^k
    \qquad &\\
    + 2(a^1+a^2) -(\tfrac{1}{2}+\lambda)^2 a^i
    \,{W}_i \bigr) &> 0 \ ,
  \end{split}
\end{equation}
respectively. The Calabi--Yau volume modulus $V$ is defined in terms of the $a^{i}$ moduli in \eqref{10}, and $\Rhat$ is the independent $S^{1}/{\mathbb{Z}}_{2}$ length modulus defined in \eqref{case1}. Note that the coefficient $a$ defined in \eqref{26A} enters both expressions via the five-brane class $ {W}_i$ and independently occurs in the second term of \eqref{69AA}.

\section{A Solution of the \texorpdfstring{$B-L$}{B-L} MSSM Vacuum Constraints}\label{sec:constraints}
  
In this section, we will present a simultaneous solution to all of the $B-L$ MSSM vacuum constraints listed above -- namely: 1) the slope-stability conditions given in \eqref{51A} for the $SU(4)$ observable sector gauge bundle, 2) the anomaly cancellation condition with an effective five-brane class presented in \eqref{33A},  and, finally, 3) the conditions for positive squared gauge couplings in both the observable and hidden sectors, presented in \eqref{68AA} and \eqref{69AA}. The slope-stability conditions for the $SU(4)$ observable sector gauge bundle is independent of the choice of the hidden-sector gauge bundle and any bulk-space five-branes. However, the remaining constraints depend strongly upon the specific choice of the line bundle $L$ in the hidden sector, its exact embedding in the hidden sector $E_{8}$ gauge group and, finally, on the location $\lambda$ and the effective class of the five-brane in the $S^{1}/{\mathbb{Z}}_{2}$ interval. For this reason, we first consider the $SU(4)$ slope-stability conditions.

\subsection{The \texorpdfstring{$SU(4)$}{SU(4)} Slope-Stability Solution}

The region of K\"ahler moduli space satisfying the slope-stability conditions \eqref{51A} was shown to be a three-dimensional subcone of the positive octant of the the full K\"ahler cone.
This subcone was displayed above as a two-dimensional ``star map'' in Figure 1. Here, to be consistent with the solution and graphical display of the remaining sets of constraints, 
we present a portion of the solution space of slope-stability conditions \eqref{51A} as a three-dimensional figure in a positive region of K\"ahler moduli space -- restricted, for specificity, to $ 0 \leq a^i \leq 10~{\rm for}~ i=1,2,3$. This is shown in Figure 2.
\begin{figure}[t]
   \centering
\includegraphics[width=0.4\textwidth]{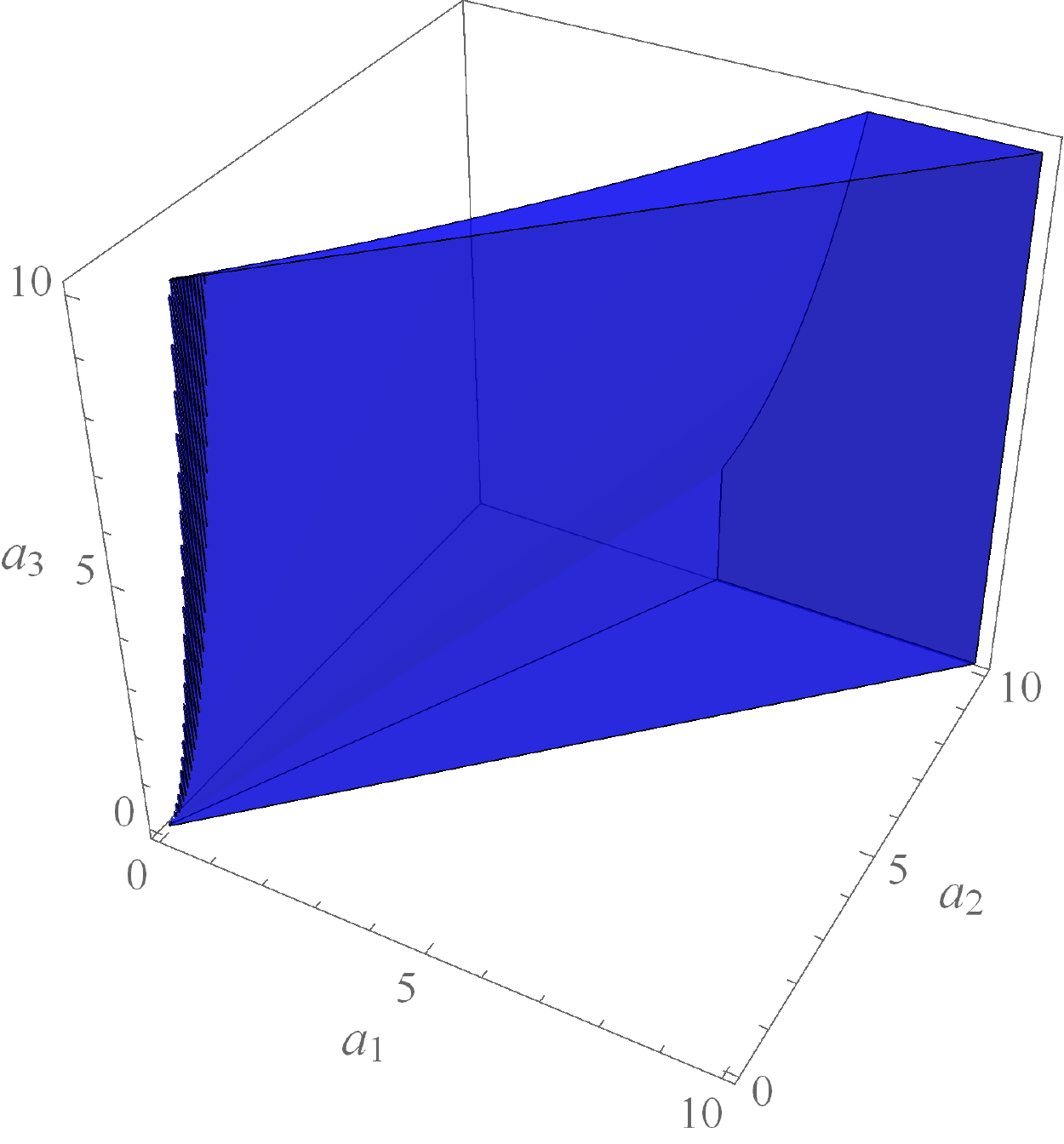}
\caption{The region of slope-stability for the $SU(4)$ observable-sector bundle, restricted to $ 0 \leq a^i \leq 10~{\rm for}~ i=1,2,3$.}
\label{fig:SU4}
\end{figure}
Before continuing, we note that the slope-stability constraint regions in \eqref{51A} are invariant under scaling $a^{i}\to \mu a^{i}$, where $\mu$ is a positive real number. Figure 2 includes \emph{all} K\"ahler moduli in the restricted region satisfying the $SU(4)$ slope-stability constraints.

\subsection{An Anomaly Cancellation Solution}\label{sec:anomaly_cancellation}

Unlike the $SU(4)$ slope-stability constraints, the condition \eqref{33A} for anomaly cancellation depends on the explicit choice of the hidden sector line bundle $L$, as well as on the parameter $a$ defined in \eqref{26A}. Hence, one must specify the exact embedding of this line bundle into the hidden sector $E_{8}$ gauge group. Here, for specificity, we will choose the line bundle to be
\begin{equation}
  L=\Ocal_X(2, 1, 3)  \ .
  \label{red1}
\end{equation}
Note that each entry is an integer and that $l^1=2$ and $l^2=1$ satisfy the equivariance condition \eqref{22}, as they must. The reason for this choice of line bundle, and the presentation of several other line bundles that lead to acceptable results, will be discussed below. Here, we focus exclusively on line bundle \eqref{red1}. Generically, there are numerous distinct embeddings of an given arbitrary line bundle into an $E_{8}$ gauge group, each with its own commutant subgroup and $a$ parameter. In this section, to be concrete, we will choose a particular embedding of $L=\Ocal_X(2, 1, 3)$ into $E_{8}$ and, having done so, explicitly calculate its $a$ parameter. 

The explicit embedding of $L$ into $E_8$ is chosen as follows. First, recall that
\begin{equation}
  SU(2) \times E_7 \subset E_8
  \label{red2}
\end{equation}
is a maximal subgroup. With respect to $SU(2) \times E_7$, the
$\Rep{248}$ representation of $E_8$ decomposes as
\begin{equation}
  \Rep{248} \to 
  (\Rep{1}, \Rep{133}) \oplus (\Rep{2}, \Rep{56}) \oplus (\Rep{3}, \Rep{1})\  .
  \label{red3}
\end{equation}
Now choose the generator of the $U(1)$ structure group of $L$ in the fundamental representation of $SU(2)$ to be $(1,-1)$. It follows that
under $SU(2) \rightarrow U(1)$ 
\begin{equation}
  \Rep{2} \to 1 \oplus -1  \ ,
  \label{red4}
\end{equation}
and, hence, under $U(1) \times E_7$
\begin{equation}
  \Rep{248} \to 
  (0, \Rep{133}) \oplus 
  \bigl( (1, \Rep{56}) \oplus (-1, \Rep{56})\bigr) \oplus 
  \bigl( (2, \Rep{1}) \oplus (0, \Rep{1}) \oplus (-2, \Rep{1}) \bigr) .
\label{red5}
\end{equation}
The generator $Q$ of this embedding of the line bundle $L$ can be read off
from expression \eqref{red5}. Inserting this into \eqref{26A}, we find that
\begin{equation}
  a=1 .
  \label{red6}
\end{equation}
We note in passing that the four-dimensional effective theory associated with choosing this explicit embedding has gauge symmetry
\begin{equation}
H=E_{7} \times U(1) \ ,
\label{red7}
\end{equation}
where the second factor is an ``anomalous''  $U(1)$. It is identical to the structure group of $L$ and arises in the low-energy theory since $U(1)$ commutes with itself. This will be discussed in detail later in this paper.

An important consequence of the explicit embedding \eqref{red2}, \eqref{red4} and, hence, \eqref{red5} is the following. To begin with, we note that $ L=\Ocal_X(2, 1, 3)$ is, indeed, a sub-line bundle of the hidden sector $E_{8}$ gauge group. However, ``embedding'' this line bundle into $E_{8}$ means that the single gauge connection associated with the $U(1)$ structure group of $L$ must also be a subset of the $\bf 248$ indexed non-abelian connection of $E_{8}$. Since the slope of this $E_{8}$ representation vanishes, it follows that the slope of the line bundle $L$ must also vanish. Generically, however, this will not be the case. It follows from \eqref{50} and \eqref{23} that the slope of $L$ is proportional to its first Chern class $c_{1}(L)=\frac{1}{v^{1/3}}(2\omega_{1}+\omega_{2}+3\omega_{3})$ and, hence, its slope does not vanish anywhere in K\"ahler moduli space. Therefore, to ``embed'' $L$ into $E_{8}$ as specified by \eqref{red4}, it is necessary to extend the hidden sector gauge bundle to the ``induced'' rank 2 bundle
\begin{equation}
\mathcal{V}=L \oplus L^{-1} \ .
\label{ind}
\end{equation}
The first Chern class of this induced bundle necessarily vanishes and, hence, the associated abelian connection can be appropriately embedded into the hidden sector {\bf 248}-valued $E_{8}$ gauge connection. We want to emphasize that this induced line bundle was implicitly used in both the anomaly constraint \eqref{33A} and the gauge coupling constraints \eqref{68AA} and \eqref{69AA} since the parameter $a=1$ was computed using the generator $Q$ of $L \oplus L^{-1}$ derived from \eqref{red5}.

Having discussed this in detail, let us now insert $L=\Ocal_X(2, 1, 3)$ and $a=1$ into the anomaly cancellation constraint \eqref{33A}. Using \eqref{4}, we find that 
\begin{equation}
W_{i}=(9 ,17 , 0)|_{i} \geq 0 \quad\text{for each } i=1,2,3  \ .
\label{red8}
\end{equation}
Hence, the anomaly cancellation condition is satisfied. 

\subsection{Moduli Scaling: Simplified Gauge Parameter Constraints}

Before presenting solutions to the gauge coupling positivity constraints \eqref{68AA} and \eqref{69AA}, we observe the following important fact. Note, using expression \eqref{10} for the volume modulus $V$, that both of these constraints remain invariant under the scaling
\begin{equation}
a^{i} \to \mu a^{i} \ ,\qquad  \epsilon'_{S}\Rhat \to \mu^{3} \epsilon'_{S}\Rhat \ ,
\label{wall1}
\end{equation}
where $\mu$ is any positive real number. It follows that the coefficient $\epsilon_S'\Rhat/V^{1/3}$ in front of the $\kappa_{11}^{4/3}$ terms in each of the two constraint equations can be set to unity by choosing the appropriate constant $\mu$; that is
\begin{equation}
\epsilon_S'\frac{\Rhat}{V^{1/3}} \to 1 \ .
\label{wall2}
\end{equation}
We will refer to this choice of $\epsilon_S'\Rhat/V^{1/3}=1$ as the ``unity'' gauge. Working in unity gauge, the gauge coupling positivity constraints \eqref{68AA} and \eqref{69AA} simplify to %
\begin{equation}
  \label{68A}
  \begin{split}
    d_{ijk} a^i a^j a^k- 3\bigl(
    -(\tfrac83 a^1 + \tfrac53 a^2 + 4 a^3)
    + \qquad& \\
    + 2(a^1+a^2) -(\tfrac{1}{2}-\lambda)^2 a^i \,{W}_i 
    \bigr) &> 0  \ ,
  \end{split}
\end{equation}
and
\begin{equation}
  \label{69A}
  \begin{split}
    d_{ijk} a^i a^j a^k- 3
    \bigl(a\,d_{ijk}a^i l^j l^k
    + \qquad& \\
    + 2(a^1+a^2) -(\tfrac{1}{2}+\lambda)^2 a^i
    \,{W}_i \bigr) &> 0 \ .
  \end{split}
\end{equation}
In the following subsections, we will solve these constraints in unity gauge. Before doing so, however, we wish to emphasize again that the $SU(4)$ slope-stability constraints discussed above are also invariant under $a^{i} \to \mu a^{i}$ scaling and, hence, the results shown in Figure 2 remain unchanged. Of course, to be consistent with the solution of the anomaly cancellation constraint presented in \eqref{red8}, we will solve the gauge coupling positivity constraints for 1) the explicit choice of line bundle $L=\Ocal_X(2, 1, 3)$, 2) the explicit embedding \eqref{red5} and 3) the associated embedding parameter $a=1$. In this specific case, the constraints \eqref{68A} and \eqref{69A} become
\begin{equation}
  \begin{split}
	({a^1})^2a^2+a^1({a^2})^2+6a^1a^2a^3&+2a^1-a^2+12a^3+\\
	&+3\left(\frac{1}{2}-\lambda \right)^2(9a^1+17a^2)>0
  \end{split}
\label{clip1}
\end{equation}
and 
\begin{equation}
  \begin{split}
	({a^1})^2a^2+a^1({a^2})^2+6a^1a^2a^3&-29a^1-50a^2-12a^3+\\
	&+3\left(\frac{1}{2}+\lambda \right)^2(9a^1+17a^2)>0
  \end{split}
\label{clip2}
\end{equation}
where we have used \eqref{4}.

\subsection{Five-Brane Location}

It is also clear from the gauge coupling constraints \eqref{clip1} and \eqref{clip2} that, even in unity gauge, it is necessary to explicitly fix the location of the bulk space five-brane by choosing its location parameter $\lambda$. As can be seen in \eqref{clip2}, the condition $(g^{(2)})^2>0$ is most easily satisfied when the value of $\lambda$ is as large as possible; that is, for the five-brane to be near the hidden wall.
For concreteness, we will take 
\begin{equation}
\lambda=0.49 \ .
\label{cup1}
\end{equation}
Note that we do not simply set $\lambda=\frac{1}{2}$, so as to avoid unwanted ``small instanton'' transitions of the hidden sector~\cite{Ovrut:2000qi}; that is, to keep the five-brane as an independent entity.

\subsection{Gauge Couplings Solution}

In unity gauge, choosing $L=\Ocal_X(2, 1, 3)$  with $a=1$, and \eqref{red8} and \eqref{cup1}, one can solve the positive gauge coupling constraints \eqref{clip1} and \eqref{clip2} simultaneously. The results are presented in Figure 3, again restricted to the region $ 0 \leq a^i \leq 10~{\rm for}~ i=1,2,3$ of K\"ahler moduli space. 
\begin{figure}
\centering
\includegraphics[width=0.4\textwidth]{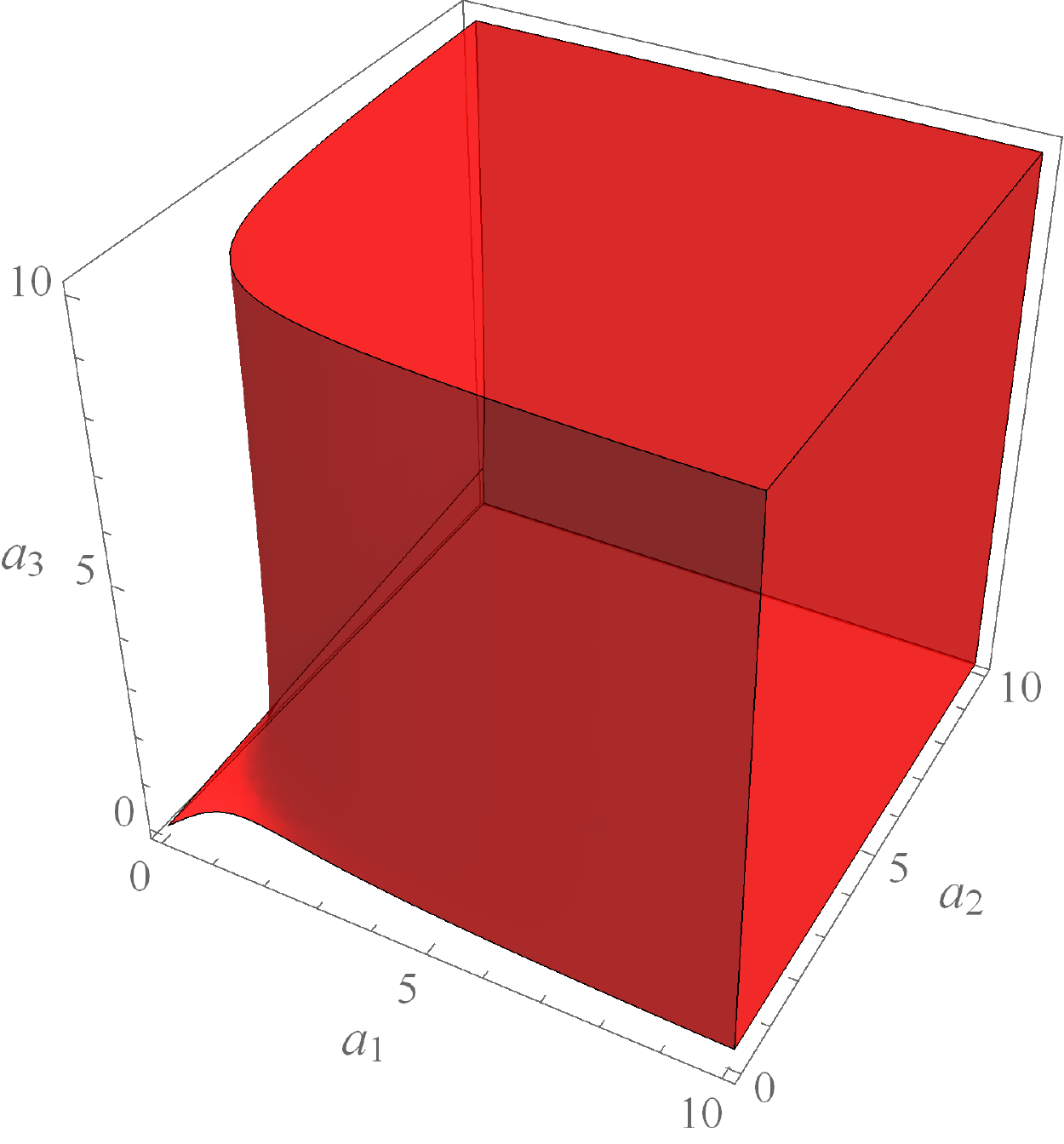}
\caption{Simultaneous solution to both $(g^{(1)})^2 >0$ and $(g^{(2)})^2>0$ gauge coupling constraints \eqref{clip1} and \eqref{clip2} in unity gauge with $\lambda=0.49$, restricted to the region $ 0 \leq a^i \leq 10$ for $i=1,2,3$.}
\label{fig:pos_couplings}
\end{figure}

\subsection{The Simultaneous Solution of All Required Constraints}

Intersecting the results of the previous subsections, we can now present a simultaneous solution to all of the $B-L$ MSSM vacuum constraints listed above -- that is, 1) the solution for the $SU(4)$ slope-stability conditions given in Figure 2, 2) the solution for the anomaly cancellation condition with an effective five-brane class presented in \eqref{red8}  and 3) the conditions for positive squared gauge couplings in both the observable and hidden sectors shown in Figure 3. We reiterate that this is a very specific solution to these constraints, with the hidden sector line bundle chosen to be $L=\Ocal_X(2, 1, 3)$, with its specific embedding into the hidden sector $E_{8}$ given in \eqref{red5} leading to parameter $a=1$  and, finally, the location of the five-brane fixed at $\lambda=0.49$. The intersecting region of K\"ahler moduli space satisfying all of these constraints computed in unity gauge is shown in Figure 4.
\begin{figure}[ht]
   \centering
\includegraphics[width=0.4\textwidth]{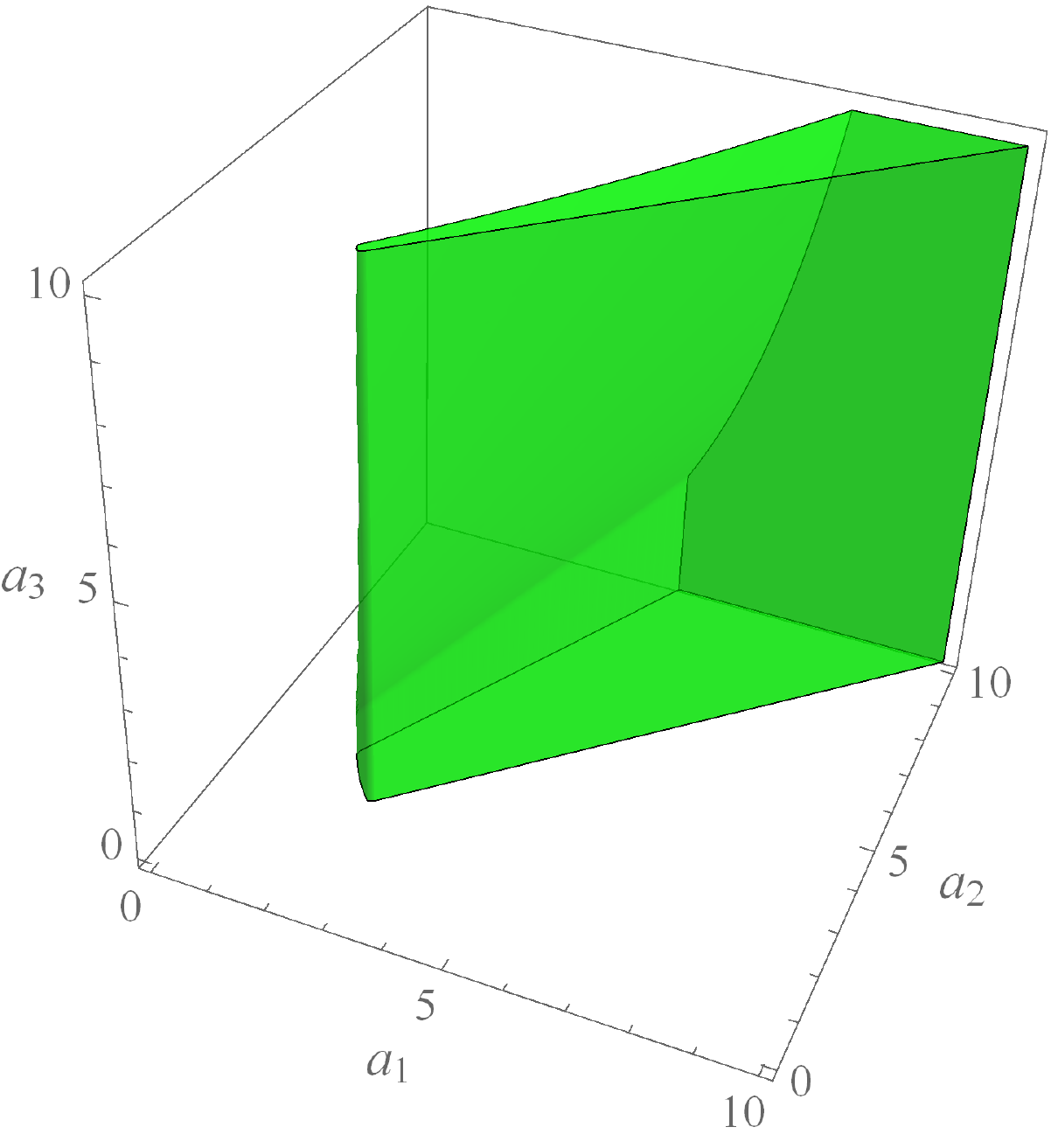}
\caption{ The region of K\"ahler moduli space where the $SU(4)$ slope-stability conditions, the anomaly cancellation constraint, and the positive squared gauge coupling constraints with $\lambda=0.49$ are simultaneously satisfied in unity gauge, restricted to $ 0 \leq a^i \leq 10$ for $i=1,2,3$. This amounts to the intersection of Figures 2 and 3.}
\label{fig:all_constr}
\end{figure}

To conclude, we must emphasize a subtle but important point. The use of unity gauge, defined by \eqref{wall2}, disguises the fact that the constraint equations actually contain the expression $\epsilon'_{S} \Rhat$, where $\epsilon'_{S}$ is a coupling parameter and $\Rhat$ is the modulus for the $S^{1}/{\mathbb{Z}}_{2}$ orbifold interval. Under the scaling $a^{i} \rightarrow \mu a^{i}$ of the K\"ahler moduli, it is the {\it product} $\epsilon'_{S} \Rhat$ that scales as $\epsilon'_{S} \Rhat \rightarrow \mu^{3} \epsilon'_{S} \Rhat$. However, the invariance of the constraint equations under this scaling does {\it not} specify the scaling behavior of either $\epsilon'_{S}$ or $\Rhat$ individually -- only their product. It follows that, in principle, $\epsilon'_{S} \rightarrow \mu^{A}\epsilon'_{S}$ and $ \Rhat \rightarrow \mu^{B} \Rhat$ for any values of $A$ and $B$ as long as $A+B=3$.  We will therefore interpret Figure 4 to be such that: 1) although every point $a^i$, $i=1,2,3$ contained in it satisfies all vacuum constraints, a given point can depend arbitrarily on any value of $\epsilon'_{S}$, 2) under $\mu$ scaling, all that is required is that $\epsilon'_{S} \Rhat \rightarrow \mu^{3} \epsilon'_{S} \Rhat$, but the degree of scaling of $\epsilon'_{S}$ and 
$\Rhat$ individually remains undetermined. This interpretation will become important below, when we impose additional constraints in the following two sections.

\section{Dimensional Reduction and Physical Constraints}

In addition to the three ``vacuum'' constraints discussed above -- and solved in unity gauge for a specific choice of line bundle, line bundle embedding and five-brane location -- there are three additional conditions that must be satisfied for the $B-L$ MSSM theory to be viable. These can be broken into two categories. First, there is a new ``reduction'' constraint required for the consistency of the $d=11$ to $d=5$ heterotic M-theory dimensional reduction. Second, there are two purely ``phenomenological'' constraints. They are that the 
$Spin(10)$ grand unification scale, $M_{U}$, and the associated unified gauge coupling, $\alpha_{u}$, in the observable sector be consistent with the phenomenologically viable values for these quantities~\cite{Deen:2016vyh,Ovrut:2015uea,Ovrut:2012wg}.

\subsection{The Reduction Constraint}

We begin by discussing the constraint required for a consistent dimensional reduction on a Calabi--Yau threefold $X$ from the $d=11$ Hořava--Witten orbifold to five-dimensional heterotic M-theory. In order for this reduction to be viable, the averaged Calabi--Yau radius must, when calculated using the eleven-dimensional M-theory metric, be sufficiently smaller than the physical length of the $S^{1}/ \mathbb{Z}_{2}$ interval. That is, one must have
\begin{equation}
\frac{\pi \rho {\Rhat} V^{-1/3}}{(vV)^{1/6}} > 1 \ ,
\label{sun1}
\end{equation}
where the constant parameters $v$ and $\rho$ were introduced above and the moduli $V$ and $\Rhat$ are defined in \eqref{10} and \eqref{case1} respectively. The extra factor of $V^{-1/3}$ in the numerator arises because the $S^{1}/\mathbb{Z}_{2}$ interval length must be computed with respect to the eleven-dimensional metric. To see this, recall from \cite{Lukas:1998tt} that the eleven-dimensional metric ansatz for the reduction to five dimensions is given by
\begin{equation}
\dd s_{11}^{2}=V^{-2/3}g_{\alpha\beta} \dd x^{\alpha} \dd x^{\beta}+g_{AB} \dd x^{A}\dd x^{B}\ ,\label{eq:11d_metric}
\end{equation}
where $g_{\alpha\beta}$ is the five-dimensional metric and $g_{AB}$ is the metric of the Calabi–Yau threefold. Note that the factor of $V^{-2/3}$ is chosen so that $g_{\alpha\beta}$ is in the five-dimensional Einstein frame. To further reduce to four dimensions, one takes
\begin{equation}
g_{\alpha\beta} \dd x^{\alpha}\dd x^{\beta}=\Rhat^{-1}g_{\mu\nu} \dd x^{\mu} \dd x^{\nu}+\Rhat^{2}(\dd  x^{11})^{2} \ ,\label{eq:5d_metric}
\end{equation}
where $x^{11}$ runs from $0$ to $\pi\rho$ and $g_{\mu\nu}$ is the four-dimensional Einstein frame metric. Note that, following the convention outlined in Appendix A and used throughout the text, we denote all moduli averaged over the $S^{1}/ \mathbb{Z}_{2}$ orbifold interval  without the subscript ``$0$''. As measured by the five-dimensional metric, the $S^{1}/ \mathbb{Z}_{2}$ orbifold interval has length $\pi\rho\Rhat$. However, if one wants to compare the scale of the orbifold interval with that of the  Calabi–Yau threefold, one must use the eleven-dimensional metric. Substituting (\ref{eq:5d_metric}) into (\ref{eq:11d_metric}) and averaging the value of $V$ over the orbifold interval, we find
\begin{equation}
\dd s_{11}^{2}=V^{-2/3}\Rhat^{-1}g_{\mu\nu} \dd x^{\mu} \dd x^{\nu}+V^{-2/3}\Rhat^{2}( \dd x^{11})^{2}+g_{AB} \dd x^{A} \dd x^{B}\ .\label{eq:11d_metric-1}
\end{equation}
From this we see that, in eleven dimensions, the orbifold interval has length $\pi\rho\Rhat V^{-1/3}$, as used in \eqref{sun1}. It is helpful to note that \eqref{sun1} can be written as
\begin{equation}
 \frac{\Rhat}{\epsilon_{R}V^{1/2}} > 1 \ ,\quad \text{where } \epsilon_{R}=\frac{v^{1/6}}{\pi \rho}  \ .
\label{soc1}
\end{equation}

\subsection{The Phenomenological Constraints}

Thus far, with the exception of subsection 2.2, the main content of this text has been exploring the mathematical constraints required for the theory to be anomaly free with a hidden sector containing a single line bundle $L$ and a single bulk space five-brane. The content of subsection 2.2, however, was more phenomenological. The K\"ahler moduli space constraints were presented so that a specific $SU(4)$ holomorphic vector bundle in the observable sector would be slope-stable and preserve $N=1$ supersymmetry. Furthermore, important phenomenological properties of the resultant effective theory were presented; specifically, that the low-energy gauge group, after turning on both $\mathbb{Z}_{3}$ Wilson lines, is that of the Standard Model augmented by an additional gauge $U(1)_{B-L}$ factor, and that the particle content of the effective theory is precisely that of the MSSM, with three right-handed neutrino chiral multiplets and a single Higgs-Higgs conjugate pair, and no exotic fields.

That being said, for the $B-L$ MSSM to be completely realistic there are additional low-energy properties that it must possess. These are: 1) spontaneous breaking of the gauged $B-L$ symmetry at sufficiently high scale, 2) spontaneous breaking of electroweak symmetry with the measured values of the $W^{\pm}$ and $Z^0$ masses, 3) the Higgs mass must agree with its measured value, and 4) all sparticle masses must exceed their current experimental lower bounds. In a series of papers \cite{Ovrut:2014rba,Ovrut:2015uea,Deen:2016vyh}, using generic soft supersymmetry breaking terms added to the effective theory, scattering the initial values of their parameters statistically over various physically interesting regions and running all parameters of the effective theory to lower energy using an extensive renormalization group analysis, it was shown that there is a wide range of initial conditions that completely solve all of the required phenomenological constraints. These physically acceptable initial conditions were referred to as ``viable black points''. Relevant to this paper is the fact that, for two distinct choices of the mass scales of the two $\mathbb{Z}_{3}$ Wilson lines, the four gauge parameters associated with the $SU(3)_{C} \times SU(2)_{L} \times U(1)_{Y} \times U(1)_{B-L}$ group were shown to grand unify -- albeit at different mass scales. Let us discuss these two choices in turn.

\subsubsection{Split Wilson Lines}
The first scenario involved choosing one of the Wilson lines to have a mass scale identical to the $Spin(10)$ breaking scale and to fine-tune the second Wilson line to have a somewhat lower scale, chosen so as to give exact gauge coupling unification. The region between the two Wilson line mass scales can exhibit either a ``left-right'' scenario or a Pati--Salam scenario depending on which Wilson line is chosen to be the lightest.  We refer the reader to \cite{Ovrut:2012wg} for details. Here, to be specific, we will consider the ``left-right'' split Wilson line scenario. For a given choice of viable black point, the gauge couplings unify at a specific mass scale $M_{U}$ with a specific value for the unification parameter $\alpha_{u}$. It was shown in \cite{Deen:2016vyh} that there were 53,512 phenomenologically viable black points. The results for $M_{U}$ and the associated gauge parameter $\alpha_{u}$ are plotted statistically over these viable black points in Figures 5 and 6 respectively. The average values for the unification scale and gauge parameter, $\langle M_U\rangle$ and $\langle \alpha_u \rangle$ respectively, are indicated.
\begin{figure}
\centering
\includegraphics[scale=0.9]{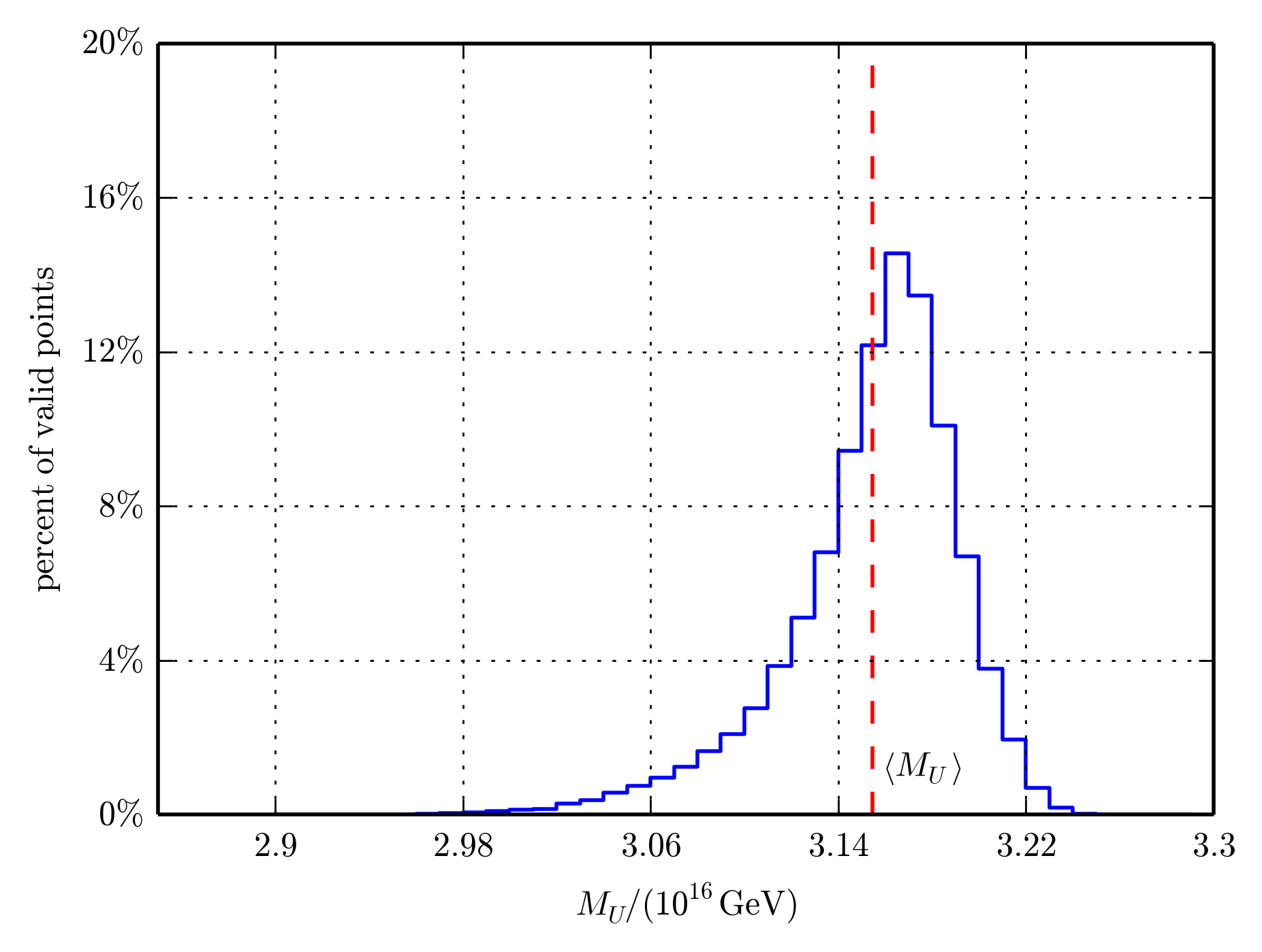}
\caption{A histogram of the unification scale for the 53,512 phenomenologically viable black points in the split Wilson line ``left-right'' unification scheme. The average unification scale is $\langle M_U\rangle=3.15\times10^{16}~\text{GeV}$.}
\label{fig:a}
\end{figure}
\\
\begin{figure}
\centering
\includegraphics[scale=0.9]{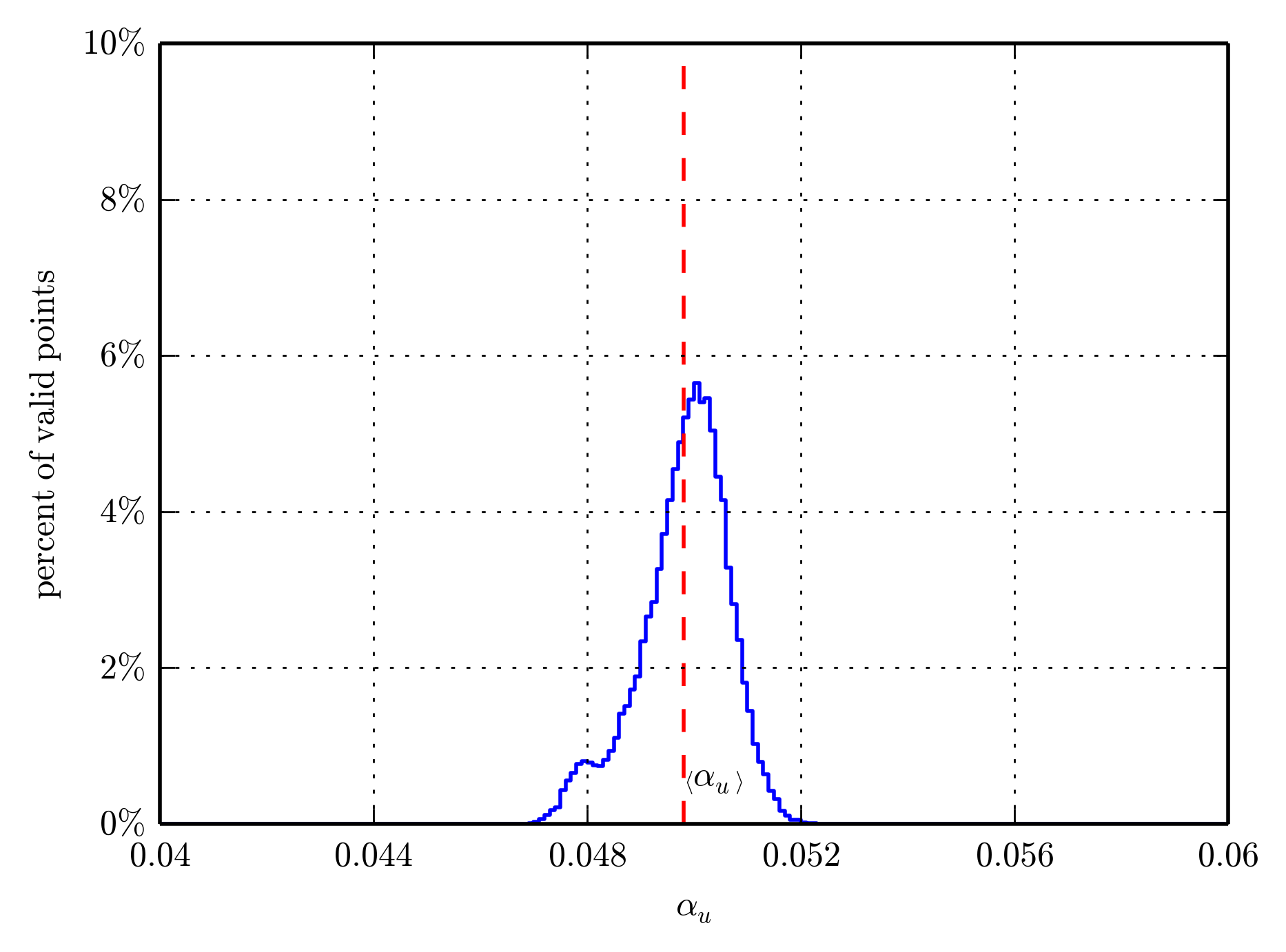}
\caption{A histogram of the unification scale for the 53,512 viable black points in the split Wilson line ``left-right'' unification scheme. The average value of the unified gauge coupling is $\langle \alpha_u\rangle=0.0498=\frac{1}{20.08}$.}
\label{fig:b}
\end{figure}
\indent The results presented in Figures 5 and 6 lead us to postulate two new ``phenomenological'' constraints on our $B-L$ MSSM vacuum. The first constraint, arising from Figure 5, is that
\begin{equation}
\langle M_{U}\rangle=3.15 \times 10^{16}~\text{GeV} \equiv\frac{1}{\boldsymbol{V}^{1/6}}=\frac{1}{v^{1/6}V^{1/6}} \ .
\label{jack1}
\end{equation}
Hence, given a point in the unity gauge K\"ahler moduli space shown in Figure 4 -- and using \eqref{10} to compute the value of $V$ at that point -- it follows from \eqref{jack1} that one can determine the required value of $v$. Up to this point in the paper, $v$ was unconstrained. To elucidate the second constraint, we must present the explicit expression for the $D=4$ effective Lagrangian for the observable and hidden sector gauge field kinetic terms. This was calculated in \cite{Lukas:1997fg,Lukas:1998yy,Lukas:1998tt} and, ignoring gravitation, was found to be
\begin{equation}
\mathcal{L}=\dots-\frac{1}{16\pi\hat{\alpha}_{GUT}}(\re f_{1}  \tr_{E_{8}}F_{1}^{\mu\nu}F_{1\mu\nu}+\re f_{2}  \tr_{E_{8}}F_{2}^{\mu\nu}F_{2\mu\nu})+\dots\label{eq:het_lagrangian}
\end{equation}
where $\hat{\alpha}_{GUT}$ is a parameter given by\footnote{As discussed in \cite{Conrad:1997ww}, the expression for $\ah$ presented here is two times larger than the result given in \cite{Lukas:1997fg,Lukas:1998yy,Lukas:1998tt}.}
\begin{equation}
\hat{\alpha}_{GUT}=\frac{\kappa_{11}^{2}}{v}\left( \frac{4\pi}{\kappa_{11}} \right)^{2/3} \ .
\label{bag1}
\end{equation}
For the specific choice of the  hidden sector line bundle $L=\Ocal_X(2, 1, 3)$  with embedding coefficient $a=1$, the functions $\re f_{1}$ and $\re f_{2}$ in unity gauge are found to be
\begin{equation}
\re f_{1} = V +\tfrac{1}{3}a^1- \tfrac{1}{6}a^2 +2a^3+\tfrac{1}{2}(\tfrac{1}{2}-\lambda)^2 (9a^1+17a^2)  \ ,
\label{bag2}
\end{equation}
and
\begin{equation}
  \label{bag2A}
 \re f_{2}= V-\tfrac{29}{6}a^1-\tfrac{25}{3}a^2-2a^3+\tfrac{1}{2}(\tfrac{1}{2}+\lambda)^2 (9a^1+17a^2)  \ ,
\end{equation}
where $W_{i}$ and $\lambda$ are given in \eqref{red8} and \eqref{cup1} respectively. It then follows from Figure 6 and \eqref{eq:het_lagrangian} that
\begin{equation}
\langle \alpha_{u} \rangle = \frac{1}{20.08}=\frac{\hat{\alpha}_{GUT}}{\re f_{1}}  \ .
\label{bag3}
\end{equation}
Hence, given a point in the unity gauge K\"ahler moduli space shown in Figure 4 -- and using \eqref{bag2} to compute the value of $ f_{1} $ at that point -- it follows from \eqref{bag3} that one can determine the required value of $\hat{\alpha}_{GUT}$, which, up to this point, was unconstrained.

Using the relations 
\begin{equation}
\kappa_{4}^{2}= \frac{8\pi}{M_{P}^{2}} =\frac{\kappa_{11}^{2}}{v2\pi\rho} \ ,
\label{bag4}
\end{equation}
where $\kappa_{4}$ and $M_{P}=1.221 \times 10^{19}~\text{GeV}$ are the four-dimensional Newton's constant and Planck mass respectively, it follows from \eqref{bag1} that
\begin{equation}
 \rho=\left( \frac{\hat{\alpha}_{GUT}}{8 \pi^{2}} \right)^{3/2}v^{1/2}M_{P}^{2} \ .
\label{bag5}
\end{equation}
Finally, using these relations, the expression for $\epsilon_S'$ in  \eqref{40AA} can be rewritten as 
\begin{equation}
 \epsilon_S' =\frac{2\pi^{2}\rho^{4/3}}{v^{1/3}M_{P}^{2/3}} \ .
\label{soc2}
\end{equation}
That is, at any given fixed point in the unity gauge K\"ahler moduli space shown in Figure 4, which, by definition, satisfies all of the $B-L$ MSSM ``vacuum constraints'' listed in Sections 3 and 4, as well as the two ``phenomenological'' constraints presented in this subsection, one can determine {\it all} constant parameters of the theory -- that is, $v$, $\hat{\alpha}_{GUT}$, $\rho$, $ \epsilon_S^\prime$ and $\epsilon_{R}$. We again emphasize, as discussed above, that the unity gauge solution space of the vacuum constraints is valid for any arbitrary choice of $ \epsilon_S^\prime$.

\subsubsection{Simultaneous Wilson lines}

In the previous subsection, we presented the phenomenological constraints for the ``left-right'' split Wilson line scenario. Here, we will again discuss the two phenomenological constraints, but this time in the scenario where the mass scales of the two Wilson lines and the ``unification'' scale are approximately degenerate. Although somewhat less precise than the split Wilson line scenario, this ``simultaneous'' Wilson line scenario is more natural in the sense that less fine-tuning is required. We refer the reader to \cite{Deen:2016vyh} for details. In this new scenario, we continue to use the previous mass scale $\langle M_{U}\rangle=3.15 \times 10^{16}~\text{GeV}$ as the $SO(10)$ ``unification'' scale -- since its mass is set by the scale of the gauge bundle -- even though when the Wilson lines are approximately degenerate the low-energy gauge couplings no longer unify there. Rather, they are split at that scale by individual ``threshold'' effects. Since the full $B-L$ MSSM low energy theory now exists at $\langle M_{U} \rangle$, we will assume that soft supersymmetry breaking also occurs at that scale. As shown in \cite{Deen:2016vyh}, we find that there are 44,884 valid black points which satisfy all low-energy physical requirements -- including the correct Higgs mass. Rather than statistical plots over the set of all phenomenological black points, as we did in Figures 5 and 6 for the previous scenario, here we present a single figure showing the running of the inverse $\alpha$ parameters for the $SU(3)_{C}$, $SU(2)_{L}$, $U(1)_{3R}$ and $U(1)^{'}_{B-L}$  gauge couplings. This is presented in Figure 7. 
 \begin{figure}
\centering
\includegraphics[scale=0.9]{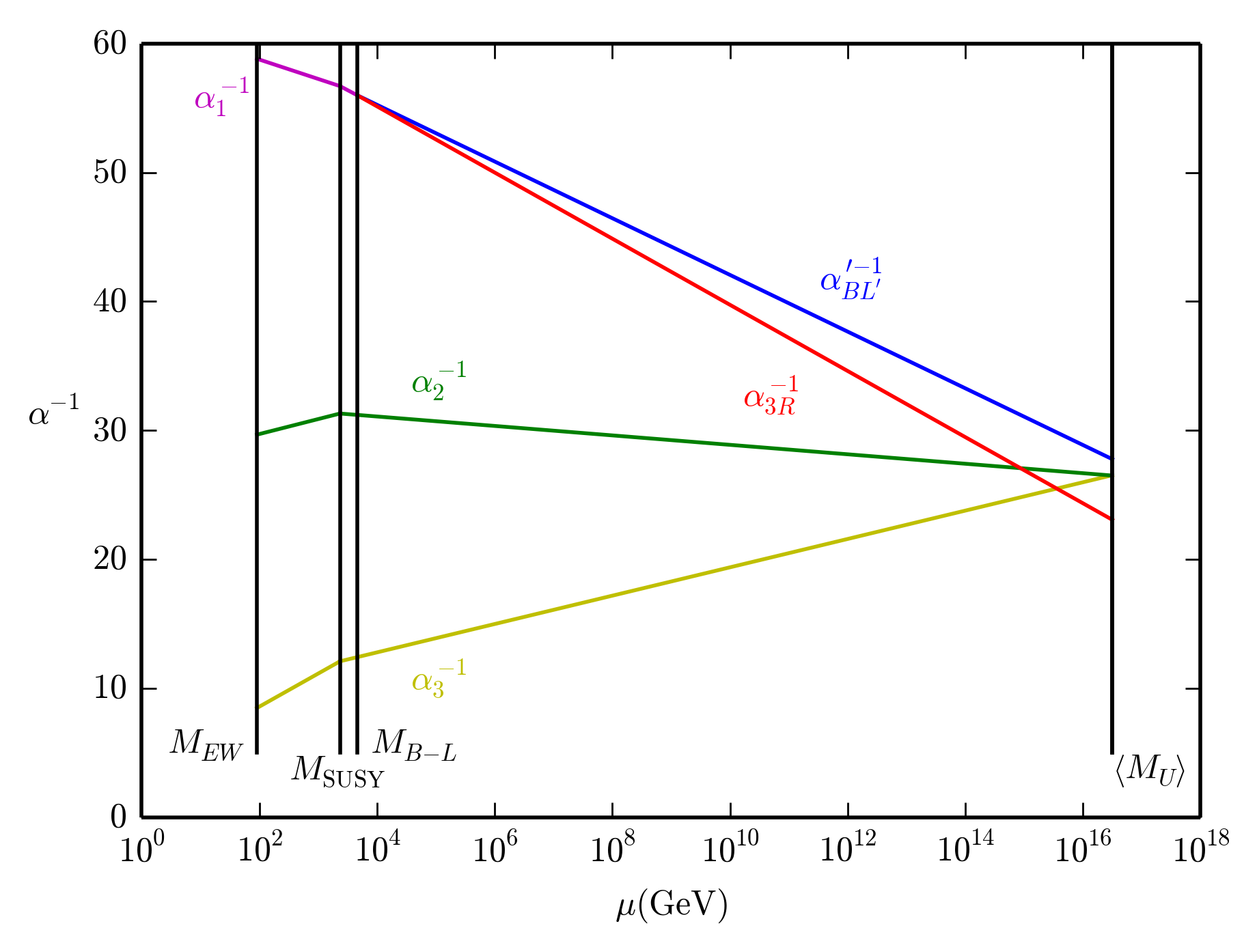}
\caption{Running gauge couplings for a sample ``valid black point''  with $M_{SUSY}=2350$ GeV, $M_{B-L}=4670$ GeV and $\sin^{2}\theta_R = 0.6$. In this example, $\alpha_3(\langle M_U \rangle)=0.0377$, $\alpha_2(\langle M_U \rangle )=0.0377$, $\alpha_{3R}(\langle M_U \rangle)=0.0433$, and $\alpha_{BL^\prime}(\langle M_U \rangle)=0.0360$.}
\label{fig:c}
\end{figure}
Note that in the analysis of Figure 7, we use the $U(1)_{3R}$ gauge group instead of $U(1)_{Y}$ and $U(1)^{'}_{B-L}$ instead of  $U(1)_{B-L}$, which is a minor redefinition of the $B-L$ charges, since this simplifies the renormalization group analysis. However, the averages over their gauge thresholds differs only minimally from the basis used in this paper. Furthermore, we will augment the results of Figure 7 with a more detailed discussion below which uses our standard basis.

As discussed in the previous paragraph, the first constraint in this new scenario is identical to constraint \eqref{jack1} above. That is,
\begin{equation}
\langle M_{U}\rangle=3.15 \times 10^{16}~\text{GeV} \equiv\frac{1}{\boldsymbol{V}^{1/6}}=\frac{1}{v^{1/6}V^{1/6}} \ .
\label{jack1A}
\end{equation}
Hence, given a point in the unity gauge K\"ahler moduli space shown in Figure 4 -- and using \eqref{10} to compute the value of $V$ at that point -- it follows from \eqref{jack1A} that one can determine the required value of $v$. To elucidate the second phenomenological constraint in this scenario, however, requires a further analysis. First note from Figure 7, which is computed for a {\it single} initial valid black point, that at $ \langle M_{U} \rangle$ the values of the $\alpha$ parameters for each of the four gauge couplings are given by
\begin{gather}
\alpha_3(\langle M_U\rangle)=0.0377\ ,\qquad \alpha_2(\langle M_U\rangle)=0.0377\ , \\
\alpha_{3R}(\langle M_U\rangle)=0.0433\ ,\qquad \alpha_{BL^\prime}(\langle M_U\rangle)=0.0360 \ ,
 \label{JFK1}
 \end{gather}
 respectively. Taking the average over these parameters, we find that for that specific valid black point,
 \begin{equation}
 \alpha_{u}^{\rm avg}= \frac{1}{25.87} \ .
 \label{JFK2}
 \end{equation}
 However, to get a more generic value for the average $\langle \alpha_{u} \rangle$ at the unification scale $\langle M_{U} \rangle$, one can either: 1) repeat the same analysis as in Figure 7, statistically calculating over all 44,884 valid black points and finding the average of the results, or 2) use the following technique, which is unique to a string theory analysis. Since our observable sector comes from an $E_{8}\times E_{8}$ heterotic string theory in ten dimensions, we will use the second analysis for simplicity.
 
 To do this, we note that, at string tree level, the gauge couplings are expected to grand 
 unify to a single parameter $g_{\rm string}$ at a ``string unification'' scale
\begin{equation}
M_{\rm string}=g_{\rm string} \times 5.27 \times 10^{17}~\mbox{GeV} \ .
\end{equation}
The string coupling parameter $g_{\rm string}$ is set by the value of the dilaton, and is typically of ${\cal{O}}(1)$. A common value in the literature, see for example \cite{Dienes:1996du,Bailin:2014nna,Nilles:1998uy}, is $g_{\rm string}= 0.7$ which, for specificity, we will use henceforth. Therefore, we take $\alpha_{\rm string}$ and the string unification scale to be 
\begin{equation}
\alpha_{\rm string}=\frac{g_{\rm string}^{2}}{4\pi} = 0.0389 \ , \qquad    M_{\rm string}=3.69 \times 10^{17}~ \mbox{GeV} \ ,
\label{hani4}
\end{equation}
respectively. Note that $ M_{\rm string}$ is approximately an order of magnitude larger than $\langle M_{U}\rangle$. Below $M_{\rm string}$ however, the couplings evolve according to the renormalization group equations of $B-L$ MSSM effective field theory. This adds another scaling regime, $\langle M_{U}\rangle \rightarrow M_{\rm string}$, to those discussed previously.
The effective field theory in this regime remains that of the $B-L$ MSSM, with the same renormalization group factors  
as between the $B-L$ breaking scale and  $\langle M_{U}\rangle $.
However, the gauge coupling renormalization group equations are now altered to
\begin{equation}
4\pi {\alpha_{a}}^{-1}( p)=4\pi \alpha_{\rm string }^{-1}-b_{a}\ln\left(\frac{p^2}{M_{\rm string}^{2}}\right) \ ,
\label{hani6}
\end{equation}
where the index $a$ runs over $SU(3), SU(2), 3R, B-L$, the coefficients $b_{a}$ are given in \cite{Ovrut:2015uea} and, for simplicity, we have ignored the ``string threshold'' corrections calculated in \cite{Deen:2016vyh}.
Note that the one-loop running gauge couplings do not unify exactly at $\langle M_U\rangle $. Rather, they are ``split'' by dimensionless threshold effects. Using \eqref{hani4} and taking $p^2=\langle M_U\rangle ^{2}$, one can evaluate the $\alpha_{a}$ parameter for each of the four gauge couplings at the  scale $\langle M_U\rangle $. We find that
\begin{gather}
\alpha_{SU(3)}(\langle M_U\rangle )=0.0430\ ,\qquad\alpha_{SU(2)}(\langle M_U\rangle )=0.0383\ ,\\
\alpha_{3R}(\langle M_U\rangle )=0.0351\ ,\qquad\alpha_{B-L}(\langle M_U\rangle )=0.0356\ ,
\label{tr1}
\end{gather}
and, hence, the average ``unification'' parameter at $\langle M_U\rangle $ is given by
\begin{equation}
\langle  \alpha_{u}\rangle =\frac{1}{26.46} \ .
\label{tr2}
\end{equation}
It follows that for the ``simultaneous'' Wilson line scenario, the second phenomenological constraint is altered to become
\begin{equation}
\langle  \alpha _{u}\rangle =\frac{1}{26.46}=\frac{\hat{\alpha}_{GUT}}{\re f_{1}}  \ .
\label{tr3}
\end{equation}
Hence, given a point in the unity gauge K\"ahler moduli space shown in Figure 4 -- and using \eqref{bag2} to compute the value of $\re f_{1} $ at that point -- it follows from \eqref{tr3} that one can determine the required value of $\hat{\alpha}_{GUT}$, which, up to this point, was unconstrained. 

As with the ``left-right'' Wilson line scenario in the previous subsection, given the values for $v$ and $\hat{\alpha}_{GUT}$ from \eqref{jack1A} and \eqref{tr3}, one can then compute the parameters $\rho$, $\epsilon_S'$ and $\epsilon_{R}$ using \eqref{bag5}, \eqref{soc2} and \eqref{soc1} respectively.

\section{A Solution of All the Constraints}

In Section 4, we displayed the solutions to all of the $B-L$ MSSM ``vacuum'' constraints, valid for a hidden sector line bundle $L=\Ocal_X(2, 1, 3)$ embedded as in \eqref{red5} with $a=1$ for a single five-brane with its bulk space location fixed to be at $\lambda=0.49$. The intersecting region of K\"ahler moduli satisfying all of these constraints, computed in unity gauge, are shown in Figure 4. In Section 5, we introduced three {\it additional} constraints. One of these, which we refer to as  the ``reduction'' constraint, is required for the consistency of the dimensional reduction from the eleven-dimensional theory down to five dimensions. The other two, which we call the ``phenomenological'' constraints, are necessary so that the $SO(10)$ grand unified group in the observable sector has the correct unification scale, $\langle M_{U}\rangle $, and the right value of the physical unified gauge parameter, $\langle \alpha_{u}\rangle $, determined from the $B-L$ MSSM via a renormalization group analysis. In this section we want to find the subspace of the solution space presented in Figure 4, which, in addition, is consistent with the reduction constraint and the two phenomenological constraints presented in Section 5. 

We begin by imposing the physical constraints. We demand that \emph{at every point} in the region of K\"ahler moduli space shown in Figure 4, all parameters of the theory are adjusted so that $\langle  M_U \rangle $  and $\langle  \alpha_u \rangle $ are fixed at the physical values presented in Section 5 -- that is, the unification scale is always set to $\langle M_U\rangle =3.15\times 10^{16}~\text{GeV}$, whereas for $SO(10)$ breaking with split Wilson lines, $\langle  \alpha_{u}\rangle =1/20.08$, while for $SO(10)$ breaking with simultaneous Wilson lines $\langle  \alpha_{u}\rangle =1/26.46$. It follows from the physical constraint equations \eqref{jack1}, and \eqref{bag3} and \eqref{tr3} that the values of $v$ and ${\hat{\alpha}}_{GUT}$ -- and, hence, the remaining parameters $\rho$, $\epsilon_S^\prime$ and $\epsilon_R$ -- can be always be chosen so as to obtain the required values of $\langle M_U\rangle $ and $\langle  \alpha_{u}\rangle $. However, different points in Figure 4 will, in general, require {\it different} values of these parameters to satisfy the physical constraints. In particular, this means that different points will correspond to different values of $\epsilon_S^\prime$. As discussed at the end of Section 4, this interpretation is completely consistent with the moduli shown in Figure 4 solving all of the ``vacuum'' constraints. To make this explicit, one can invert constraint equations \eqref{jack1}, \eqref{bag3} and \eqref{tr3} so as to express $v$ and ${\hat{\alpha}}_{GUT}$ explicitly as functions of $\langle M_{U}\rangle $, $\langle  \alpha_{u}\rangle $ and the K\"ahler moduli. That is, expression \eqref{jack1} can be inverted to give
\begin{equation}
v=\frac{1}{\langle M_U\rangle ^{6}V} \ ,
\label{door1}
\end{equation}
while \eqref{bag3} and \eqref{tr3} give
\begin{equation}
{\hat{\alpha}}_{GUT}=\frac{1}{\langle  \alpha_{u} \rangle \re f_{1}} \ .
\label{door2}
\end{equation}
As first presented in \eqref{bag2}, the function $\re f_{1}$ is given by
\begin{equation}
\re f_{1} = V +\tfrac{1}{3}a^1- \tfrac{1}{6}a^2 +2a^3+\tfrac{1}{2}(\tfrac{1}{2}-\lambda)^2(9a^{1}+17a^{2}) \ ,
\label{door3}
\end{equation}
where $\lambda=0.49$ and $\langle  \alpha_{u} \rangle =\{\frac{1}{20.08}, \frac{1}{25.87}\}$ for split and simultaneous Wilson lines respectively. Inserting these expressions into \eqref{bag5}, \eqref{soc2} and \eqref{soc1}, one obtains the following expressions for $\rho$, $\epsilon_S^\prime$ and $\epsilon_R$ respectively. We find that
\begin{equation}
 \rho= \left(\frac{\langle  \alpha_{u} \rangle}{16\pi^2}\right)^{3/2}\frac{M_P^2}{\langle M_{U}\rangle ^3}  \frac{ ( \re f_{1} )^{3/2}}{V^{1/2}}
\label{seb6}
\end{equation}
and
\begin{equation}
 \epsilon_S' =\frac{\langle  \alpha_{u} \rangle^2}{128\pi^2}\frac{M_P^2}{\langle M_U\rangle ^2}\frac{(\re f_{1})^2}{V^{1/3}}\ ,\qquad \epsilon_{R}=\frac{64\pi^{2}}{\langle \alpha_{u}\rangle ^{3/2} }\frac{\langle M_{U}\rangle^{2}}{M_{P}^{2}}\frac{V^{1/3}}{(\re f_{1})^{3/2}}  \ .
\label{seb7}
\end{equation}
Using these expressions, the parameters at any fixed point of the moduli space in Figure 4 can be calculated. Again, we note that these parameters change from point to point in Figure 4.

Next, we impose the dimensional reduction constraint. We require that \eqref{sun1} be valid; that is, the length of the $S^{1}/{\mathbb{Z}}_{2}$ orbifold interval should be larger than the average Calabi--Yau radius 
\begin{equation}
\frac{\pi \rho {\Rhat} V^{-1/3}}{(vV)^{1/6} } > 1 \ .
\label{seb3}
\end{equation}
Choosing any point $a^{i}$ in the K\"ahler moduli space of Figure 4, one can use \eqref{door1}, \eqref{door2}, \eqref{seb6} and \eqref{seb7} to determine the parameters $v$, ${\hat{\alpha}}_{GUT}$, $\rho$ and $\epsilon'_{S}$ that satisfy the phenomenological constraints at that point. Also note, since we are working in unity gauge \eqref{wall2}, that
\begin{equation}
\Rhat=\frac{V^{1/3}}{\epsilon'_{S}} \ .
\label{exam1}
\end{equation}
Now, at that chosen point in Figure 4, insert the calculated values of $v$, ${\hat{\alpha}}_{GUT}$, $\rho$ and $\epsilon'_{S}$, as well the value of $\Rhat$ computed from \eqref{exam1}, into the inequality \eqref{seb3} for the ratio of the dimensions. Scanning over all points in Figure 4,
we will be able to find the subspace of that region of K\"ahler moduli space in which condition  \eqref{seb3} is satisfied. That is, at any such a point, not only are all the ``vacuum'' constraints satisfied, but the ``reduction'' and ``phenomenological'' constraints are as well. There, of course, will be two such regions -- one corresponding to the ``split'' Wilson line scenario and a second corresponding to the ``simultaneous'' Wilson line scenario. These regions are shown as the brown subspaces of Figure 8 (a) and (b) respectively.
\begin{figure}[t]
   \centering
\begin{subfigure}[c]{0.47\textwidth}
\includegraphics[width=1.0\textwidth]{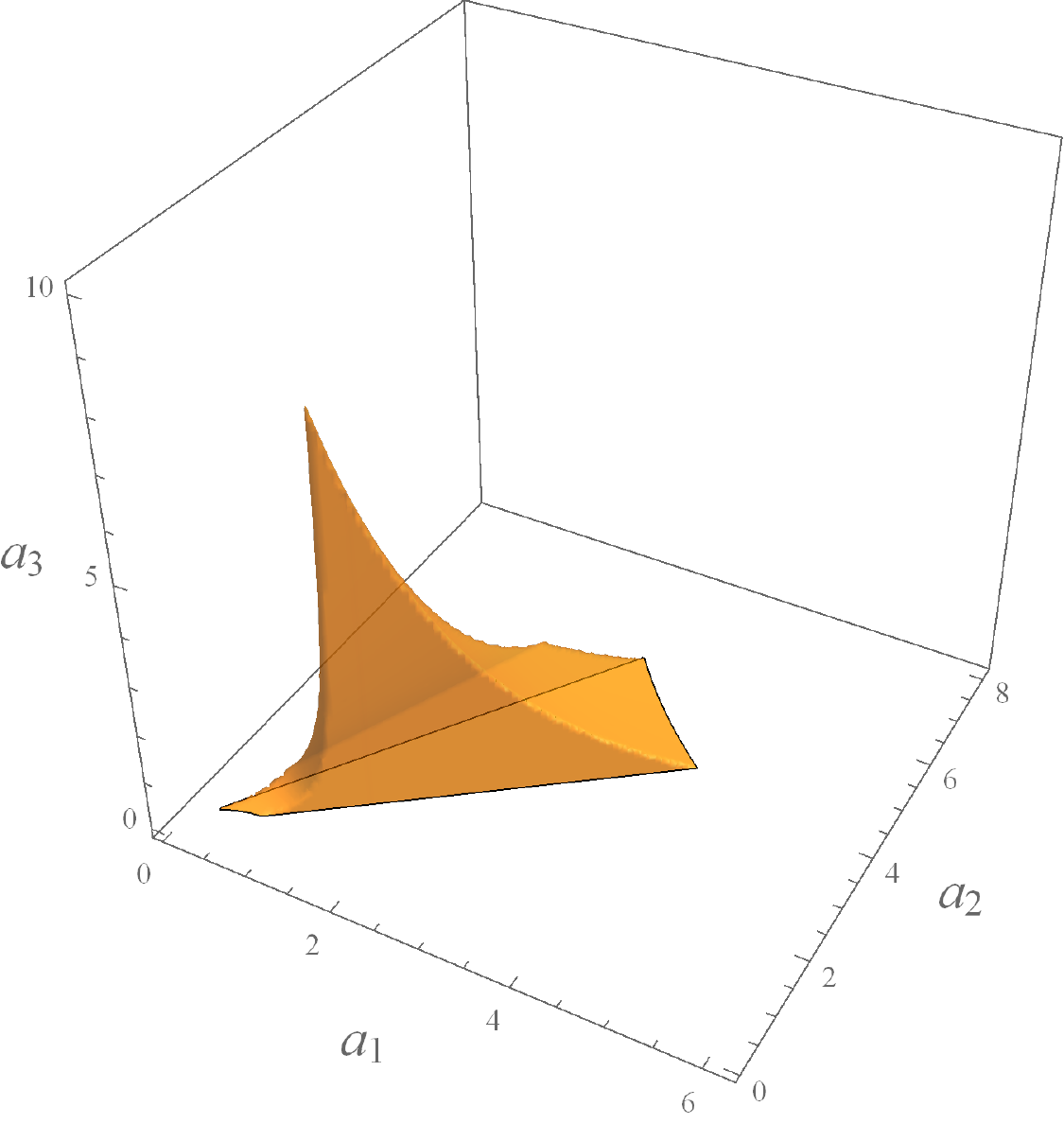}
\caption{$\langle  \alpha_{u}\rangle =\frac{1}{20.08}$}
\end{subfigure}
\begin{subfigure}[c]{0.47\textwidth}
\includegraphics[width=1.0\textwidth]{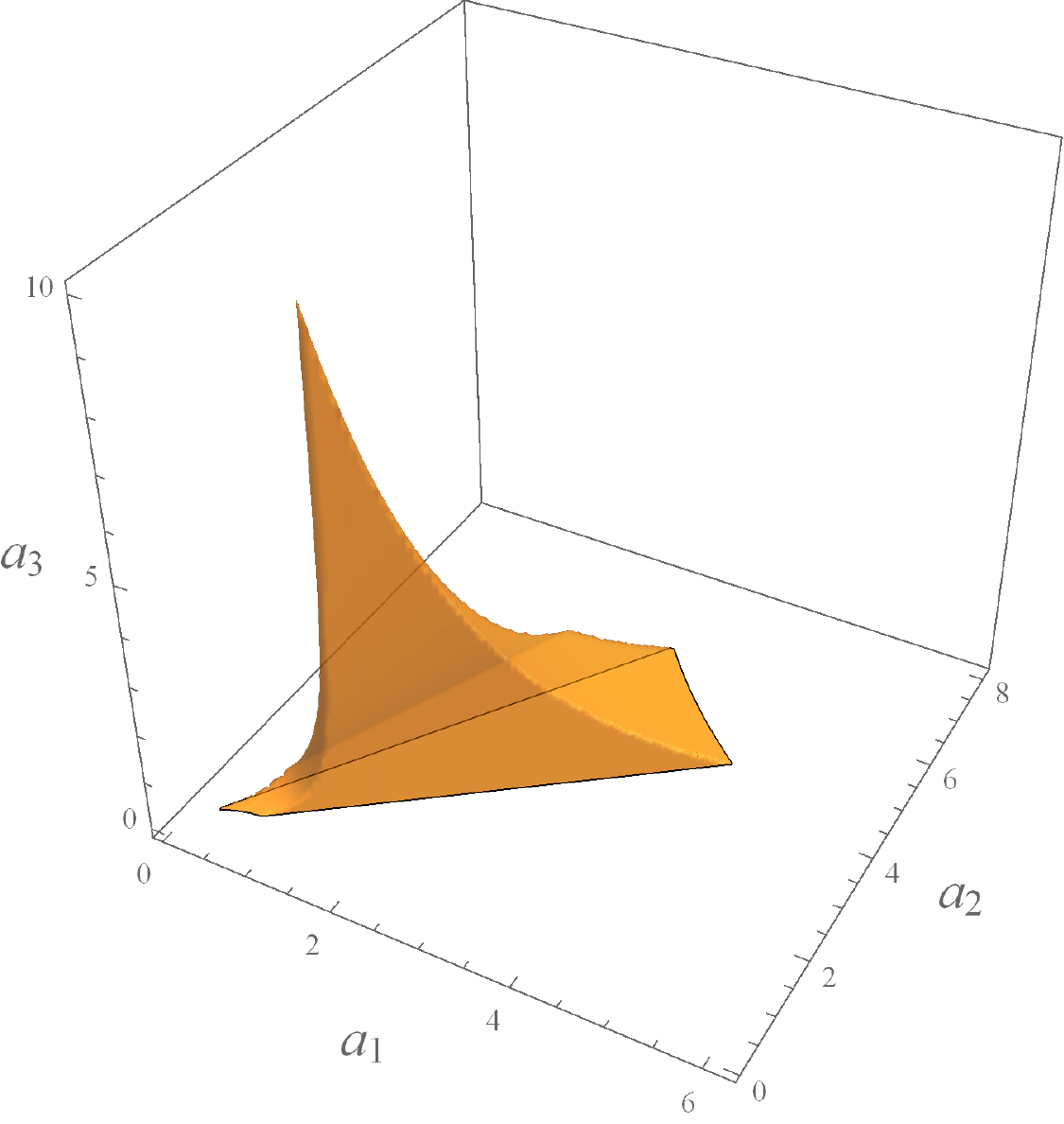}
\caption{$\langle  \alpha_{u}\rangle =\frac{1}{26.46}$}
\end{subfigure}
\caption{The region of K\"ahler moduli space where the $SU(4)$ slope-stability conditions, the anomaly cancellation constraint and the positive squared gauge coupling constraint from Figure 4 are satisfied, in addition to the dimensional reduction and the phenomenological constraints introduced in Section 5. The results are valid for a hidden sector line bundle $L=\Ocal_X(2, 1, 3)$  with $a=1$ and for a single five-brane located at $\lambda=0.49$. We study both cases of split Wilson lines, with $\langle  \alpha_{u}\rangle =\frac{1}{20.08}$, and simultaneous Wilson lines with $\langle  \alpha_{u}\rangle =\frac{1}{26.46}$. Note that reducing the size of $\langle  \alpha_u\rangle $ increases the space of solutions.}
\label{fig:PhysContraint}
\end{figure}
One can go further and, by scanning over the brown subspace associated with each Wilson line scenario, find the numerical range of the ratio $\frac{\pi \rho {\Rhat} V^{-1/3}}{(vV)^{1/6}}$ in each case. We find that
\begin{equation}\
1 \lesssim \frac{\pi \rho {\Rhat} V^{-1/3}}{(vV)^{1/6}} \lesssim 17.4
\label{er1}
\end{equation}
for the split Wilson line scenario and 
\begin{equation}\
1 \lesssim \frac{\pi \rho {\Rhat} V^{-1/3}}{(vV)^{1/6}} \lesssim 19.8
\label{er2}
\end{equation}
for the simultaneous Wilson lines.

Finally, but importantly, we want to emphasize that all of the results of this and preceding sections have, thus far, been calculated in ``unity'' gauge; that is, choosing
\begin{equation}
\frac{\epsilon'_{S}{\Rhat}}{V^{1/3}}=1 \ .
\label{er33}
\end{equation}
This was possible because, as discussed in Section 4, all ``vacuum'' constraints remained form invariant under the scaling
\begin{equation}
a^{i} \rightarrow \mu a^{i}\ , \qquad\epsilon'_{S} \Rhat \rightarrow \mu^{3} \epsilon'_{S}\Rhat \ , 
\label{er4}
\end{equation}
where $\mu > 0$. However, it follows from this that if any point \{$a^{i}$\} in Figure 4 satisfies all vacuum constraints, so will any point \{$\mu a^{i}$\}. Do any of these ``scaled'' moduli carry new information concerning both the reduction and the physical constraints? The answer to this is no, as we will now demonstrate.

Let us pick any point \{$a^{i}$\} in Figure 4 and assume that the values of the parameters at that point are given by $v$, ${\hat{\alpha}}_{GUT}$, $\rho$ and $\epsilon'_{S}$ obtained from expressions \eqref{door1}, \eqref{door2}, \eqref{seb6} and \eqref{seb7} respectively evaluated at this point. We now want to determine how each of these parameters changes under the $\mu$ scaling given in \eqref{er4}. To do this, one can again use the same equations  \eqref{door1}, \eqref{door2}, \eqref{seb6} and \eqref{seb7}, but now scaling the original point as in \eqref{er4}. To do this, we must know the scaling behavior of both $V$ and $\re f_{1}$ respectively. It follows from \eqref{10} that $V \rightarrow \mu^{3} V$. However, to obtain the scaling behavior of $\re f_{1}$, one must go back to \eqref{69AA} and recall that the terms in $Ref_{1}$ linear in the K\"ahler moduli are, generically, multiplied by the factor $\epsilon'_{S} \Rhat/ V^{1/3}$. Since this is set to 1 in unity gauge, it does not appear in expression \eqref{door3}. Hence, it follows from \eqref{er4} that under $\mu$ scaling $\re f_{1} \rightarrow \mu^{3} \re f_{1}$. Using these results, we find that
\begin{gather}
v \rightarrow \mu^{-3}v\ ,\quad{\hat{\alpha}}_{GUT} \rightarrow \mu^{3}{\hat{\alpha}}_{GUT}\ ,\quad\rho \rightarrow \mu^{3} \rho\ ,\\
\epsilon'_{S} \rightarrow  \mu^{5}\epsilon'_{S}\ ,\quad\epsilon_{R} \rightarrow \mu^{-7/2} \epsilon_{R} \ . 
\label{sf1}
\end{gather}
Note that, until now, we knew that scaling invariance required $\epsilon'_{S}\Rhat \rightarrow \mu^{3} \epsilon'_{S}\Rhat$, but could not specify the scaling of $\epsilon'_{S}$ parameter and the modulus $\Rhat$ individually. However, from the last term in \eqref{sf1} it follows that 
\begin{equation}
\Rhat \rightarrow \mu^{-2} \Rhat \ .
\label{sf2}
\end{equation}
It is now straightforward to insert these results into the expression for the ratio of the orbifold interval length/average Calabi--Yau radius. We find that the scaling of the individual parameters and moduli {\it exactly cancel}. That is, under the scaling given in \eqref{er4} and \eqref{sf1}
\begin{equation}
\frac{\pi \rho {\Rhat} V^{-1/3}}{(vV)^{1/6} }  \rightarrow \frac{\pi \rho {\Rhat} V^{-1/3}}{(vV)^{1/6} } \ .
\label{sf3}
\end{equation}
We conclude from this that the $\mu$-scaled point $\{\mu a^{i}\}$ of any point $\{a^{i}\}$ in the brown regions of Figure 8 (a) and (b) continues to satisfy all the ``vacuum'' and ``phenomenological'' constraints and and has identical values for the orbifold interval length/average Calabi--Yau radius. For this reason, we find it sufficient to display the final results as the brown regions in Figure 8 (a) and (b) only.

\section{Slope-Stability and Supersymmetry}

Thus far, we have found the region of K\"ahler moduli space in which the $\text{\ensuremath{\SU 4}}$ bundle is slope-stable with vanishing slope, the five-brane class is effective, the squares of both gauge couplings are positive, the length of the orbifold is larger than the characteristic length scale of the Calabi--Yau threefold and the vacuum is consistent with both the mass scale and gauge coupling of $SO(10)$ grand unification in the observable sector. 

Importantly, however, we still must satisfy two remaining conditions. First, it is necessary that the gauge connection associated with a hidden sector line bundle on the Calabi--Yau threefold be a solution of the Hermitian Yang--Mills (HYM) equations~\cite{UY,Donaldson} and, second, that the line bundle be such that the low-energy effective theory admits an $N=1$ supersymmetric vacuum. We will now analyze both of these remaining constraints. First, for specificity, we restrict the analysis to the particular line bundle discussed above; that is,  $L=\Ocal_X(2, 1, 3)$ embedded into $SU(2) \subset E_{8}$ as in \eqref{red4} with coefficient $a=1$. In addition, we choose $\lambda=.49$ as in \eqref{cup1}. Following that, however, we will present a discussion of these constraints for a ``generic'' line bundle with the same embedding \eqref{red2} into $SU(2) \subset E_{8}$ and $\lambda=.49$. To carry out these analyses, it is first necessary to introduce the Fayet--Iliopoulos term associated with the hidden sector $U(1)$ gauge group and to discuss the $\kappa_{11}^{2/3}$ correction to both the Fayet--Iliopoulos term and the slope.

\subsection{A Fayet--Iliopoulos Term and the \texorpdfstring{$\kappa_{11}^{2/3}$}{kappa 2/3} Slope Correction} 

 It follows from \eqref{59} that the Fayet--Iliopoulos term associated with a generic single line bundle $L=\Ocal_X(l^{1}, l^{2}, l^{3})$ and a single five-brane located at $\lambda\in[-1/2,1/2]$ is given in ``unity'' gauge by
\begin{equation}
  \label{again2A} 
FI= \frac{a}{2} \frac{ \epsilon_S \epsilon_R^2}{\kappa_{4}^{2}}
  \frac{1} {\Rhat V^{2/3}} \bigl(d_{ijk} l^i a^j a^k - a\,d_{ijk}l^il^jl^k 
  - l^i(2,2,0)|_i
  +(\tfrac{1}{2}+\lambda)^2l^iW_i \bigr) \ ,
\end{equation}
with the volume modulus $V$ and $W_{i}$ presented in \eqref{10} and \eqref{33A} respectively and $\Rhat$ defined in \eqref{case1}.  Note that the coefficient $a$ defined in \eqref{26A} enters this expression both explicitly and via the five-brane class $ {W}_i$.

It is important to note -- using \eqref{3}, \eqref{50} and \eqref{23} -- that the ``classical'' slope of the line bundle $L=\Ocal_X(l^{1}, l^{2}, l^{3})$ is given by\footnote{Note that this is not the same as the scaling factor $\mu$ of the previous section. From here onwards, $\mu$ will denote the slope.}
\begin{equation}
\mu(L)=d_{ijk} l^ia^j a^k \ ,
\label{river1}
\end{equation}
that is, the first term in the bracket of \eqref{again2A}. It follows that the remaining terms in the bracket, specifically
\begin{equation}
- a\,d_{ijk}l^il^jl^k - l^i(2,2,0)|_i+(\tfrac{1}{2}+\lambda)^2l^iW_i \ ,
\label{river2}
\end{equation}
are the strong coupling $\kappa_{11}^{4/3}$ corrections to the slope of $L$. For the remainder of this paper, we will take the slope of the line bundle $L=\Ocal_X(l^{1}, l^{2}, l^{3})$ to be the 
$\kappa_{11}^{4/3}$, genus-one corrected expression
\begin{equation}
\mu(L)=d_{ijk} l^ia^j a^k- a\,d_{ijk}l^il^jl^k - l^i(2,2,0)|_i+(\tfrac{1}{2}+\lambda)^2l^iW_i \ .
\label{trenton1}
\end{equation}

\subsection{Slope-Stability of the Hidden Sector Bundle \texorpdfstring{$L=\Ocal_X(2, 1, 3)$}{L = O(2,1,3)} } 

Although any line bundle $L$ is automatically slope-stable, since it has no sub-bundles, in order for its gauge connection to ``embed'' into the hidden sector $E_{8}$ gauge connection it is necessary to extend the bundle to $L \oplus L^{-1}$, as discussed in subsection 4.2. However, even though the connection associated with the bundle $L \oplus L^{-1}$ can, in principle, embed properly into the $\bf 248$ gauge connection of the hidden sector $E_{8}$, it remains necessary to show that $L \oplus L^{-1}$ is ``slope-stable''; that is, that its associated connection satisfies the Hermitian Yang--Mills equations. More properly stated, since $L \oplus L^{-1}$ is the Whitney sum of two line bundles, it was shown in \cite{UY, Donaldson} that it will admit a connection that uniquely satisfies the Hermitian Yang--Mills equations if and only if it is ``polystable''; that is, if and only if
\begin{equation}
\mu(L)=\mu(L^{-1})=\mu(L \oplus L^{-1}) \ .
\label{poly1}
\end{equation}
Since $\mu(L \oplus L^{-1})$ must vanish by construction, it follows that $L \oplus L^{-1}$ is polystable if and only if $\mu(L)=0$.

Let us now consider the specific line bundle $L=\Ocal_X(2, 1, 3)$ embedded into $SU(2) \subset E_{8}$ as in \eqref{red4} with coefficient $a=1$, and take $\lambda=0.49$. It follows from \eqref{again2A} that in this case
\begin{equation}
FI= \frac{ \epsilon_S \epsilon_R^2}{2\kappa_{4}^{2}}
  \frac{1} {\Rhat V^{2/3}} \left(   \tfrac{1}{3}(a^1)^{2}+\tfrac{2}{3}(a^2)^{2} +8a^1a^2+4a^2a^3 +2a^1a^3 -13.35 \right) 
\label{trenton2}
\end{equation}
and from \eqref{trenton1} that the associated genus-one corrected slope is
\begin{equation}
\mu(L)= \tfrac{1}{3}(a^1)^{2}+\tfrac{2}{3}(a^2)^{2} +8a^1a^2+4a^2a^3 +2a^1a^3 -13.35 \ .
\label{trenton3}
\end{equation}
Hence, this specific hidden sector bundle will be slope polystable -- and, therefore, admit a gauge connection satisfying the corrected Hermitian Yang--Mills equations -- if and only if the K\"ahler moduli $a^{i}, i=1,2,3$ satisfy the condition that
\begin{equation}
 \tfrac{1}{3}(a^1)^{2}+\tfrac{2}{3}(a^2)^{2} +8a^1a^2+4a^2a^3 +2a^1a^3 -13.35 = 0 \ .
\label{trenton4}
\end{equation}
The region of K\"ahler  moduli space satisfying this condition is the two-dimensional surface displayed in Figure 9. 

\begin{figure}[t]
   \centering
\begin{subfigure}[c]{0.6\textwidth}
\caption*{}
\end{subfigure}\\
\begin{subfigure}[c]{0.49\textwidth}
\includegraphics[width=1.0\textwidth]{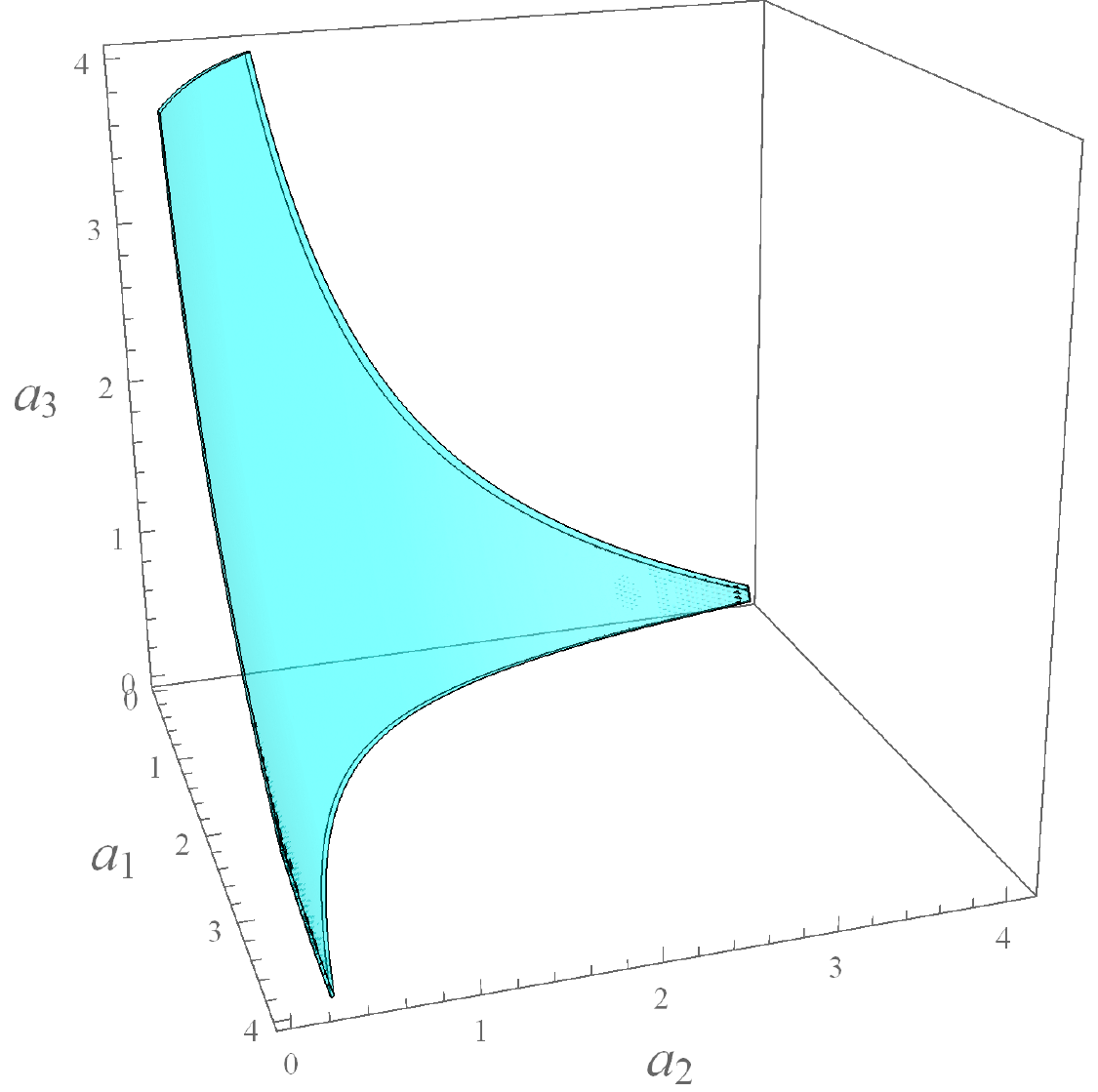}
\end{subfigure}
\caption{The surface in K\"ahler moduli space where the genus-one corrected slope of the hidden sector line bundle $L=\Ocal_X(2, 1, 3)$ vanishes.}
\label{fig:ZeroSlope}
\end{figure}

Recall that for hidden sector line bundle $L=\Ocal_X(2, 1, 3)$  with $a=1$ and for a single five-brane located at $\lambda=0.49$, the region of K\"ahler  moduli space satisfying all {\it previous} constraints -- that is, the $SU(4)$ slope-stability conditions, the anomaly cancellation constraint, the positive squared gauge coupling constraints, in addition to the dimensional reduction and the phenomenological constraints -- are shown as the brown regions in Figure 8 (a) and (b), for the split Wilson lines and the simultaneous Wilson line scenarios respectively. It follows that the intersection of the brown regions of Figure 8 (a) and (b) with the the two-dimensional surface in Figure 9 will further constrain our theory so that the hidden sector gauge connection satisfies the corrected Hermitian Yang--Mills equations -- as it must. The regions of intersection are displayed graphically in Figure 10. We emphasize that although the brown regions of Figure 8 (a) and (b) overlap in this region of K\"ahler moduli space, each point in their overlap region has a somewhat different set of parameters associated with it. Hence, in discussing a point in the magenta region of Figure 10, for example, it is necessary to state whether it is arising from the split Wilson line or simultaneous Wilson line scenario.

\begin{figure}[t]
   \centering
\begin{subfigure}[c]{0.6\textwidth}
\caption*{}
\end{subfigure}\\
\begin{subfigure}[c]{0.49\textwidth}
\includegraphics[width=1.0\textwidth]{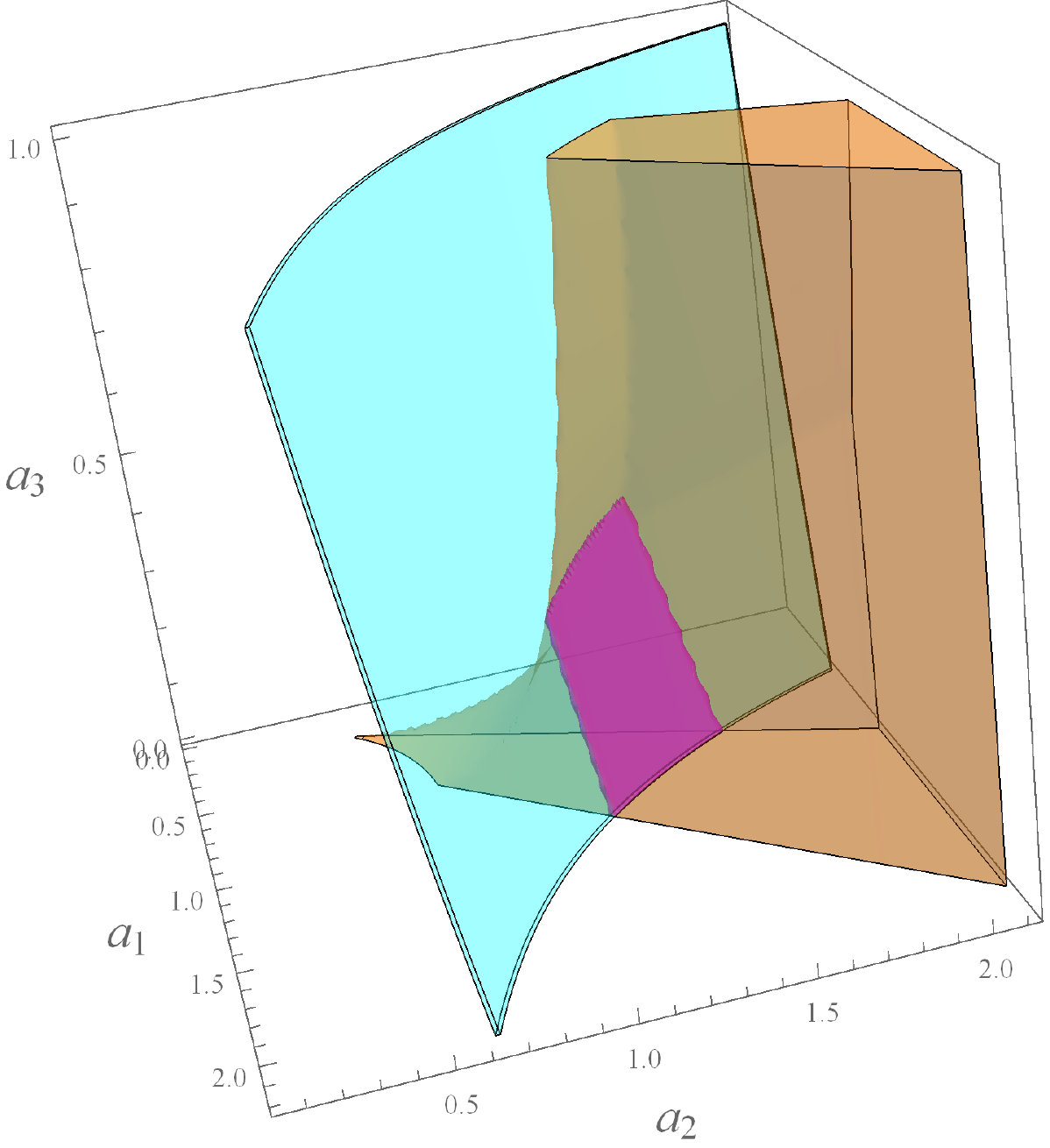}
\end{subfigure}
\caption{ The magenta region shows the intersection between the brown regions of Figure 8 (a) and (b) and the two-dimensional cyan surface in Figure 9. Therefore, the magenta region represents the sub-region of the vanishing, genus-one corrected slope surface, each point of which satisfies all the necessary constraints discussed in Section 6. The size of the magenta region is the same for both the split and simultaneous Wilson lines scenarios. However, the values of the coupling parameters differ slightly for these two cases, at any point in this intersection subspace. }
\label{fig:Intersection}
\end{figure}

As we did previously for the brown regions presented in Figure 8 (a) and (b), it is of interest to scan over the magenta subspace of Figure 10 to find the numerical range of the ratio $\frac{\pi \rho {\Rhat} V^{-1/3}}{(vV)^{1/6}}$. We find that
\begin{equation}\
6.4 \lesssim \frac{\pi \rho {\Rhat} V^{-1/3}}{(vV)^{1/6}} \lesssim 12.9
\label{er1A}
\end{equation}
for the split Wilson line scenario and 
\begin{equation}\
7.3 \lesssim \frac{\pi \rho {\Rhat} V^{-1/3}}{(vV)^{1/6}} \lesssim 14.7
\label{er2A}
\end{equation}
for simultaneous Wilson lines. In fact, one can go further and present a histogram of the percentage versus the ratio $\frac{\pi \rho {\Rhat} V^{-1/3}}{(vV)^{1/6}}$ for each scenario. These histograms are shown in Figure 11 (a) and (b) for the split Wilson line and simultaneous Wilson line scenarios respectively.

\begin{figure}[t]
   \centering
\begin{subfigure}[c]{0.49\textwidth}
\includegraphics[width=1.0\textwidth]{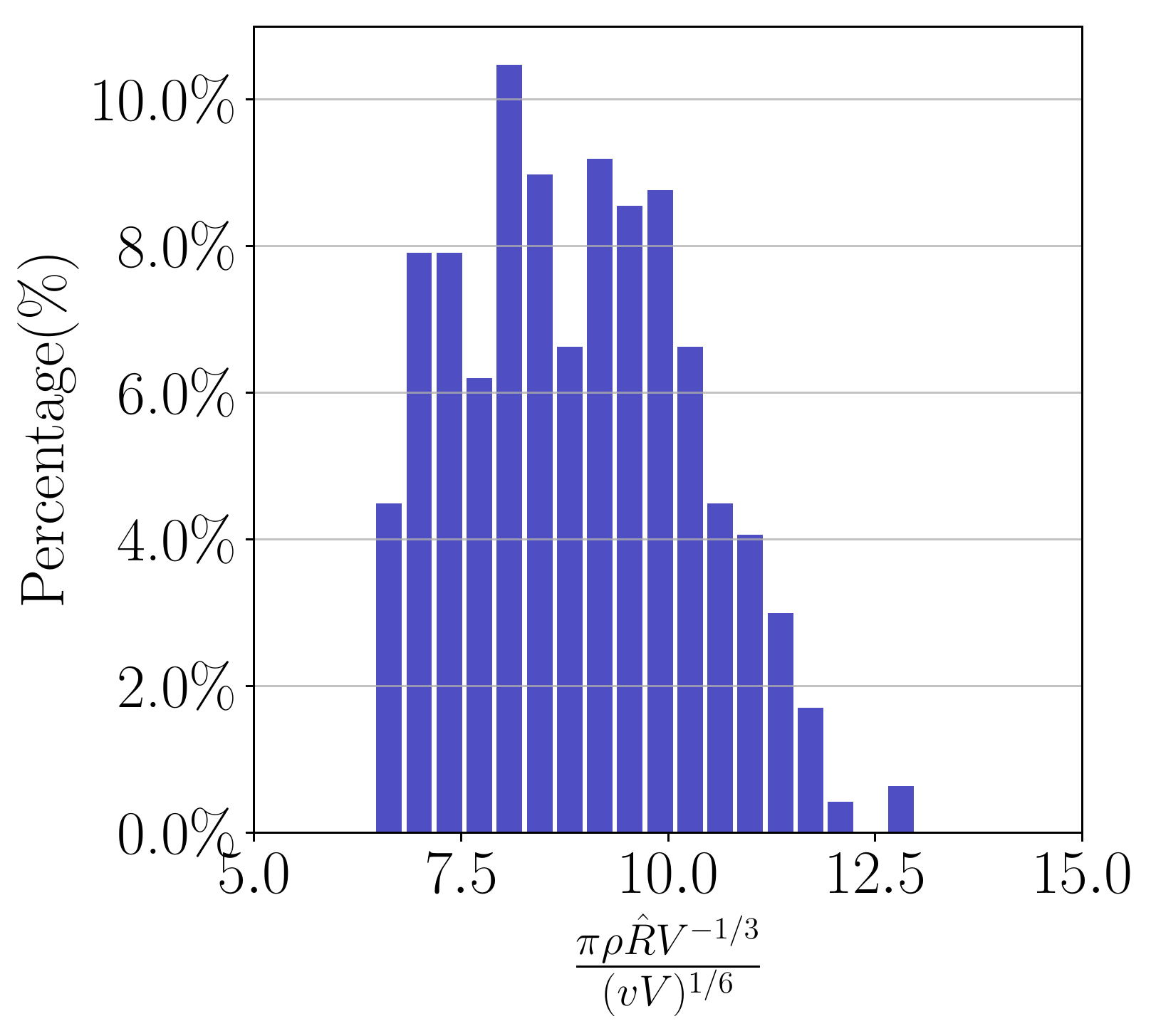}
\caption{Split Wilson lines: $\langle  \alpha_{u}\rangle =\frac{1}{20.08}$.}
\end{subfigure}
\begin{subfigure}[c]{0.49\textwidth}
\includegraphics[width=1.0\textwidth]{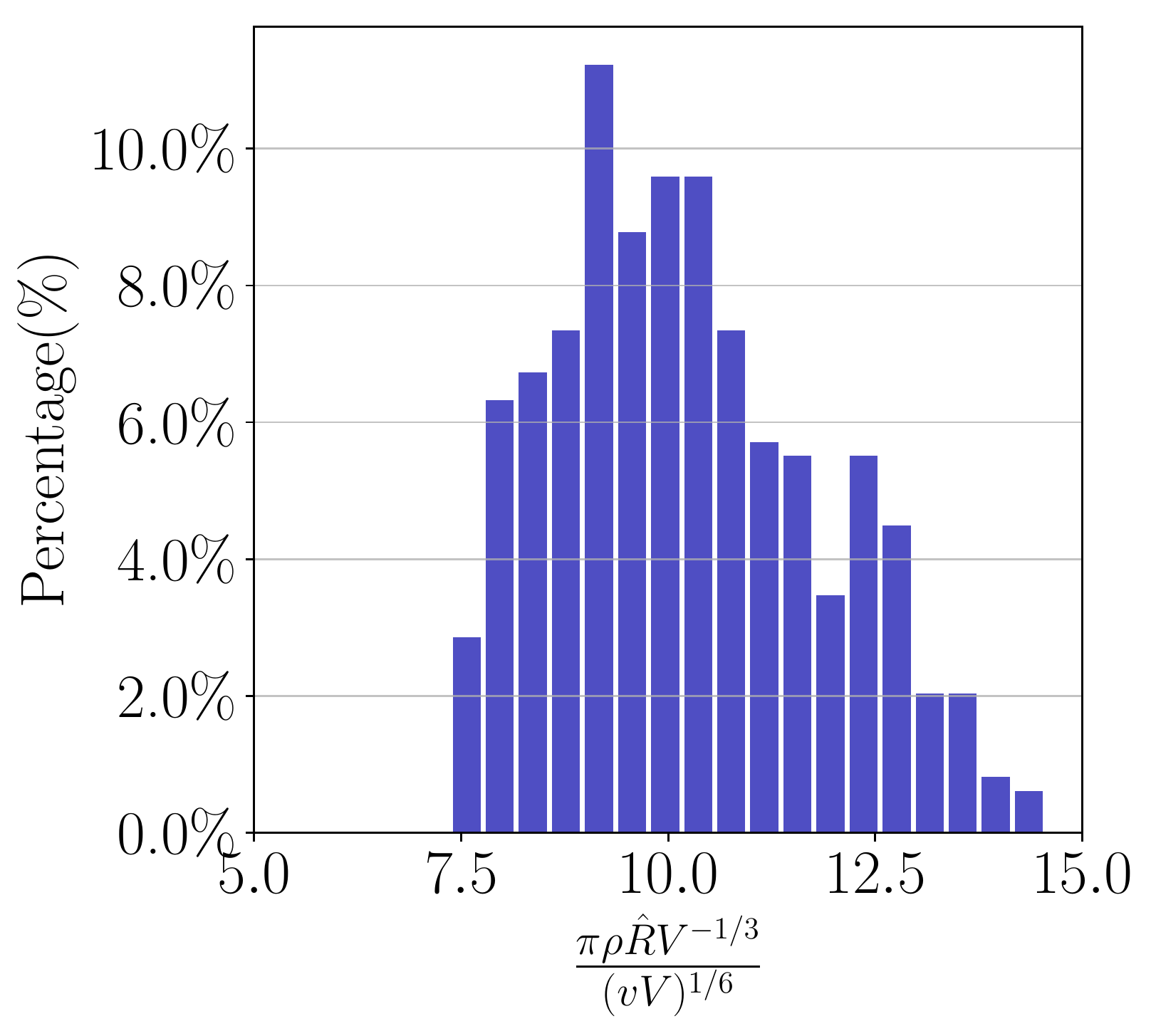}
\caption{Simultaneous Wilson lines: $\langle  \alpha_{u}\rangle =\frac{1}{26.46}$.}
\end{subfigure}
\caption{Plots of the percentage of occurrence  versus the ratio of the orbifold interval length to the average Calabi--Yau radius; for the split Wilson line scenario in (a) and for the simultaneous Wilson line scenario in (b). The results shown in (a) and (b) represent a scan over the magenta region of K\"ahler moduli space displayed in Figure 10, where all vacuum, reduction and physical constraints are satisfied and the line bundle  $L=\Ocal_X(2, 1, 3)$ is slope polystable.}
\label{fig:2Dplots}
\end{figure}

Before proceeding to the discussion of $N=1$ supersymmetry in the $D=4$ effective theory, it will be useful to present the formalism for computing the low energy matter spectrum associated with a given hidden sector line bundle. We do this in the next subsection, displaying the formalism and low energy spectrum within the context of the line bundle $L=\Ocal_X(2, 1, 3)$ for specificity.

\subsection{The Matter Spectrum of the \texorpdfstring{$L=\Ocal_X(2, 1, 3)$}{L = O(2,1,3)}, \texorpdfstring{$D=4$}{D = 4} Effective Theory} 

Having found the explicit sub-region of K\"ahler moduli space that satisfies all required constraints for the $L=\Ocal_X(2, 1, 3)$ line bundle, in this subsection we will discuss the computation of the vector and chiral matter content of the $D=4$ low-energy theory of the hidden sector. Generically,
the low-energy matter content depends on the precise hidden sector line bundle under consideration, as well as its embedding into $E_{8}$. In this subsection, we again choose the line bundle to be $L=\mathcal{O}_{X}(2,1,3)$,  embedded  into $E_{8}$ as in \eqref{red5} with $a=1$. However, the formalism presented is applicable to any line bundle with any embedding into $E_{8}$. The commutant of the $U(1)$ structure group of our specific embedding is $\Uni1 \times E_{7}$. As discussed in Section 4, the $\Rep{248}$ decomposes under $U(1) \times\Ex 7$ as
\begin{equation}
\Rep{248} \to 
(0, \Rep{133}) \oplus 
\bigl( (1, \Rep{56}) \oplus (-1, \Rep{56})\bigr) \oplus 
\bigl( (2, \Rep{1}) \oplus (0, \Rep{1}) \oplus (-2, \Rep{1}) \bigr)\ .
\label{red55}
\end{equation}
The $(0,\Rep{133})$ corresponds to the adjoint representation of $\Ex 7$, while the $(\pm1,\Rep{56})$ give rise to chiral matter superfields with  $\pm1$ $U(1)$ charges transforming in the $\underline{\bf56}$ representation of $\Ex 7$ in four dimensions. 
The $(\pm2,\Rep 1)$ are $E_7$ singlet chiral superfields fields with charges $\pm 2$ under $U(1)$. Finally, the $(0,\Rep{1})$ gives the one-dimensional adjoint representation of the $\Uni 1$ gauge group. The embedding of the line bundle is such that fields with $U(1)$ charge $-1$ are counted by $H^{*}(X,L)$, charge $-2$ fields are counted by $H^{*}(X,L^{2})$ and so on.\footnote{This is due to the form of the gauge transformation of the matter fields specified in \eqref{eq:delta_C}. This was chosen so as to agree with \cite{Wess:1992cp,Anderson:2009nt}.}

The low-energy massless spectrum can be determined by examining the chiral fermionic zero-modes of the Dirac operators for the various representations in the decomposition of the $\Rep{248}$. Generically, the Euler characteristic $\chi(\mathcal{F})$ counts $n_{\text{R}}-n_{\text{L}}$, where $n_{R}$ and $n_{L}$ are the number of right- and left-chiral zero-modes respectively transforming under the representation associated with the bundle $\mathcal{F}$. With the notable exception of $\mathcal{F}=\mathcal{O}_{X}$, which is discussed below, paired right-chiral and left-chiral zero-modes are assumed to form a massive Dirac fermion and are integrated out of the low-energy theory. Therefore, it is precisely the difference of the number of right-chiral fermions minus the left-chiral fermions, counted by the Euler characteristic $\chi$, that give the massless zero-modes of the $D=4$ theory. On a Calabi–Yau threefold $X$, $\chi(\mathcal{F})$ can be computed by the Atiyah–Singer index theorem as
\begin{equation}
\chi(\mathcal{F})=\sum_{i=0}^{3}(-1)^{i}h^{i}(X,\mathcal{F})=\int_{X}\op{ch}(\mathcal{F})\wedge\op{Td}(X)\ ,
\end{equation}
where $h^{i}$ are the dimensions of the $i$-th cohomology group, $\op{ch}(\mathcal{F})$ is the Chern character of $\mathcal{F}$, and $\op{Td}(X)$ is the Todd class of the tangent bundle of $X$. When $\mathcal{F}=L=\mathcal{O}_{X}(l^{1},l^{2},l^{3})$ is a line bundle, this simplifies to
\begin{equation}\label{eq:chi}
\chi(L)=\tfrac{1}{3}(l^{1}+l^{2})+\tfrac{1}{6}d_{ijk}l^{i}l^{j}l^{k}\ .
\end{equation}
Unlike the case of an $\SU N$ bundle, when $L$ is a line bundle with non-vanishing first Chern class, $\chi$ can receive contributions from all \emph{four} $h^i$, $i=0,1,2,3$. For example, $h^1(X,L)+h^3(X,L)$ then counts the number of (left-handed) chiral multiplets while $h^0(X,L)+h^2(X,L)$ counts (right-handed) anti-chiral multiplets, both transforming in the $(-1,\Rep{56})$ representation. Note that the multiplets counted by $h^0(X,L)+h^2(X,L)$ are simply the CPT conjugate partners of those already counted by $h^1(X,L^{-1})+h^3(X,L^{-1})$. Since it is conventional to give a supersymmetric matter spectrum in terms of (left-handed) chiral supermultiplets, it is sufficient to compute $h^1+h^3$ for the various bundles under consideration.

Using \eqref{eq:chi}, it is straightforward to compute the value of $\chi$ for the powers of $L$ associated with the decomposition \eqref{red55}. These are presented in Table \ref{tab:chiral_spectrum}. Having done this, let us discuss the spectrum in more detail.\footnote{See \cite{Braun:2005ux}, for example, for a similar discussion of the hidden-sector spectrum for an $\SU 2$ bundle.} 
\begin{table}
	\noindent \begin{centering}
		\begin{tabular}{rrr}
			\toprule 
			$U(1) \times \Ex 7$ & Cohomology & Index $\chi$\tabularnewline
			\midrule
			\midrule 
			$(0,\Rep{133})$ & $H^{*}(X,\mathcal{O}_{X})$ & $0$\tabularnewline
			\midrule 
			$(0,\Rep 1)$ & $H^{*}(X,\mathcal{O}_{X})$ & $0$\tabularnewline
			\midrule 
			$(-1,\Rep{56})$ & $H^{*}(X,L)$ & $8$\tabularnewline
			\midrule 
			$(1,\Rep{56})$ & $H^{*}(X,L^{-1})$ & $-8$\tabularnewline
			\midrule 
			$(-2,\Rep 1)$ & $H^{*}(X,L^{2})$ & $58$\tabularnewline
			\midrule 
			$(2,\Rep 1)$ & $H^{*}(X,L^{-2})$ & $-58$\tabularnewline
			\bottomrule
		\end{tabular}
		\par\end{centering}
	\caption{The chiral spectrum for the hidden sector $\protect\Uni 1\times\protect\Ex 7$ with a single line bundle $L=\mathcal{O}_{X}(2,1,3)$. The Euler characteristic (or index) $\chi$ gives the difference between the number of right- and left-chiral fermionic zero-modes transforming in the given representation. We denote the line bundle dual to $L$ by $L^{-1}$ and the trivial bundle $L^{0}$ by $\mathcal{O}_{X}$.\label{tab:chiral_spectrum}}
	
\end{table}
\begin{itemize}
	\item The index of the bundle $\mathcal{O}_X$ associated with the $(0,\Rep{133})$ and $(0,\Rep{1})$ representations vanishes, so the corresponding fermionic zero-modes must be \emph{non-chiral}. As discussed in \cite{Braun:2013wr}, since the trivial bundle $\mathcal{O}_{X}$ has $h^{0}(X,\mathcal{O}_{X})=h^{3}(X,\mathcal{O}_{X})=1$ and zero otherwise, there is a single right-chiral fermionic zero-mode (counted by $h^{0}$) and a single left-chiral fermionic zero-mode (counted by $h^{3}$), which combine to give the conjugate gauginos in a massless vector supermultiplet. In other words, the low-energy theory has one vector supermultiplet transforming in the $(0,\Rep{133})$ adjoint representation of $E_{7}$ and one vector supermultiplet in the $(0,\Rep{1})$ adjoint representation of $\Uni 1$. 
	
	\item The $(1, \Rep{56})$ multiplets are counted by $H^{*}(X,L^{-1})$. Since $\chi(L^{-1})=-8$, there are 8 unpaired left-chiral fermionic zero-modes that contribute to 8 chiral matter supermultiplets transforming in the $(1,\Rep{56})$ of $U(1) \times E_{7}$.
        \item Similarly, the $(-1, \Rep{56})$ multiplets are counted by $H^{*}(X,L)$. Since $\chi(L)=8$, there are 8 unpaired right-chiral fermionic zero-modes that contribute to 8 anti-chiral matter supermultiplets transforming in the $(-1,\Rep{56})$ of $U(1) \times E_{7}$.  However, these do not give extra fields in the spectrum: they are (right-handed) anti-chiral $(-1, \Rep{56})$ supermultiplets which are simply the CPT conjugate partners of the 58 chiral $(1, \Rep{56})$ supermultiplets already counted above~\cite{Green:1987mn}.
	\item Since $\chi(L^{-2})=-58$, there are 58 unpaired left-chiral fermionic zero-modes that contribute to 58 chiral matter supermultiplets transforming in the $(2,\Rep{1})$ representation of $U(1) \times E_{7}$.

	\item Similarly, the  $(-2,\Rep{1})$ multiplets are counted by $H^{*}(X,L^{2})$. Since $\chi(L^{2})=58$, there are 58 unpaired right-chiral fermionic zero-modes that contribute to 58 charged anti-chiral matter supermultiplets transforming in the $(2,\Rep{1})$ representation of $U(1) \times E_{7}$. However, as discussed above, these do not give extra fields in the spectrum: they are (right-handed) anti-chiral $(-2,\Rep{1})$ supermultiplets which are simply the CPT conjugate partners of the 58 chiral $(2,\Rep{1})$ supermultiplets already counted above. 
	
\end{itemize}
In summary, the $\Uni 1\times\Ex 7$ hidden sector massless spectrum for $L=\mathcal{O}_{X}(2,1,3)$ is
\begin{equation}
1\times(0,\Rep{133})+1\times(0,\Rep{1})+8\times(1,\Rep{56})+58\times(2,\Rep{1})\ ,\label{eq:matter}
\end{equation}
corresponding to one vector supermultiplet transforming in the adjoint representation of $\Ex7$, one $\Uni1$ adjoint representation vector supermultiplet, eight chiral supermultiplets transforming as $(1,\Rep{56})$ and 58 chiral supermultiplets transforming as $(2,\Rep{1})$. 

Note that since we have a chiral spectrum charged under $\Uni1$ with all positive charges, the $\Uni1$ gauge symmetry will be anomalous. As we discuss in the next subsection, this anomaly is canceled by the four-dimensional version of the Green--Schwarz mechanism which, in addition, gives a non-zero mass to this ``anomalous'' hidden sector $U(1)$.

\subsection{\texorpdfstring{$D=4$}{D = 4} Effective Lagrangian and the Anomalous \texorpdfstring{$U(1)$}{U(1)} Mass}

Before proceeding to the discussion of $N=1$ supersymmetry, it will be useful to present the $D=4$ effective theory for the hidden sector and to explicitly compute the anomalous mass of the $U(1)$ gauge boson. We present the results for a generic hidden sector line bundle $L=\mathcal{O}_{X}(l^1,l^2,l^3)$ with an arbitrary embedding into the hidden sector $E_{8}$. However, we conclude subsection 7.4.2 by computing the anomalous mass associated with the specific line bundle $L=\mathcal{O}_{X}(2,1,3)$  embedded  into $E_{8}$ as in \eqref{red5} with $a=1$.  

\subsubsection{\texorpdfstring{$D=4$}{D = 4} Effective Lagrangian}
Following the conventions of \cite{Brandle:2003uya,Freedman:2012zz}, the relevant terms in the four-dimensional effective action for the hidden sector of the strongly coupled heterotic string are
\begin{equation}
\mathcal{L}=\ldots -G_{LM}D_{\mu}C^{L}D^{\mu}{\bar{C}}^{M}-\tfrac{1}{2}g_{ij}D_{\mu}T^i D^{\mu}\bar{T}^{j}-\frac{4a\re f_{2}}{16\pi\ah}F_{2}^{\mu\nu}F_{2\mu\nu}-\frac{\pi\ah}{2a\re f_{2}}D_{\Uni 1}^{2}\ , \label{eq:het_lagrangian-4} 
\end{equation}
where $C^L$ denote the scalar components of the charged zero-mode chiral superfields, generically with different $U(1)$ charges $Q^{L}$ discussed in the previous subsection, $T^{i}$ are the {\it complex} scalar components 
\begin{equation}
T^{i}=t^{i}+\ii\,2\chi^{i}\qquad i=1,2,3 \ ,
\label{gd1}
\end{equation}
of the Kähler moduli superfields, where the $t^{i}$ are defined in \eqref{47} and $\chi^{i}$ are the associated axions, and $F_{2\mu\nu}$ is the hidden sector four-dimensional $\Uni 1$ field strength. The K\"ahler metrics $G_{LM}$ and $g_{ij}$ are functions of the dilaton and K\"ahler moduli with positive eigenvalues. As we will see below, the exact form of $G_{LM}$ is not important in this paper, whereas the exact form of $g_{ij}$ will be essential in the calculation of the anomalous $U(1)$ vector superfield mass. An explicit calculation of $g_{ij}$ is presented in Appendix C. Note that we have written the kinetic term for the hidden sector gauge field as a trace over $\Uni1$ instead of $\Ex8$ using \eqref{eq:trace_identity}, so that $\tr_{E_{8}} F_2^{\mu\nu} F_{2\mu\nu}=4a\,F_2^{\mu\nu} F_{2\mu\nu}$. The final term in \eqref{eq:het_lagrangian-4} is the potential energy, where $D_{\Uni 1}$ is proportional to the solution of the auxiliary D-field equation of motion and is given by
\begin{equation}
D_{\Uni 1}=FI-Q^{L}C^{L}G_{LM}{\bar{C}}^{M} \ .
\label{again1}
\end{equation}
The complex scalar fields $C^{L}$ enter the expression for $D_{U(1)}$ since they transform linearly under $U(1)$ with charge $Q^{L}$. Following   \eqref{again2A},  $FI$ is the genus-one corrected Fayet--Iliopoulos term, which is associated with a single line bundle and a single five-brane located at $\lambda\in[-1/2,1/2]$. In ``unity'' gauge, where ${\epsilon_S^\prime \hat R}/{V^{1/3}}=1$, it is given by
\begin{equation}
  \label{again2} 
FI= \frac{a}{2} \frac{ \epsilon_S \epsilon_R^2}{\kappa_{4}^{2}}
  \frac{1} {\Rhat V^{2/3}} \bigl(d_{ijk} l^i a^j a^k - a\,d_{ijk}l^il^jl^k 
  - l^i(2,2,0)|_i
  +(\tfrac{1}{2}+\lambda)^2l^iW_i \bigr) \ ,
\end{equation}
with the volume modulus $V$ and $W_{i}$ presented in \eqref{10} and \eqref{33A} respectively. 

\subsubsection{The Anomalous \texorpdfstring{$U(1)$}{U(1)} Mass}

As is commonly known, a $\Uni 1$ symmetry that appears in the both the internal and four-dimensional gauge groups is generically anomalous~\cite{Dine:1986zy,Dine:1987xk,Lukas:1999nh,Blumenhagen:2005ga}. Hence, there must be a Green–Schwarz mechanism in the original heterotic M-theory which will cancel this anomaly in the effective field theory. Importantly, however, in addition to canceling this anomaly, the Green--Schwarz mechanism will give a mass for the $U(1)$ vector superfield~\cite{Green:1984sg}. This occurs as follows. The Green--Schwarz mechanism  leads to a non-linear $U(1)$ action on the $\chi^{i}$ axionic partners of the $a^{i}$ K\"ahler moduli. That is, under a $U(1)$ gauge transformation, one finds that 
\begin{equation}
\delta\chi^{i}=-a\epsilon_{S}\epsilon_{R}^{2}\varepsilon l^{i} \ ,
\label{again5}
\end{equation}
where the $\epsilon_{S}$ and $\epsilon_{R}$ parameters are defined in \eqref{40AA} and \eqref{soc1} respectively, $a$ is the parameter associated with the embedding of the line bundle into $E_{8}$ and $\varepsilon$ is a gauge parameter. It follows that to preserve $U(1)$ gauge invariance, the kinetic energy term for the complex K\"ahler moduli must be written with a covariant derivative of the form
\begin{equation}
D_{\mu}T^i =\partial_{\mu} T^{i}+i2a\epsilon_{S}\epsilon_{R}^{2}l^{i}A_{\mu} \ .
\label{again6}
\end{equation}
Inserting this into the kinetic energy term for the $T^{i}$ moduli in \eqref{eq:het_lagrangian-4}, and scaling the gauge connection $A_{\mu}$ so that its kinetic energy term is in the canonical form $-\frac{1}{4}F_{\mu \nu}F^{\mu \nu}$, generates a mass for the $U(1)$ vector superfield given by
\begin{equation}
m_{A}^{2}=\frac{\pi\ah}{a\re f_{2}}2a^{2}\epsilon_{S}^{2}\epsilon_{R}^{4}g_{ij} l^{i}l^{j} \ .
\label{again7}
\end{equation}
The subscript $A$ refers to the fact that this mass arises from the Green--Schwarz mechanism required to cancel the gauge anomaly in the effective field theory. Using \eqref{sun11}, \eqref{sun2} and \eqref{sun3A}, one can evaluate the metric $g_{ij}$, which is presented in \eqref{pen1}. Inserting this into \eqref{again7} leads to an expression 
for $m_A^2$ of the form
\begin{equation}
m_A^2 =\frac{\pi\hat\alpha_{GUT}}{a \re f_2}\frac{a^2\epsilon_S^2\epsilon_R^4}{\kappa_{4}^{2}\Rhat^{2}}
\left(\frac{1}{8V^{4/3}}\mu(L)^{2}-\frac{1}{2V^{1/3}}d_{ijk}l^{i}l^{j}a^{k}\right) \ ,
\label{eq:anomalous_massA}
\end{equation}
which is valid for a generic line bundle $L=\mathcal{O}_{X}(l^1,l^2,l^3)$ embedded arbitrarily into the hidden sector $E_{8}$. 
\begin{figure}[t]
   \centering
\begin{subfigure}[c]{0.49\textwidth}
\includegraphics[width=1.0\textwidth]{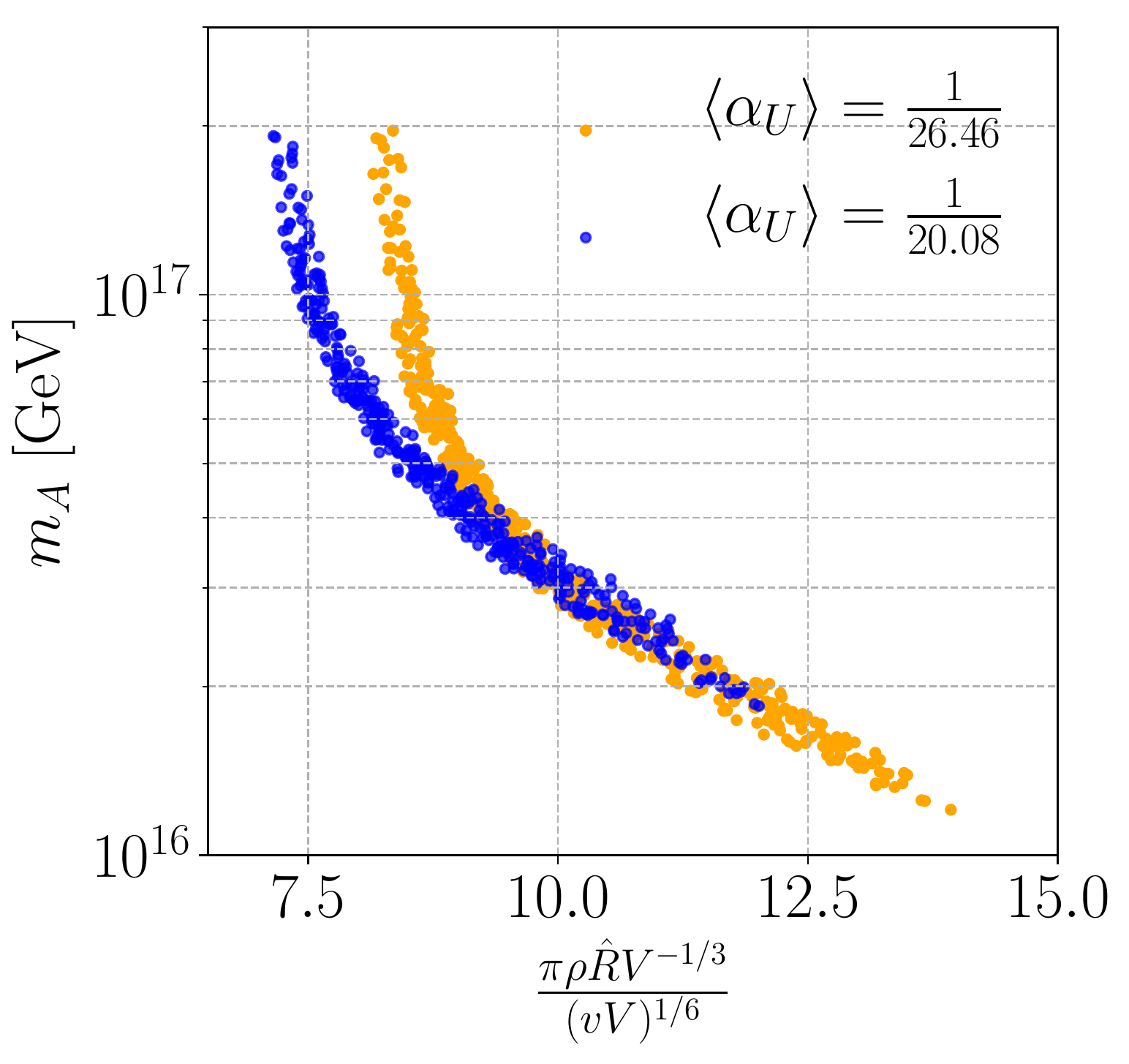}
\end{subfigure}
\caption{Value of $m_{A}$ versus the ratio ${\pi \rho {\Rhat} V^{-1/3}}/{(vV)^{1/6} }$ of the five-dimensional orbifold length to the averaged Calabi--Yau radius at different points across the magenta region shown in Figure 10, for both the split (blue) and simultaneous (orange) Wilson line scenarios.}
\label{fig:2DAnomalyPlots}
\end{figure}

We conclude this subsection, by evaluating \eqref{eq:anomalous_massA} for the specific line bundle $L=\mathcal{O}_{X}(2,1,3)$  embedded  into $E_{8}$ as in \eqref{red5} with $a=1$. We display in Figure \ref{fig:2DAnomalyPlots} the value of $m_{A}$ versus the ratio ${\pi \rho {\Rhat} V^{-1/3}}/{(vV)^{1/6} }$ of the five-dimensional orbifold length to the average Calabi--Yau radius at different points across the magenta region shown in Figure \ref{fig:Intersection} for both the split and simultaneous Wilson line scenarios.

\subsection{Supersymmetric Vacuum Solutions in Four Dimensions}

The generic form of the $U(1)$ D-term in the four-dimensional effective theory for an arbitrary hidden sector line bundle $L=\Ocal_X(l^{1}, l^{2}, l^{3})$ was presented in \eqref{again1}.
Using this result, we will now discuss the conditions for unbroken $N=1$ supersymmetry in the four-dimensional theory.

$N=1$ supersymmetry will be preserved in the $D=4$ effective theory only if the $D_{\Uni 1}$ term presented in \eqref{again1} vanishes at the minimum of the potential energy; that is
\begin{equation}
\langle D_{\Uni 1} \rangle=0 \ .
\end{equation}
Whether or not the $D=4$ effective theory can satisfy this condition, and the exact details as to how it does so, depends strongly on the value of the Fayet--Iliopoulos term. There are two generic possibilities.
\begin{enumerate}
\item[](i) The genus-one corrected FI-term vanishes. In this case, the VEVs of the scalar fields either all vanish or the VEVs of those fields with opposite charge, should they exist, cancel against each other.
\item[](ii) The genus-one corrected FI-term is non-vanishing. In this case, non-zero VEVs of the scalar fields $C^L$ with the same sign as $FI$ turn on to cancel the non-vanishing FI-term.
\end{enumerate}
Each of the two scenarios comes with its own conditions which have to be met. In the first case, in order to obtain a vanishing FI-term, the strong coupling $\kappa_{11}^{2/3}$ corrections to the slope need to cancel the tree-level ``classical'' slope in \eqref{again2}. For that to happen, one needs to be in a very strongly coupled regime, where working only to order $\kappa_{11}^{2/3}$ may be a poor approximation. We provide a detailed discussion about the strong coupling expansion to first and higher order in Appendix D. In the second case, the low-energy spectrum needs to contain scalars $C^L$ with the correct charge $Q^L$ under $U(1)$, such that their VEVs can cancel the non-zero $FI$ contribution. In such a scenario, one can move in K\"ahler moduli space--while still satisfying all the vacuum and phenomenological constraints (that is, generically, outside the magenta region of Figure 10 while remaining within the brown region)--to a less strongly coupled regime in which the first-order expansion to $\kappa_{11}^{2/3}$ is more accurate. However, as we will show in Section 7.5.2, the VEVs of these scalar fields may deform the hidden sector line bundle to an $SU(2)$ bundle, which might or might not be slope-stable. 

\subsubsection{Vanishing FI}\label{sec:vanishing_FI}

Let us start by analyzing the first case. A simple way to ensure unbroken $N=1$ supersymmetry in $D=4$ and slope stability of the hidden sector bundle is to require that
\begin{equation}
FI=0 \ .
\label{lamp1}
\end{equation}
There are then two scenarios in which supersymmetry can remain unbroken in the low energy theory. These are the following:
\begin{enumerate}
	\item The first, and simplest, possibility is that the charges $Q^{L}$ of the scalar fields $C^L$ are all of the same sign. It follows that the potential energy will set all VEVs to zero, $\langle C^{L} \rangle=0$, and hence $D_{\Uni 1}$ will vanish at this {\it stable} vacuum. Thus $N=1$ supersymmetry will be unbroken.
	\item A second possibility is that some of the $Q^{L}$ signs may differ. This will lead to unstable, flat directions where both $D_{\Uni 1}$ and the potential energy vanish. If one is at an arbitrary point, away from the origin, in a flat direction, then at least two VEVs will be non-vanishing $\langle C^{L} \rangle \neq 0$ and, hence, although preserving $N=1$ supersymmetry, such a vacuum would also spontaneously break the $U(1)$ symmetry. In such a scenario, the non-zero VEVs of the $C^L$ scalars give a mass to the $\Uni1$ vector field via the super-Higgs effect.  Scaling the gauge connection $A_{\mu}$ so that its kinetic energy term is in the canonical form $-\frac{1}{4}F_{\mu \nu}F^{\mu \nu}$, the value of this mass is easily computed and found to be
	\begin{equation}
		m_{D}^{2}=\frac{\pi\ah}{a\re f_{2}} Q^{L} {Q}^{M}G_{LM}\langle C^{L}\rangle\langle {\bar{C}}^{M}\rangle \ .
		\label{again4A}
	\end{equation}
\end{enumerate}
Having discussed this second possibility, we note again that the associated potential energy must have at least one flat direction and, hence, is an {\it unstable} vacuum state. For this reason, we will ignore such vacua in this paper. However, the first scenario is easily satisfied, as we now demonstrate with an explicit example.

\subsubsection*{$FI=0$ Example: $N=1$ Supersymmetry for $L=\mathcal{O}_X(2,1,3)$}

We now discuss $N=1$ supersymmetry in the example introduced above, where the line bundle is taken to be $L=\mathcal{O}_X(2,1,3)$ and embedded into $SU(2) \subset E_{8}$ as in \eqref{red4} with coefficient $a=1$, and the location of the single five-brane is at $\lambda=0.49$.
Recall that the genus-one corrected FI-term for this line bundle and embedding was presented in \eqref{trenton2} and given by
\begin{equation}
FI= \frac{ \epsilon_S \epsilon_R^2}{2\kappa_{4}^{2}}
  \frac{1} {\Rhat V^{2/3}} \left(  \tfrac{1}{3}(a^1)^{2}+\tfrac{2}{3}(a^2)^{2} +8a^1a^2+4a^2a^3 +2a^1a^3 -13.35 \right) \ .
\label{bird1}
\end{equation}
In Figure 10, the region of K\"ahler moduli space which satisfies all of the required constraints, including slope stability, was presented. As discussed in detail in subsection 7.2, in order for the bundle $L \oplus L^{-1}$ to be polystable, it was necessary to restrict this region of moduli space to points that set the genus-one corrected slope \eqref{trenton3} -- and, hence, the FI-term in \eqref{bird1} -- to zero. Furthermore, the low-energy scalar spectrum carrying non-vanishing $U(1)$ charge was determined in subsection 7.3. It was shown there that the low-energy scalar spectrum of the hidden sector -- specifically $8 \times (1,\Rep{56})+58 \times (2,\Rep{1})$ -- each had charges $Q^{L}$ of the same sign. It then follows from the above discussion that the potential energy must have a unique minimum where the VEVs vanish, $\langle C^{L} \rangle=0$, such that $\langle D_{\Uni 1} \rangle=0$ at this minimum. Hence, $N=1$ supersymmetry is {\it unbroken} in the vacuum state of the $D=4$ effective theory. Since the VEVs of all light $U(1)$ charged scalar fields vanish for this explicit example, it follows from \eqref{again4A} that
\begin{equation}
m_{D}=0 \ .
\end{equation}
However, as discussed above, since the $U(1)$ symmetry is anomalous, the mass $m_{A}$ presented in \eqref{eq:anomalous_massA} is non-vanishing and, for this explicit example, plotted in Figure 12.

We would like to point out that $L=\mathcal{O}_X(2,1,3)$ is not the only hidden sector line bundle which, if embedded into $SU(2) \subset E_{8}$ as in \eqref{red5} with $a=1$, has a region of K\"ahler moduli space where all required constraints are satisfied, $FI=0$ and the $D=4$ vacuum preserves $N=1$ supersymmetry. However, any such line bundle $L$ must be ``ample''--that is, each of its defining integers $l^{i}, i=1,2,3$ where $l^{1}+l^{2}=0~ mod ~3$ must either be all positive or all negative. The reason is that for the Schoen manifold  defined in Section 2, one can show that the genus-one corrected Fayet-Iliopoulos term can vanish, that is, $FI=0$, if and only if $L$ is ample. Restricting to ample line bundles, one can indeed find a significant number satisfying all required constraints. However, of these, many have a large number of equal sign zero-mode chiral multiplets -- some with large charges $Q^{L}$--making them incompatible with spontaneous supersymmetry breaking via gaugino condensation. While potentially of physical interest, we wish to focus on the subset of ample line bundles that have a sufficiently small zero-mode chiral spectrum, with sufficiently small charges, to be compatible with supersymmetry breaking via $E_{7}$ gaugino condensation. These line bundles are specified by
	\begin{gather}\label{many}
		\mathcal{O}_X(2,1,3)\ , \qquad \mathcal{O}_X(1,2,3)\ ,\qquad \mathcal{O}_X(1,2,2)\ ,  \qquad\mathcal{O}_X(2,1,2)\ , \nonumber \\
		\mathcal{O}_X(2,1,1)\ , \qquad\mathcal{O}_X(1,2,1)\ ,  \qquad\mathcal{O}_X(2,1,0) \ ,
	\end{gather}
and their duals; that is, for example, $\mathcal{O}_X(-2,-1,-3)$. Spontaneous supersymmetry breaking via $E_{7}$ gaugino condensation in this context will be explored in a future publication.

As discussed at the beginning of this section, although this hidden sector vacuum satisfies all required physical and phenomenological constraints, setting the FI-term to zero necessitates exact cancellation of the genus-one corrected slope against the tree-level slope of the hidden sector line bundle. Unsurprisingly, this fine-tuning can only be carried out in a relatively strongly coupled regime of heterotic M-theory -- thus making the validity of the linearized approximation used in this paper uncertain. This is made more explicit and discussed in detail in Appendix D. It is, therefore, of some interest to explore vacua for which the genus-one corrections to the slope are significantly smaller than the tree level slope of the hidden sector bundle. In this case, one expects the effective coupling parameter to be smaller than in the previous scenario and, hence, the linearized results used in this paper to be a better approximation. For this reason, we now introduce the basic criteria required by hidden sector vacua where the FI-term does not vanish. Explicit examples of such hidden sector vacua will be presented in a future publication.

\subsubsection{Non-vanishing FI}

We now consider what happens when the $\kappa_{11}^{2/3}$ correction to the tree-level slope is small and so cannot be used to set the FI-term to zero. The question then is, given a non-vanishing FI-term
\begin{equation}
	FI\neq0\ ,
\end{equation}
can one still preserve $N=1$ supersymmetry in the four-dimensional effective theory? 

Recall that the conditions for a supersymmetric vacuum are the vanishing of the F- and D-terms. For consistency with the previous content of this paper, let us continue to assume that the vacuum has an unbroken $U(1) \times \Ex 7$ gauge symmetry, embedded into $E_8$ as discussed in Section 4.2. It follows that the $\Ex 7$ D-terms will vanish by setting the VEVs of the non-abelian $(\pm1,\Rep{56})$ matter fields to be zero. Any F-terms involving the non-abelian matter fields will then also vanish. As discussed in \cite{Anderson:2010ty}, the F-term conditions for the $(\pm2,\Rep 1)$ matter fields permit us to give VEVs to only one set of fields, that is, either $(2,\Rep 1)$ or $(-2,\Rep 1)$ but not both. The remaining condition for the vacuum solution to be supersymmetric is the vanishing of the $\Uni 1$ D-term, $\langle D_{\Uni 1}\rangle=0$. Since the $FI$ term does \emph{not} vanish for any choice of line bundle when the $\kappa_{11}^{2/3}$ correction is small, one is forced to cancel the $FI$ term against the VEVs of the charged singlet fields. In other words, we want
\begin{equation}\label{eq:vevs}
	Q^{L}\langle C^{L}\rangle G_{LM}\langle\bar{C}^{M}\rangle=FI\quad\Rightarrow\quad\langle D_{\Uni 1}\rangle=0\ .
\end{equation}
Obviously, such a cancellation will depend on the relative sign of $FI$ and the charges of the scalars $C^{L}$. For example, if the $FI$ term is \emph{positive}, one needs at least one zero-mode chiral supermultiplet whose scalar component is a singlet under the non-abelian group and has \emph{positive} $\Uni 1$ charge. Whether or not such scalar fields are present will depend on the specific line bundle studied.

If one can cancel the FI-term in this way, the non-zero VEVs of the $C^{L}$ scalars give a mass to the $\Uni 1$ vector field via the super-Higgs effect:
\begin{equation}
	m_{D}^{2}=\frac{\pi\ah}{a\re f_{2}}Q^{L}Q^{M}\langle C^{L}\rangle G_{LM}\langle\bar{C}^{M}\rangle\ ,
\end{equation}
where the subscript $D$ indicates that this mass is due to the non-vanishing VEVs needed to set the D-term to zero. Note that in the case where one gives VEVs to fields of a single charge $Q^{L}$, the mass is related to the $FI$ term as
\begin{equation}
	m_{D}^{2}=\frac{\pi\ah}{a\re f_{2}}Q^{L}\,FI\ .
\end{equation}
As in the case of vanishing slope in the previous subsection, the $\Uni 1$ vector field mass also receives a contribution from the Green–Schwarz mechanism. Hence the total mass of the vector field is given by the sum of $m_{D}^{2}$ above and $m_{A}^{2}$ from \eqref{eq:anomalous_massA}.

As we discussed in Section \ref{sec:anomaly_cancellation}, the embedding of $\Uni 1$ inside $\Ex 8$ that we have considered for much of this paper factors through the $\SU 2$ subgroup of $\Ex 8$ that commutes with $\Ex 7$. The $\Uni 1$ gauge connection $A$ for the line bundle $L$ can be thought of as defining an $\Ex 8$ connection in two equivalent ways: either embedding directly in $\Ex 8$ via the generator $Q$ discussed around \eqref{red5}, or first embedding in $\SU 2$ as $\text{diag}(A,-A)$ and then embedding $\SU 2$ in $\Ex 8$ via \eqref{red3}. The second of these two pictures is helpful for understanding the effect of allowing non-zero VEVs for the charged singlets, $\langle C^{L}\rangle\neq0$. 

First note that the connection $\text{diag}(A,-A)$ is a connection for an $\SU 2$ bundle
which splits as a direct sum 
\begin{equation}
	\mathcal{V}=L\oplus L^{-1}
\end{equation}
of line bundles. How does this relate to supersymmetry? The induced $\Ex 8$ connection will solve the (genus-one corrected) Hermitian Yang–Mills equation, and so give a supersymmetric solution, if the $\SU 2$ connection itself solves the (genus-one corrected) Hermitian Yang–Mills equation. This is guaranteed if the rank two $L\oplus L^{-1}$ bundle is polystable with vanishing slope.\footnote{Here, slope is taken to mean the genus-one corrected slope. The same comments apply if one considers only the tree-level expression.} Since $\mu(\mathcal{V})=0$ by construction, the remaining conditions for polystability are
\begin{equation}
	\mu(L)=\mu(L^{-1})=0\ .
\end{equation}
This is exactly the vanishing $FI$ case studied in \ref{sec:vanishing_FI}, where the corrected slope of $L$ is set to zero and the VEVs of the charged singlet matter fields vanish.

When $\mu(L)\neq0$, the $\SU 2$ bundle $\mathcal{V}$ is no longer polystable and so its connection does not solve the Hermitian Yang–Mills equation. The four-dimensional consequence of this is that the $FI$ term no longer vanishes. However, we might be able to turn on VEVs for appropriate charged singlet matter fields in order to cancel the $FI$ term and set the D-term to zero, thus preserving superysymmetry. One might wonder: what is the bundle interpretation of turning on VEVs for these charged singlet matter fields? As discussed in \cite{0905.1748,1012.3179,1506.00879}, these VEVs should be seen as deforming the gauge bundle away from its split form $\mathcal{V}=L\oplus L^{-1}$ to a \emph{non-split} $\SU 2$ bundle $\mathcal{V}'$ which admits a connection that \emph{does} solve the Hermitian Yang--Mills equations.

Consider the case where $\mu(L)>0$ (equivalent to $FI>0$) in some region of K\"ahler moduli space where the constraints of Section \ref{sec:constraints} are all satisfied. From \eqref{eq:vevs} we see that one can set $\langle D_{\Uni 1}\rangle=0$ provided we have charged scalars $C^{L}$ with positive charge, $Q^{L}>0$. From the generic form of the cohomologies in Table \ref{tab:chiral_spectrum}, the required scalars are those transforming in $(2,\Rep 1)$, with the chiral superfields which contain these scalars counted by $h^{1}(X,L^{-2})+h^{3}(X,L^{-2})$. Hence, giving VEVs to $(2,\Rep 1)$ scalars corresponds to allowing non-trivial elements of $H^{1}(X,L^{-2})\oplus H^{3}(X,L^{-2})$. The first summand has an interpretation as the space of extensions of $L^{-1}$ by $L$, with the exact sequence
\begin{equation}
	0\to L^{-1}\to\mathcal{V}'\to L\to0
\end{equation}
defining an $\SU 2$ bundle $\mathcal{V}'$. This extension can be non-trivial ($\mathcal{V}\neq\mathcal{V}'$) provided
\begin{equation}
	\op{Ext}^{1}(L^{-1},L) = H^{1}(X,L^{-2})\neq0\ .
\end{equation}
Choosing a non-zero element of this space then corresponds to turning on VEVs for some set of $(2,\Rep 1)$ scalars. Thus we see that giving VEVs to positively charged singlet scalars arising from $H^1(X,L^{-2})$ amounts to deforming the induced $L \oplus L^{-1}$ bundle $\mathcal{V}$ to the $\SU 2$ bundle $\mathcal{V}'$. Note that if $H^1(X,L^{-2})=0$, the VEVs of positively charged matter coming from $H^3(X,L^{-2})$ cannot be interpreted as deforming to a new $SU(2)$ bundle. Hence, the bundle remains  $L \oplus L^{-1}$ which is unstable and, therefore, its gauge connection does not solve the Hermitian Yang--Mills equation.

Assuming one can show for a given line bundle $L$ that $H^1(X,L^{-2}) \neq 0$, it might seem that we are done – the $\Uni 1$ D-term vanishes and supersymmetry appears to have been restored. However, the four-dimensional analysis is insensitive to whether the new bundle $\mathcal{V}'$ is slope stable and thus actually admits a solution to Hermitian Yang–Mills. Unfortunately, checking slope stability is a difficult calculation that one must do explicitly for each example. As a preliminary check, one can first see whether $\mathcal{V}'$ satisfies some simpler \emph{necessary} conditions for slope stability. First, the obvious subbundle $L^{-1}$ should not destabilise $\mathcal{V}'$. In our case this is guaranteed as we have assumed $\mu(L)>0$, so that $L^{-1}$ has negative slope.\footnote{If instead $\mu(L)<0$, one simply swaps the roles of $L$ and $L^{-1}$ in the above discussion and instead considers the extension of $L^{-1}$ by $L$.} Second, $\mathcal{V}'$ must satisfy the Bogomolov inequality~\cite{huybrechts2010geometry}. For a bundle with vanishing first Chern class, this states that if $\mathcal{V}'$ is slope stable with respect to some choice of K\"ahler class $\omega=a^{i}\omega_{i}$, then
\begin{equation}
	\int_{X}c_{2}(\mathcal{V}')\wedge\omega\geq0\ .
\end{equation}
Since $\mathcal{V}'$ is constructed as an extension of line bundles, we have
\begin{equation}
	c_{2}(\mathcal{V}')\equiv c_{2}(L\oplus L^{-1})=-\tfrac{1}{2}c_{1}(L)\wedge c_{1}(L)\ ,
\end{equation}
with $c_{1}(L)={v^{-1/3}}l^{i}\omega_{i}$. Thus if $\mathcal{V}'$ is to be slope stable, we must be in a region of K\"ahler moduli space where
\begin{equation}\label{eq:Bolg}
	-\tfrac{1}{2}\int_{X}c_{1}(L)\wedge c_{1}(L)\wedge\omega\geq0\quad\Rightarrow\quad d_{ijk}l^{i}l^{j}a^{k}\leq0\ .
\end{equation}
Note that this is a necessary but not sufficient condition. However, it is often the case that the Bogomolov inequality is the only obstruction to finding stable bundles~\cite{Braun:2005zv}.

We thus have a new set of necessary conditions on $L$ (in addition to the physically and mathematically required constraints presented in Section \ref{sec:constraints}) for there to be a supersymmetric vacuum after turning on the VEVs to cancel the FI-term. These are
\begin{enumerate}
	\item Singlet matter with the correct charge must be present, so that $FI$ can be canceled and the D-term set to zero.
	\item $H^1(X,L^{-2})$ must not vanish.
	\item The Bogomolov inequality, $d_{ijk}l^{i}l^{j}a^{k}\leq0$, must be satisfied.
\end{enumerate}

Does our previous choice of $L=\mathcal{O}_{X}(2,1,3)$ satisfy these conditions? Note that $\mu(L)>0$ everywhere in the K\"ahler cone for this line bundle. From its low-energy spectrum in Table \ref{tab:chiral_spectrum}, we see we have 58 massless positively charged singlets transforming in the $(2,\Rep 1)$ representation, and so we do indeed have the correct matter to cancel the FI-term and set the D-term to zero. However, as discussed in \cite{Braun:2013wr}, if $L$ is an {\it ample} line bundle then
\begin{equation}
 H^1(X,L)=0 \ .
 \label{snow1}
 \end{equation}
 Since $L=\mathcal{O}_{X}(2,1,3)$--and, hence, $L^{-2}$--is ample, it follows that $H^1(X,L^{-2})=0$. Therefore, condition 2 above implies that $L \oplus L^{-1}$ cannot admit an extension to an $SU(2)$ bundle $\mathcal{V}'$. Ignoring this for a moment, and assuming that there did exist an $SU(2)$ extension, we would still have to check whether or not the Bogomolov inequality, a necessary condition for $\mathcal{V}'$ to be slope stable, is satisfied.
However, from \eqref{eq:Bolg} and the positivity of the K\"ahler moduli, we see that it is {\it impossible} to satisfy this inequality, implying (again) that the split bundle constructed from $L=\mathcal{O}_{X}(2,1,3)$ cannot be deformed to admit a solution to the Hermitian Yang–Mills equation. Moreover, we see the same will be true for any ample line bundle -- the $l^{i}$ are positive and $d_{ijk}l^{i}l^{j}a^{k}\leq0$ is not satisfied anywhere in the positive K\"ahler cone.

What about other choices of line bundle? It turns out that of the three conditions, the Bogomolov inequality is the more difficult to satisfy. Scanning over different choices of $L$, one finds that in the region of K\"ahler moduli space where the $\SU 4$ bundle is stable, the only line bundles that are equivariant with $\mu(L)>0$,\footnote{We restrict to $\mu(L)>0$ in our scan to match our analysis above. Including bundles with $\mu(L)<0$ would give the reverse extension sequence with the bundle and its dual swapped, leading to the same $\SU 2$ bundles that were already captured by restricting to positive slope.} allow for anomaly cancellation and satisfy the Bogomolov inequality are
\begin{equation}
	\mathcal{O}_{X}(1,2,-1)\ ,\qquad\mathcal{O}_{X}(2,1,-1)\ ,\qquad\mathcal{O}_{X}(7,2,-2)\ ,\qquad\mathcal{O}_{X}(7,5,-3)\ .
\end{equation}
Do any of these have positively charged singlet matter in their low-energy spectrum to allow for a non-trivial extension? That is, do we have $H^{1}(X,L^{-2})>0$ for any of these candidate line bundles? For $\mathcal{O}_{X}(1,2,-1)$ and $\mathcal{O}_{X}(2,1,-1)$, it is simple to show using a Leray spectral sequence that the answer is no. For a definitive answer in the remaining two cases, one must extend the analysis of Appendix A of \cite{Braun:2005zv} to higher degree line bundles on dP$_9$. This is beyond the scope of the present paper. Therefore, for now, we content ourselves with noting that $\chi(L^{-2})$ is positive for both remaining line bundles, which is consistent with $H^{1}(X,L^{-2})=0$ and the absence of an extension to a $SU(2)$ bundle $\mathcal{V}'$.

As exploited by a number of other works~\cite{Anderson:2010ty,Anderson:2012yf,Nibbelink:2015ixa}, moving from a single line bundle to two or more such bundles provides a richer low-energy spectrum, making it much easier to find examples which possess the correct charged matter and satisfy both the phenomenological constraints and the Bogomolov inequality. We intend to pursue this in detail in future work.

\section{Conclusions}

In this paper, we have explicitly chosen the hidden sector line bundle $\mathcal{O}_X(2,1,3)$, embedded in a specific way with embedding coefficient $a=1$ into the $E_{8}$ gauge group, and studied its phenomenological properties. This choice of hidden sector was shown to satisfy all ``vacuum'' constraints required to be consistent with present low-energy phenomenology, as well as both the ``reduction'' and ``physical'' constraints required to be a ``strongly-coupled'' heterotic vacuum consistent with both the mass scale and gauge coupling of a unified $SO(10)$ theory in the observable sector. Additionally, we showed that the induced $\SU2$ bundle $L \oplus L^{-1}$ is polystable after including genus-one corrections, and that the effective low-energy theory admits an $N=1$ supersymmetric vacuum. We pointed out that there are actually a large number of different line bundles that one could choose, and a large number of inequivalent embeddings of such line bundles into $E_{8}$. An alternative choice of hidden sector bundle could lead to: 1) a different commutant subgroup $H$ and hence a different low-energy gauge group, 2) a different spectrum of zero-mass particles transforming under $H \times U(1)$, 3) a different value for the associated Fayet--Iliopoulos term and, hence, a different D-term mass for the $U(1)$ vector superfield, and so on. Furthermore, a richer zero-mode spectrum could open the door to mechanisms for arbitrary size spontaneous $N=1$ supersymmetry breaking, new dark matter candidates and other interesting phenomena. We will explore all of these issues in several upcoming papers.

\subsection*{Acknowledgements}

We would like to thank Yang-Hui He and Fabian Ruehle for helpful discussions. Anthony Ashmore and Sebastian Dumitru are supported in part by research grant DOE No.~DESC0007901. Burt Ovrut is supported in part by both the research grant DOE No.~DESC0007901 and SAS Account 020-0188-2-010202-6603-0338.

%%%%%%%%%%%%%%%%%%%%%%

\appendix

\section{Vacuum Constraints for Strongly Coupled Heterotic M-Theories}

The required vacuum constraints for heterotic M-theories, in the context of both weak and strong coupling, have been discussed in several previous papers \cite{Braun:2013wr,Ovrut:2018qog}. In particular, the explicit constraints involving both the observable and hidden sectors for phenomenologically realistic strongly coupled heterotic M-theory vacua were presented in detail in \cite{Ovrut:2018qog}. Since these form the starting point for the analysis in the present work, in this Appendix we will briefly list the important definitions, summarize the relevant results and emphasize the physically pertinent conclusions contained in \cite{Ovrut:2018qog}.

\subsection{The Hidden Sector Bundle}

In \cite{Ovrut:2018qog}, the hidden sector vector bundle was chosen to have the generic form of a Whitney sum 
\begin{equation}
V^{(2)}={\cal{V}}_{N} \oplus {\cal{L}}~, \qquad {\cal{L}}=\bigoplus_{r=1}^R L_r 
\end{equation}
where ${\cal{V}}_{N}$ is a slope-stable, non-abelian bundle and 
each $L_{r}$, $r=1,\dots,R$ is a holomorphic line bundle with  structure group $U(1)$. However, in this Appendix, we will restrict our discussion to hidden sector gauge bundles of the form
\begin{equation}
\qquad {V^{(2)}}={\cal{L}}=\bigoplus_{r=1}^R L_r 
\label{dude1bAA}
\end{equation}
Being a Whitney sum of line bundles,  $V^{(2)}$ must be polystable -- that is, each line bundle must have the same slope which is, however, prior to a discussion of $N=1$ supersymmetry in the low energy theory, otherwise unrestricted. 

\subsubsection{Properties of Hidden Sector Bundles}

It follows that form of the hidden sector bundle will be
\begin{equation}
  {\cal{L}} = \bigoplus_{r=1}^R L_r
  ,\quad 
  L_r=\Ocal_X(l^1_r, l^2_r, l^3_r)
  \label{21}
\end{equation}
where
\begin{equation}
  (l^1_r+l^2_r) \tmod 3 = 0
  , \quad r=1,\dots,R 
\end{equation}
for any positive integer $R$. The structure group is $U(1)^R$, where
each $U(1)$ factor has a specific embedding into the hidden sector
$E_8$ gauge group. It follows from the definition that
$\rank({\cal{L}})=R$ that the first Chern class is
\begin{equation}
  c_1({\cal{L}})
  =\sum_{r=1}^{R}c_1(L_r), \quad c_{1}(L_{r})=
  \frac{1}{v^{1/3}}  (l^1_r \omega_1 + l^2_r \omega_2 + l^3_r \omega_3) 
  .
\label{23}
\end{equation}
Note that since ${\cal{L}}$ is a sum of holomorphic line bundles,
$c_2({\cal{L}})=c_3({\cal{L}})=0$. However, the relevant quantity for the
hidden sector vacuum is the second Chern character defined in
\cite{Ovrut:2018qog}. For ${\cal{L}}$ this becomes
\begin{equation}
  ch_2({\cal{L}})
  = \sum_{r=1}^R ch_2(L_r) \ .
\label{24}
\end{equation}
Since $c_2(L_r)=0$, it follows that
\begin{equation}
  ch_2(L_r)=a_rc_1(L_r) \wedge c_1(L_r) 
  \label{25}
\end{equation}
where
\begin{equation}
  a_r=\tfrac{1}{4} \tr_{E_{8}} Q_r^2
  \label{26}
\end{equation}
with $Q_r$ the generator of the $r$-th $U(1)$ factor embedded into
the $\Rep{248}$ adjoint representation of the hidden sector $E_8$, and the trace is taken over the $\Rep{248}$ of $\Ex8$ (including a conventional factor of $1/30$).

\subsection{Anomaly Cancellation} 

As discussed in~\cite{Lukas:1998tt,Donagi:1998xe}, anomaly cancellation in heterotic M-theory requires that
\begin{equation}
  \sum_{n=0}^{N+1}J^{(n)}=0 ,
  \label{27}
\end{equation}
where 
\begin{equation}
  \label{28}
  \begin{split}
    J^{(0)}=&\;
    -\frac{1}{16 \pi^2}
    \Big( \tr_{E_8} F^{(1)} \wedge F^{(1)}
    -\frac{1}{2}\tr_{SO(6)} R \wedge R \Big) \\[1ex]
    J^{(n)}=&\;
    W^{(n)}, \quad n=1,\dots,N, \\[1ex]
    J^{(N+1)}=&\;
    -\frac{1}{16 \pi^2}
    \Big( \tr_{E_8} F^{(2)} \wedge F^{(2)}
    -\frac{1}{2}\tr_{SO(6)} R \wedge R \Big) \\
  \end{split}
\end{equation}
Note that the indices $n=0$ and $n=N+1$ denote the observable and hidden sector domain walls respectively, and {\it not} the location of a five-brane.
Using the definitions of the associated Chern characters, the anomaly cancellation condition can be
expressed as
\begin{equation}
  c_2(TX)-c_2(\Vvis)
  +\sum_{r=1}^R  a_r c_1(L_r) \wedge c_1(L_r) - W 
  = 0 ,
  \label{29}
\end{equation}
where we have restricted the hidden sector bundle to be of the form \eqref{21} and $W=\sum_{n=1}^N W^{(n)}$ is the total five-brane class. Furthermore, it follows from the properties of the Chern characters, and defining
\begin{equation}
  W_i = \frac{1}{v^{1/3}} \int_X W \wedge \omega_i \ ,
  \label{32}
\end{equation}
that the anomaly condition \eqref{29} can be expressed as 
\begin{equation}
  W_i= \big( \tfrac{4}{3},\tfrac{7}{3},-4\big)\big|_i
  +\sum_{r=1}^R a_r d_{ijk} l^j_r l^k_r \geq 0 \ , \quad i=1,2,3  .
\label{33}
\end{equation}
The positivity constraint on $W$ follows from the requirement that it
be an effective class to preserve $N=1$ supersymmetry.

Finally, it is useful to define the charges
\begin{equation}
  \beta^{(n)}_i = 
  \frac{1}{v^{1/3}}
  \int_X J^{(n)} \wedge \omega_i \ , \quad i=1,2,3  .
\label{34}
\end{equation}
For example, it follows from \eqref{28}, using results for the second Chern class of the observable sector gauge bundle given in \cite{Ovrut:2018qog} and the intersection numbers \eqref{3} and
\eqref{4}, we find that
\begin{equation}
  \beta^{(0)}_i = 
  \big( \tfrac{2}{3},-\tfrac{1}{3},4 \big)\big|_i \ .
  \label{35}
\end{equation}

\subsection{The Linearized Double Domain Wall}

The five-dimensional effective theory of heterotic M-theory, obtained
by reducing Hořava--Witten theory on the above Calabi--Yau
threefold, admits a BPS double domain wall solution with five-branes
in the bulk space \cite{Lukas:1998yy,Donagi:1999gc,Lukas:1998tt,Lukas:1998hk,Lukas:1999kt,Lukas:1997fg}. This solution depends on 
the moduli $V$ and $b^i$ defined in the text, as well as the $a$, $b$ functions of the
five-dimensional metric
\begin{equation}
 \dd s_5^2=a^2\dd x^{\mu}\dd x^{\nu}\eta_{\mu\nu}+b^2(\dd x^{11})^{2} \ ,
  \label{37}
\end{equation}
all of which are dependent on the five coordinates $x^{\alpha}$,
$\alpha=0,\dots,3,11$ of $M_4 \times S^1/\Z_2$. Denoting the reference radius of $S^{1}$ by $\rho$, then $x^{11} \in [0,\pi \rho]$.
The detailed structure of the linearized double domain wall depends on the solution of three non-linear equations discussed in \cite{Lukas:1998tt}. These can be approximately solved by expanding to linear order in the quantity $\epsilon_S'\beta^{(0)}_i\big(z-\frac{1}{2}\big)$, where we define  $z={x^{11}}/{\pi\rho}$ with $z \in [0,1]$, $\beta^{(0)}_i$ is given in \eqref{35} and 
\begin{equation}
  \epsilon'_S = \pi \epsilon_{S}\ , \qquad
  \epsilon_{S}= \left(\frac{\kappa_{11}}{4\pi} \right)^{2/3}\frac{2\pi\rho}{v^{2/3}} \ .
  \label{40}
\end{equation}
The parameters $v$ and $\rho$ are defined in the text and $\kappa_{11}$ is the dimensionful eleven-dimensional  Planck constant. It is also convenient to express the moduli of the theory in terms of orbifold average functions defined as follows. For an arbitrary dimensionless function $f$ of the five $M_4
\times S^1/\Z_2$ coordinates, define its average over the $S^1/\Z_2$
orbifold interval as
\begin{equation}
\langle f \rangle_{11}=\frac{1}{\pi \rho}\int_0^{\pi\rho}{\dd x^{11}f} \ ,
\label{42}
\end{equation}
where $\rho$ is the reference length. Then $\langle f \rangle_{11}$ is
a function of the four coordinates $x^{\mu}$, $\mu=0,\dots,3$ of $M_4$
only. The linearized solution is expressed in terms of orbifold
average functions
\begin{equation}
  V_0=\langle V \rangle_{11}\  , \quad 
  b^i_0=\langle b^i \rangle_{11} \ , \quad 
  \left(\frac{\Rhat_0}{2}\right)^{-\frac{1}{2}}=\langle a \rangle_{11}\  , \quad
  \frac{\Rhat_0}{2}=\langle b \rangle_{11} \ .
\label{42a}
\end{equation}
The fact that they are averaged is indicated by the subscript $0$.

The solution to these linearized equations depends on the number of five-branes located within the fifth-dimensional interval. Here, for simplicity, we will consider the vacuum to contain a single five-brane, wrapped on a holomorphic curve, and located at the fifth-dimensional coordinate $z_{1} \in [0,1]$. It was then shown in \cite{Lukas:1998tt} that the 
conditions for the validity of the linear approximation then break
into two parts. Written in terms of the averaged moduli, these are
\begin{equation}
  2\epsilon_S'\frac{\Rhat_0}{V_0}
  \left|
    \beta_i^{(0)} \big(z-\tfrac{1}{2}\big)
    -\frac{1}{2}\beta_i^{(1)}(1-z_1)^2
  \right|
  \ll 
  \left| d_{ijk} b_0^j b_0^k \right|
  \ , \quad z \in [0,z_1]\ ,
\label{45B}
\end{equation}
and 
\begin{equation}
  2\epsilon_S'\frac{\Rhat_0}{V_0}
  \left|
    (\beta_i^{(0)}+\beta_i^{(1)})
    \big(z-\tfrac{1}{2}\big)
    -\frac{1}{2}\beta_i^{(1)}z_1^2
  \right| 
  \ll 
  \left| d_{ijk} b_0^j b_0^k \right|
 \  , \quad z \in [z_1,1] \ .
\label{45C}
\end{equation}

When dimensionally reduced on this linearized BPS solution, the
four-dimensional functions $a_0^i$, $V_0$, $b_0^i$ and $\Rhat_0$
will become moduli of the $D=4$ effective heterotic M-theory. The
geometric role of $a_0^i$ and $V_0, b_0^i$ will remain the same as
above -- now, however, for the averaged Calabi--Yau threefold. For
example, the dimensionful volume of the averaged Calabi--Yau manifold
will be given by $vV_0$. The new dimensionless quantity $\Rhat_0$
will be the length modulus of the orbifold. The dimensionful length of
$S^1/\Z_2$ is given by $\pi \rho \Rhat_0$. Finally, since the
remainder of this Appendix will be within the context of the $D=4$
effective theory, we will, for simplicity, {\it drop the subscript $``0"$
on all moduli} henceforth -- as well as everywhere in the text of this paper. 

\subsection{The \texorpdfstring{$D=4$}{D=4} \texorpdfstring{$E_8 \times E_8$}{E8 x E8} Effective Theory}

When $d=5$ heterotic M-theory is dimensionally reduced to four
dimensions on the linearized BPS double domain wall with five-branes,
the result is, modulo the discussion below, an $N=1$ supersymmetric effective four-dimensional
theory with (potentially spontaneously broken) $E_8 \times E_8$ gauge group. The
Lagrangian will break into two distinct parts. The first contains
terms of order $\kappa_{11}^{2/3}$ in the eleven-dimensional Planck
constant $\kappa_{11}$, while the second consists of terms of order
$\kappa_{11}^{4/3}$.

\subsubsection{The \texorpdfstring{$\kappa_{11}^{2/3}$}{Order 2/3} Lagrangian}

This Lagrangian is well-known and was presented in~\cite{Lukas:1997fg}. Here we
discuss only those properties required in this paper. In four
dimensions, the moduli must be organized into the lowest components of
chiral supermultiplets. Here, we need only consider the real part of
these components. Additionally, one specifies that these chiral
multiplets have \emph{canonical} K\"ahler potentials in the effective
Lagrangian. The dilaton is simply given by
\begin{equation}
  \re S=V \ .
  \label{46}
\end{equation}
However, neither $a^i$ nor $b^i$ -- defined in \eqref{8} and \eqref{11} respectively -- have canonical kinetic energy. To
obtain this, one must define the re-scaled moduli
\begin{equation}
  t^i = 
  \Rhat b^i = 
  \Rhat V^{-1/3} a^i \ ,
  \label{47}
\end{equation}
where we have used \eqref{11} in the text, and choose the complex K\"ahler moduli
$T^i$ so that
\begin{equation}
  \re T^i = t^i \ .
\label{48}
\end{equation}
Denote the real modulus specifying the location of the $n$-th
five-brane in the bulk space by $z_n={x^{11}_n}/{\pi \rho}$ where
$n=1,\dots,N$. As with the K\"ahler moduli, it is necessary to define
the fields
\begin{equation}
  \re Z^n = \beta_i^{(n)} t^i z_n \ .
\label{49}
\end{equation}
These rescaled $Z^n$ five-brane moduli have canonical kinetic
energy.

\subsubsection{The \texorpdfstring{$\kappa_{11}^{4/3}$}{Order 4/3} Lagrangian}

The terms in the BPS double domain wall solution proportional to
$\epsilon_S'$ lead to order $\kappa_{11}^{4/3}$ additions to the $D=4$
Lagrangian. These have several effects. The simplest is that the
five-brane location moduli now contribute to the definition of the
dilaton, which becomes
\begin{equation}
  \re S = V +\frac{\epsilon_S'}{2}\sum_{n=1}^{N}\beta_i^{(n)}t^iz_n^2 \ ,
  \label{52}
\end{equation}
where the fields $t^{i}$ are defined in \eqref{47}.
More profoundly, these $\kappa_{11}^{4/3}$ terms lead, first,  to threshold
corrections to the gauge coupling parameters and, second, to additions
to the Fayet--Iliopoulos (FI) term associated with any anomalous $U(1)$
factor in the low-energy gauge group. Let us analyze these in
turn.

\subsection{Gauge Threshold Corrections}

The gauge couplings of the non-anomalous components of the $D=4$ gauge
group, in both the observable and hidden sectors, have been computed
to order $\kappa_{11}^{4/3}$ in~\cite{Lukas:1998hk}. Written in terms of the fields $b^{i}$ defined in \eqref{11} and including five-branes in the bulk
space, these are given by
\begin{equation}
  \frac{4\pi}{(g^{(1)})^2} \propto
  V \left(1+\epsilon_S' \frac{\Rhat}{2V} 
   \sum_{n=0}^{N}(1-z_n)^2 b^i \beta^{(n)}_i \right)
\label{62}
\end{equation}
and 
\begin{equation}
  \frac{4\pi}{(g^{(2)})^2} \propto
  V \left(1+\epsilon_S' \frac{\Rhat}{2V} 
  \sum_{n=1}^{N+1}z_n^2 b^i\beta^{(n)}_i\right)
  \label{63}
\end{equation}
respectively. The positive definite constant of proportionality is identical for both gauge couplings and is not relevant to the present discussion. It is important to note that the effective parameter of the $\kappa_{11}^{2/3}$ expansion in \eqref{62} and \eqref{63}, namely $\epsilon_S'{\Rhat}/{V}$, is identical to the parameter appearing in \eqref{45B} and \eqref{45C}) for the validity of the linearized approximation with a single five-brane. That is, the effective strong coupling parameter of the $\kappa_{11}^{2/3}$ expansion is given by
\begin{equation}
\epsilon_{S}^{\rm eff}= \epsilon_S' \frac{\Rhat}{V} \ .
\label{pink1}
\end{equation}
We point out that this is, up to a constant factor of order one, precisely the strong coupling parameter presented in equation (1.3) of \cite{Banks:1996ss}.

Recall that $n=0$ and $n=N+1$ correspond to the observable and hidden sector domain walls -- not to five-branes. Therefore, $z_{0}=0$ and $z_{N+1}=1$.
Using \eqref{28} and  \eqref{34}, one can evaluate the $\beta^{(n)}_i$ coefficients in terms of the the $a^{i}, i=1,2,3$ K\"ahler moduli defined in \eqref{8}. Rewrite the above expressions in terms of these moduli using \eqref{10}, \eqref{11},
\eqref{24}, \eqref{25}, and redefine the five-brane moduli to be
\begin{equation}
  \lambda_n = 
  z_n-\tfrac{1}{2}
  \ , \qquad 
  \lambda_n \in \left[-\tfrac{1}{2},\tfrac{1}{2}\right]  \ .
\label{pink2}
\end{equation}
Furthermore, choosing our hidden sector bundle to be that given in \eqref{dude1bAA}, we find that
\begin{equation}
  \label{64}
  \begin{split}
    \frac{4\pi}{(g^{(1)})^2} \propto \;&
    \frac{1}{6v}\int_X\omega \wedge \omega
    \wedge \omega 
    -\epsilon_S' \frac{\Rhat}{2V^{1/3}}
    \frac{1}{v^{1/3}}
    \\ &
    \times
    \int_X\omega \wedge 
    \left(
      -c_2(\Vvis) 
      +\tfrac{1}{2}c_2(TX)-\sum_{n=1}^{N}(\tfrac{1}{2}-\lambda_n)^2W^{(n)}  
    \right)
  \end{split}
\end{equation}
and 
\begin{equation}
  \label{65}
  \begin{split}
  \frac{4\pi}{(g^{(2)})^2} \propto \;&
  \frac{1}{6v}\int_X\omega \wedge \omega
  \wedge \omega 
  -\epsilon_S' \frac{\Rhat}{2V^{1/3}}
  \frac{1}{v^{1/3}}
  \\ &
  \times
  \int_X\omega \wedge 
  \left(
    \sum_{r=1}^{R} a_{r} c_{1}(L_{r}) \wedge c_{1}(L_{r})
    +\tfrac{1}{2}c_2(TX)-\sum_{n=1}^{N}(\tfrac{1}{2}+\lambda_n)^2W^{(n)}  
    \right)
  \end{split}
\end{equation}
where $a_r$ is given in \eqref{26}. The first term on the right-hand
side, that is, the volume $V$ defined in \eqref{10}, is the order
$\kappa_{11}^{2/3}$ result. The remaining terms are the $\kappa_{11}^{4/3}$
M-theory corrections first presented in~\cite{Lukas:1998hk}.

Clearly, consistency of the $D=4$ effective theory requires both
$(g^{(1)})^2$ and $(g^{(2)})^2$ to be positive. It follows that the
moduli of the four-dimensional theory are constrained to satisfy
\begin{multline}
  \frac{1}{v}\int_X\omega \wedge \omega \wedge \omega -3\epsilon_S'
  \frac{\Rhat}{V^{1/3}} \frac{1}{v^{1/3}}\int_X\omega \wedge
  \big(-c_2(\Vvis) 
  \\
  +\tfrac{1}{2}c_2(TX)-\sum_{n=1}^{N}(\tfrac{1}{2}-\lambda_n)^2W^{(n)}  \big) > 0
  \label{66}
\end{multline}
and 
\begin{multline}
  \frac{1}{v}\int_X\omega \wedge \omega \wedge \omega -3\epsilon_S'
  \frac{\Rhat}{V^{1/3}} \frac{1}{v^{1/3}}\int_X\omega \wedge
  \big(
    \sum_{r=1}^{R} a_{r} c_{1}(L_{r}) \wedge c_{1}(L_{r})
  \\
  +\tfrac{1}{2}c_2(TX)-\sum_{n=1}^{N}(\tfrac{1}{2}+\lambda_n)^2W^{(n)}  \big) > 0 .
  \label{67}
\end{multline}
One can use \eqref{3}, \eqref{4},
\eqref{8}, \eqref{23}  and \eqref{32} to rewrite
these expressions as
\begin{equation}
  \label{68}
  \begin{split}
    d_{ijk} a^i a^j a^k- 3 \epsilon_S' \frac{\Rhat}{V^{1/3}} \Big(
    -(\tfrac83 a^1 + \tfrac53 a^2 + 4 a^3)
    + \qquad& \\
    + 2(a^1+a^2) -\sum_{n=1}^{N}(\tfrac{1}{2}-\lambda_n)^2 a^i \;W^{(n)}_i
    \Big) &> 0 
  \end{split}
\end{equation}
and
\begin{equation}
  \label{69}
  \begin{split}
    d_{ijk} a^i a^j a^k- 3 \epsilon_S' \frac{\Rhat}{V^{1/3}}
    \Big(d_{ijk}a^i \sum_{r=1}^{R}a_r l^j_r l^k_r
    + \qquad& \\
    + 2(a^1+a^2) -\sum_{n=1}^{N}(\tfrac{1}{2}+\lambda_n)^2 a^i
    \;W^{(n)}_i \Big) &> 0
  \end{split}
\end{equation}
respectively.

\subsection{Corrections to a Fayet--Iliopoulos Term}\label{sec:FI_correc}

In the heterotic standard model vacuum, the observable sector vector
bundle $\Vvis$ has structure group $SU(4)$. Hence, it does not lead
to an anomalous $U(1)$ gauge factor in the observable sector of the
low energy theory. However, the hidden sector bundle $\Vhid$ introduced above
consists of a sum of line bundles with the additional structure group $U(1)^{R}$. Each $U(1)$ factor leads to
an anomalous $U(1)$ gauge group in the four-dimensional effective
field theory of the hidden sector and, hence, an associated $D$-term.  Let $L_r$ be
any one of the irreducible line bundles of $\Vhid$.  The string one-loop corrected Fayet--Iliopoulos (FI) term for $L_{r}$ was
computed in~\cite{Blumenhagen:2005ga} within the context of the weakly coupled heterotic string, and in \cite{Anderson:2009sw,Anderson:2009nt} for a specific embedding of $U(1)_{r}$ into $E_8$ in strongly coupled heterotic $M$-theory. Comparing various results in the literature, it is straightforward to show that to order $\kappa_{11}^{4/3}$ the one-loop corrected FI-term for $L_{r}$ in the strongly coupled heterotic string is 
\begin{equation}
 \label{54}
  FI_r =
  \frac{a_r}{2} \frac{ \epsilon_S \epsilon_R^2}{\kappa_{4}^{2}}
  \frac{1} {\Rhat V^{2/3}}
  \Big( \mu(L_r) + 
     \epsilon_S' \frac{\Rhat}{V^{1/3}} 
    \int_X c_1(L_r) \wedge 
   \big( J^{(N+1)}+\sum_{n=1}^{N} z_n^2 J^{(n)} \big) \Big) , 
\end{equation}
where $a_r$ is a group-theoretical coefficient, defined in \eqref{26}, determined by how the $\Uni1$ structure group of $L_r$ embeds in $\Ex8$, and $\mu(L_r)$ is given in \eqref{50}. We note that the $\kappa_{11}^{2/3}$ part of this expression is identical to that derived in \cite{Anderson:2009nt}.
Insert \eqref{28}  and, following the
conventions of~\cite{Blumenhagen:2005ga,Weigand:2006yj}, redefine the five-brane moduli as in \eqref{pink2}. Furthermore, choosing our hidden sector bundle to be that given in \eqref{dude1bAA}, we find that
the FI-term becomes
\begin{multline}
  \label{56}
  FI_r = 
   \frac{a_r}{2} \frac{ \epsilon_S \epsilon_R^2}{\kappa_{4}^{2}}
  \frac{1} {\Rhat V^{2/3}}
  \Big( \mu(L_r) - 
  \epsilon_S' \frac{\Rhat}{V^{1/3}} 
  \\
  \int_X c_1(L_r)\wedge
  \big(\sum_{s=1}^{R} a_{s} c_{1}(L_{s}) \wedge c_{1}(L_{s})
  +\tfrac{1}{2} c_2(TX) 
  -\sum_{n=1}^{N}(\tfrac{1}{2}+\lambda_n)^2 W^{(n)} \big) \Big) \ ,
\end{multline}
The first term on the right-hand
side, that is, the slope of $L_r$, is the
order $\kappa_{11}^{2/3}$ result. The remaining terms are the
$\kappa_{11}^{4/3}$ M-theory corrections first presented in~\cite{Lukas:1998hk}.
Note that the dimensionless parameter $ \epsilon_S'
\frac{\Rhat}{V^{1/3}}$ of the $\kappa_{11}^{4/3}$ term is identical to
the expansion coefficient of the linearized solution -- when expressed
in term of the $a^i$ moduli.
Finally, recalling definition \eqref{50} of the slope, 
using \eqref{3}, \eqref{4}, \eqref{8},
\eqref{23}, \eqref{32} and the properties of the second Chern character, it follows that for each $L_r$ the associated Fayet--Iliopoulos factor $FI_{r}$ in  \eqref{56} can be written as
\begin{multline}
  \label{59} 
FI_{r}= \frac{a_r}{2} \frac{ \epsilon_S \epsilon_R^2}{\kappa_{4}^{2}}
  \frac{1} {\Rhat V^{2/3}} \Big(d_{ijk} l_r^i a^j a^k - 
  \epsilon_S' \frac{\Rhat}{V^{1/3}} 
  \\
 \big(d_{ijk}l_r^i\sum_{s=1}^{R}a_{s}l_{s}^jl_{s}^k 
  + l^i_r(2,2,0)|_i
  -\sum_{n=1}^{N}(\tfrac{1}{2}+\lambda_n)^2l_r^iW^{(n)}_i\big) \Big)\ ,
\end{multline}
where
\begin{equation}
  V = \frac{1}{6} d_{ijk} a^i a^j a^k \ .
  \label{60}
\end{equation}

\section{Gauge Transformations and the FI-Term\label{sec:FI-term}}

\subsection{The Green--Schwarz Mechanism and the Axion Transformation}

In heterotic M-theory, a gauge transformation of the Yang--Mills field leads to a transformation of the M-theory three-form. This transformation was used in \cite{Anderson:2009nt} to derive the anomalous gauge transformation of the K\"ahler axions and the resulting D-term contribution to the four-dimensional potential. Since our $\Uni 1$ structure group embeds into $E_{8}$  in a different fashion from that in \cite{Anderson:2009nt}, it is instructive to re-derive this transformation. Following the conventions of \cite{Brandle:2003uya}, the gauge-invariant four-form field strength is given by $G=\dd C-\omega_{\text{YM}}$, where the Chern–Simons three-form is
\begin{equation}
\omega_{\text{YM}}=(4\pi\kappa_{11}^{2})^{1/3}\left[\delta(x^{11})\omega_{(0)}\wedge\dd x^{11}+\delta(x^{11}-\pi\rho)\omega_{(N+1)}\wedge\dd x^{11}\right]\ ,\label{eq:omega_YM-1}
\end{equation}
and $\omega_{(0)}$ and $\omega_{(N+1)}$ satisfy
\begin{align}
\dd\omega_{(0)} & =\frac{1}{16\pi^{2}}\left(\tr_{E_{8}} F_{1}\wedge F_{1}-\tfrac{1}{2}\tr_{SO(6)} R\wedge R\right)_{x^{11}=0}\ ,\\
\dd\omega_{(N+1)} & =\frac{1}{16\pi^{2}}\left(\tr_{E_{8}}  F_{2}\wedge F_{2}-\tfrac{1}{2}\tr_{SO(6)} R\wedge R\right)_{x^{11}=\pi\rho}\ .
\end{align}
Focusing on the second $\Ex 8$ factor and dropping the subscript ``2'', locally $\omega_{(N+1)}$ is given by the usual Chern–Simons three-form
\begin{equation}
\omega_{(N+1)}=\frac{1}{16\pi^{2}}\tr_{E_{8}}\left(A\wedge\dd A+\frac{2}{3} A \wedge A \wedge A \right)+\ldots\label{eq:omega_N+1-1} \ ,
\end{equation}
where the dots indicate terms that contribute to $\tr_{SO(6)} R\wedge R$ and are not relevant to our discussion. The four-form $G$ is clearly gauge invariant under standard shifts by closed three-forms, $\delta C=\dd\lambda$. However, it is less obvious that $G$ is invariant under Yang–Mills gauge transformations. Under a gauge transformation of $A$ by $\delta A=-\dd_{A}\varepsilon$, the four-form $G$ must be invariant. For this to be true, one must have $\dd\delta C=\delta\omega_{\text{YM}}.$ From (\ref{eq:omega_N+1-1}), it is straightforward to show that the transformation of $\omega_{(N+1)}$ is
\begin{equation}
\delta\omega_{(N+1)}=-\frac{1}{16\pi^{2}}\dd\tr_{E_{8}}(\varepsilon\dd A)\ .
\end{equation}
Using the expressions for $\omega_{\text{YM}}$ in (\ref{eq:omega_YM-1}) and $\dd\delta C=\delta\omega_{\text{YM}}$, locally the gauge transformation of $C$ must be
\begin{equation}
(\delta C)_{11,ab}=-\left(\frac{\kappa_{11}}{4\pi}\right)^{2/3}\frac{1}{4\pi}\delta(x^{11})\tr_{E_{8}}(\varepsilon F_{ab})\ ,
\end{equation}
where we have written this in components and specialized to an abelian gauge field, in agreement with \cite{Anderson:2009nt}.

The gauge transformation of the three-form leads to a gauge transformation of the Kähler axions $\chi^{i}$ via $\delta C_{11,a\bar{b}}=\delta\chi^{i}(J_{i})_{a\bar{b}}$. Integrating this over $\mathcal{C}^{i}\times\text{S}^{1}/\mathbb{Z}^{2}$, where the $\mathcal{C}^{i}$ are dual to $\omega_{i}$, then gives
\begin{equation}
\delta\chi^{i}=-\frac{\epsilon_{S}\epsilon_{R}^{2}}{8\pi}\int_{\mathcal{C}^{i}}\tr_{E_{8}}(\varepsilon F)\ .
\end{equation}
When $F$ is the field strength of a non-abelian symmetry, the above expression for $\delta\chi^{i}$ vanishes and so this does not result in a gauge transformation in four dimensions. Why is this? If $F$ is non-abelian and it has a non-zero VEV on the internal space, then the corresponding gauge group cannot appear as a four-dimensional gauge symmetry (since a non-abelian group does not commute with itself inside $\Ex 8$), and so there is no symmetry group in four dimensions to have a gauge transformation with respect to. However, for a $\Uni 1$ factor, these gauge transformations are present since the $\Uni 1$ appears as a factor in both the internal and four-dimensional groups. Written in terms of the four-dimensional $\Uni 1$, the trace reduces to
\begin{equation}
\tr_{E_{8}}(\varepsilon F)=4a\varepsilon F\ ,
\end{equation}
where $\varepsilon$ is the gauge parameter for the four-dimensional transformation and $F$ on the right-hand side is the internal $\Uni 1$ field strength. Noting that this $\Uni 1$ is identified with the structure group of the internal line bundle $L$, we have $F=2\pi\,c_{1}(L)=2\pi v^{-1/3}l^{i}\omega_{i}$, so that the gauge transformation of the axions can be written as\footnote{The four-dimensional $\Uni 1$ in \cite{Anderson:2009nt} is that which appears in the commutant of $\text{S}(\Uni 2\times\Uni 1)\subset\Ex 8$, leading to a different group-theoretic coefficient in $\delta\chi^{i}$.}
\begin{equation}
\delta\chi^{i}=-a\epsilon_{S}\epsilon_{R}^{2}\varepsilon l^{i}\ .
\end{equation}
We see that the four-dimensional gauge transformation of the K\"ahler axions is fixed by the strong coupling parameters $\epsilon_{S}$ and $\epsilon_{R}$, together with the group-theoretic factor $a$ that characterizes how the $\Uni 1$ embeds in the hidden $\Ex 8$. The gauge transformations of the axions imply a gauge transformation of the complexified K\"ahler moduli, defined by $T^{k}=t^{k}+\ii2\chi^{k}$, where the $t^i$ are defined in \eqref{47}.

The singlet matter fields $C^{L}$ also transform under the four-dimensional $\Uni1$ symmetry, with gauge transformations given by
\begin{equation}
\delta C^{L}=-\ii\varepsilon Q^{L}C^{L}\ ,\label{eq:delta_C}
\end{equation}
where, for example, $Q^{L}=2$ for the scalars of the superfields transforming as $(2,\Rep 1)$. Using these transformations rules, both the D-term and the mass of the $\Uni1$ can be derived by standard methods~\cite{Wess:1992cp,Freedman:2012zz}.

\subsection{Various Trace Relations}

The kinetic terms for the gauge fields are naturally written as traces over the $\Ex 8$ generators in the $\Rep{248}$ representation normalized with an extra factor of $1/30$. For the visible sector, the four-dimensional gauge group is $\SO{10}$ (broken down from $\Ex 8$ by the $\SU 4$ bundle). Writing the visible sector curvature as $F_{1}=F_{1}^{a}T^{a}$, the $\Ex 8$ and $\SO{10}$ traces are related via
\begin{equation}
\tr_{E_{8}} T^{a}T^{b}=\tr_{\SO{10}}T^{a}T^{b}\ ,
\end{equation}
where the $\SO{10}$ trace is taken in the fundamental $\Rep{10}$ representation, which is normalized so that $\tr_{\SO{10}}T^{a}T^{b}=\delta^{ab}$. The four-dimensional gauge group in the hidden sector is the commutant of the $\Uni 1$ inside $\Ex 8$. Generically this will be $H\times\Uni 1$, where $H$ is a non-abelian factor, chosen to be $\Ex 7$ for the majority of this paper. For the computation of the FI-term, we are interested only in the $\Uni 1$ gauge field. Therefore, we should rewrite the trace over the $\Ex 8$ generator $Q$ specifying the $U(1)$ embedding in $E_{8}$ in terms of the trace over the fundamental representation $Q_\text{fund}$ of $\Uni 1$ instead. Tracing over $Q^{2}$ we define
\begin{equation}\label{eq:trace_identity}
\tr Q^{2}=4a\tr_{\Uni 1}Q_{\text{fund}}^{2}=4a \ ,
\end{equation}
where we are taking $1$ as the generator of the fundamental representation of $\Uni 1$. For the $\Ex 7\times\Uni 1$ embedding we use in this paper, one finds that  $a=1$.

\section{The Anomalous Mass Contribution}

As we discuss in the main text, the mass of the $U(1)$ vector field is
\begin{equation}
m_{\Uni 1}^{2}=\frac{\pi\ah}{a\re f_{2}}\left(Q^{L}{Q}^{M}G_{LM}\langle C^{L}\rangle\langle{ \bar{C}}^{M}\rangle+2a^{2}\epsilon_{S}^{2}\epsilon_{R}^{4}g_{ij}l^{i}l^{j} \right) \ ,
\end{equation}
which can be expressed as 
\begin{equation}
m_{\Uni 1}^{2}=m_D^2+m_A^2\ .
\end{equation}
The first term $m_D^2$ appears because of the non-zero VEVs of the scalar fields $C_L$ which, if their charges are appropriate, could  cancel the D-term in our four-dimensional theory. The second term $m_A^2$ is induced by the Green--Schwarz mechanism.
To compute the second term, we need to know the metric $g_{ij}$. This is defined by 
\begin{equation}
g_{ij}=\partial_{T^{i}}\partial_{\bar{T}^{j}}\mathcal{K} \ ,
\label{sun11}
\end{equation}
with the K\"ahler potential $\mathcal{K}$ given by
\begin{equation}
\kappa_{4}^{2}\mathcal{K}=-\ln{\mathcal V}
\label{sun2}
\end{equation}
where
\begin{equation}
\mathcal{V}\equiv\tfrac{1}{48}d_{ijk}(T+{\bar{T}})^{i}(T+{\bar{T}})^{j}(T+{\bar{T}})^{k} \ .
\label{sun3A}
\end{equation}
Taking derivatives with respect to the complex scalars $T^i$, we then find
\begin{equation}
g_{ij}=-\frac{d_{ijk}t^{k}}{4\kappa_{4}^{2}\mathcal{V}}+\frac{d_{ikl}t^{k}t^{l}d_{jmn}t^{m}t^{n}}{16\kappa_{4}^{2}{\mathcal{V}}^{2}}.
\label{pen1}
\end{equation}
Contracting this expression with $l^i l^{j}$, we then have
\begin{equation}
\begin{split}g_{ij}l^{i}l^{j} & =-\frac{d_{ijk}l^{i}l^{j}V^{-1/3}a^{k}}{4\kappa_{4}^{2}\Rhat^{2}}+\frac{V^{-4/3}d_{ikl}a^{k}a^{l}d_{jmn}l^{i}l^{j}a^{m}a^{n}}{16\kappa_{4}^{2}\Rhat^{2}}\\
& =-\frac{1}{4\kappa_{4}^{2}\Rhat^{2}V^{1/3}}d_{ijk}l^{i}l^{j}a^{k}+\frac{1}{16\kappa_{4}^{2}\Rhat^{2}V^{4/3}}\mu(L)^{2} \ .
\end{split}
\label{eq:ll}
\end{equation}
Putting this all together, we find an expression for $m_A^2$ of the form
\begin{equation}
m_A^2 =\frac{\pi\hat\alpha_{GUT}}{a \re f_2}\frac{a^2\epsilon_S^2\epsilon_R^4}{\kappa_{4}^{2}\Rhat^{2}}
\left(\frac{1}{8V^{4/3}}\mu(L)^{2}-\frac{1}{2V^{1/3}}d_{ijk}l^{i}l^{j}a^{k}\right) \ .
\label{eq:anomalous_mass}
\end{equation}
Note that since the scale of the FI-term and, hence, for $m_D^2$, is set by $\epsilon_S \epsilon_R^2 / \kappa_4^2 \Rhat$, the anomalous mass $m_A^2$ is suppressed by a factor of $\epsilon_S \epsilon_R^2/\Rhat$ relative to $m_D^2$.

\section{The BPS Vacuum Solution and the Linearization Constraints}

\subsection{The Strong Coupling Linearization Constraints}

In \eqref{45B} and \eqref{45C} of Appendix A, the constraints required for the validity of the linearized approximation to the five-dimensional BPS solution of the Hořava--Witten theory are presented. These constraints are restricted to the case of a single five-brane located at $z_{1}$. Reproducing the equations here, we have
\begin{equation}
  2\epsilon_S'\frac{\Rhat}{V}
  \left|
    \beta_i^{(0)} \big(z-\tfrac{1}{2}\big)
    -\tfrac{1}{2}\beta_i^{(1)}(1-z_1)^2
  \right|
  \ll 
  | d_{ijk} b^j b^k |
  \ , \quad z \in [0,z_1] \ ,
\label{45BB}
\end{equation}
and 
\begin{equation}
  2\epsilon_S'\frac{\Rhat}{V}
  \left|
    (\beta_i^{(0)}+\beta_i^{(1)})
    \big(z-\tfrac{1}{2}\big)
    -\tfrac{1}{2}\beta_i^{(1)}z_1^2
  \right| 
  \ll 
  | d_{ijk} b^j b^k |
  \ , \quad z \in [z_1,1] \ ,
\label{45CC}
\end{equation}
where, as discussed in Appendix A, we have removed the subscript $``0 "$ on the orbifold averaged moduli. Note that expressions \eqref{45BB} and \eqref{45CC} depend on the observable-sector gauge bundle and the bulk-space five-brane but are, however, independent of the gauge bundle on the hidden sector. In unity gauge and in terms of the K\"ahler moduli $a^i$, the strong coupling constraints \eqref{45BB} and \eqref{45CC} simplify to
\begin{equation}
  \left|
    \beta_i^{(0)} \big(z-\tfrac{1}{2}\big)
    -\tfrac{1}{2}W_i(\tfrac{1}{2}-\lambda)^2
  \right|
  \ll 
 \tfrac{1}{2} | d_{ijk} a^j a^k |
  \ , \quad z \in [0, \lambda + \tfrac{1}{2}] \label{83A} \ ,
  \end{equation}
  and
\begin{equation}
  \left|
    (\beta_i^{(0)}+W_i)
    \big(z-\tfrac{1}{2}\big)
    -\tfrac{1}{2}W_i(\tfrac{1}{2}+\lambda)^2
  \right| 
  \ll 
 \tfrac{1}{2}   | d_{ijk} a^j a^k |
  \ , \quad z \in [\lambda + \tfrac{1}{2},1] \ .
  \label{84A}
\end{equation}
respectively, where $\lambda=z_{1}-\frac{1}{2}$.

\begin{figure}[t]
\centering
\begin{subfigure}[b]{0.49\textwidth}
\includegraphics[width=1\textwidth]{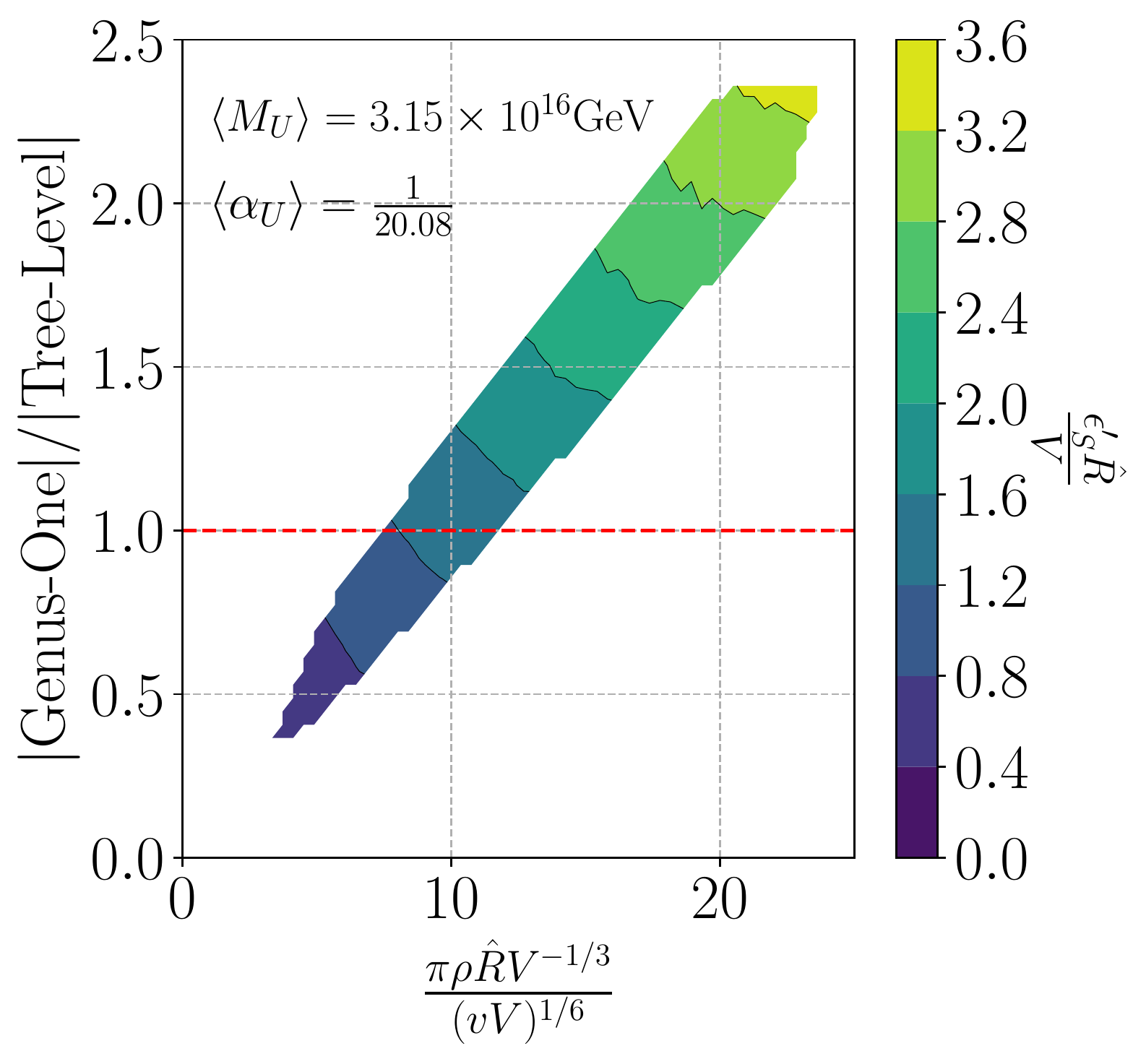}
\caption{Split Wilson lines: $\langle  \alpha_{u}\rangle =\frac{1}{20.08}$.}
\end{subfigure}
\begin{subfigure}[b]{0.49\textwidth}
\includegraphics[width=1\textwidth]{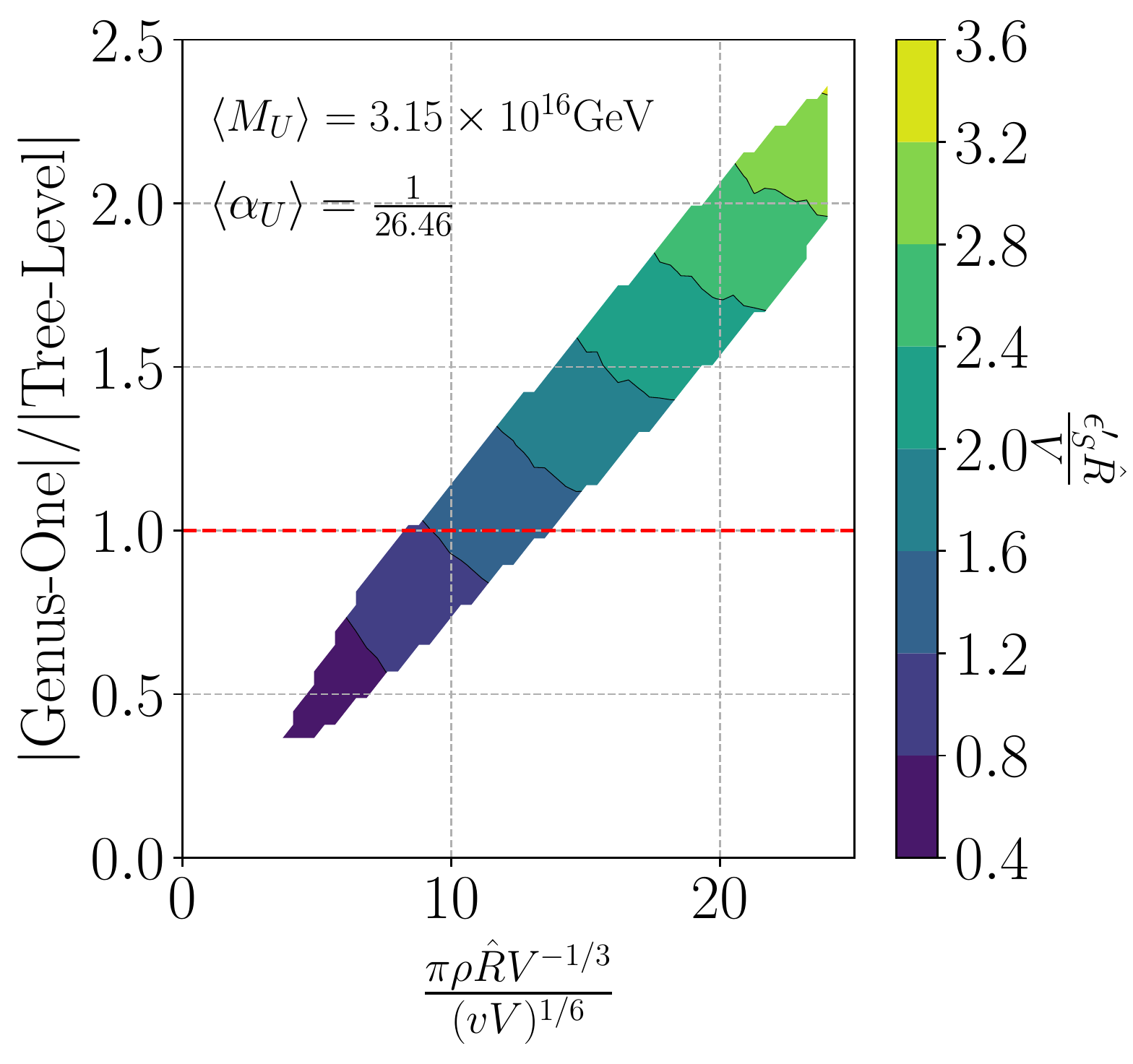}
\caption{Simultaneous Wilson lines: $\langle  \alpha_{u}\rangle =\frac{1}{26.46}$.}
\end{subfigure}
\caption{In this Figure, we graph both the ratio of the genus-one correction to the tree level value of the FI-term (shown on the $y$-axis) and the effective strong coupling parameter $\epsilon_S^{\text{eff}}=\frac{\epsilon_S^\prime \hat R}{V}$ (shown on the colorbar) against the ratio of the fifth-dimensional length $\pi \rho \hat R V^{-1/3}$ to the average Calabi--Yau radius $(vV)^{1/6}$ (displayed on the $x$-axis). The horizontal red line indicates the region in which the genus-one corrected FI-term exactly vanishes; that is, 
when $|$Genus-One$|/|$Tree-Level$|=1$. We produced the two histograms by sampling points inside the ``brown'' regions of Figures 8 (a) and (b) respectively. 
In (a) above, the genus-one corrected FI-term vanishes exactly for parameter  $\epsilon_S^{\text{eff}}\sim 1.4$ (where the distance between the hidden and observable walls is approximately 6 to 13 times larger than the Calabi--Yau length). In (b), the genus-one corrected FI-term vanishes exactly for parameter  $\epsilon_S^{\text{eff}}\sim 1.3$ (where the distance between the hidden and observable walls is approximately 7 to 15 times larger than the Calabi--Yau length). }
\label{fig:2D_expansion}
\end{figure}

As discussed in Appendix A.5 for the linearized expansion of the gauge threshold corrections, and defined in \eqref{pink1}, the effective ``strong coupling parameter'' is given by
\begin{equation}
\epsilon_S^{\text{eff}}=\frac{\epsilon_S^\prime \hat R}{V}\ .
\label{seb2}
\end{equation}
Note that this is precisely the moduli dependent parameter on the left-hand side of \eqref{45BB} and \eqref{45CC} above. By definition, the effective strong coupling expansion parameter $\epsilon_S^{\text{eff}}$  is a reasonable measure of the validity of the linearized expansion of the various quantities used in this paper. It is important to note that in heterotic M-theory, where the observable and hidden sector orbifold planes are separated by the fifth-dimension, one expects the effective coupling parameter $\epsilon_S^{\text{eff}}$ to become perturbatively small only when the walls are {\it very} close to each other. Even for a small separation on the order of a few times the average Calabi--Yau threefold radius, this parameter is fairly large, rapidly becoming a strong coupling parameter as the distance between the orbifold walls begins to significantly exceed the Calabi--Yau radius. That is, heterotic M-theory is, by definition, a strongly coupled theory. Be that as it may, very loosely speaking, when $ \epsilon_S^{\text{eff}} \lesssim.7$, one might expect the linear approximation used throughout this paper to be weakly valid. However, when $.7 < \epsilon_S^{\text{eff}}$, one is truly in a strongly coupled regime and the validity of the linearized approximation comes into doubt. As a check on this, we find that in the strongly coupled regime with  $.7 < \epsilon_S^{\text{eff}}$ the terms on the left and on the right sides of the expressions \eqref{45BB} and \eqref{45CC} are roughly of similar size. On the other hand, the left side is somewhat smaller than the right-hand side of these inequalities when $ \epsilon_S^{\text{eff}} \lesssim.7$.

The major result of this paper is the construction of a hidden sector for the ``heterotic standard model''. The hidden sector consists of a line bundle $L=\mathcal{O}_X(2,1,3)$, embedded into $SU(2) \subset E_{8}$ as in \eqref{red4} with coefficient $a=1$. In addition there is a single five-brane located at $\lambda=0.49$ in the fifth-dimension. It was shown that for a significant region of K\"ahler moduli space, the induced $L \oplus L^{-1}$ rank two bundle of this hidden sector satisfied the required ``vacuum'' constraints, the reduction and phenomenological constraints (the brown regions of Figure 8 (a) and (b)), as well as being slope polystable and $D=4$, $N=1$ supersymmetric (the magenta region of Figure 10). However, importantly, in order to satisfy the last two constraints -- that is, slope poly-stability and $D=4$, $N=1$ supersymmetry -- it was necessary to be in the region of K\"ahler moduli space in which the associated Fayet--Iliopoulos term vanished. That is, $FI=0$. For this to be the case, it is necessary to exactly cancel the genus-one correction to the FI-term against its tree level value. In addition to the fine-tuning required, this clearly suggests that the parameter $\epsilon_S^{\text{eff}}$ might be quite large. To determine this analytically, in Figure 13 (a) and (b) we plot both $\epsilon_S^{\text{eff}}$ and $\left|\frac{\text{genus-one}}{\text{tree-level}}\right|$ against the ratio of the length of the fifth-dimension to the average Calabi--Yau radius for the brown regions of Figure 8 (a) and (b) -- all of which can be determined for any point in the relevant K\"ahler moduli space. In each figure (a) and (b), the horizontal red line indicates when $|\text{genus-one}|=|\text{tree-level}|$; that is, when $FI=0$. We immediately see that to set $FI=0$, it is necessary for $\epsilon_S^{\text{eff}}$ to take the values $\sim 1.4$ and $\sim 1.3$ for the split and simultaneous Wilson line scenarios respectively. That is, as expected, in both cases $.7< \epsilon_S^{\text{eff}}$ and, hence, one is in a strongly coupled regime where the linear expansion solution to the five-dimensional BPS state might not be a good approximation.

An alternative scenario, in which the FI-term does {\it not} vanish, would allow us to be in a less strongly coupled region where the linearity constraints \eqref{45BB} and \eqref{45CC} would be well-satisfied. This scenario was discussed in detail in Section 7.5.2, but was proven to be non-applicable to a hidden sector consisting of a single line bundle. This leaves the vanishing-slope scenario as the only viable option for obtaining a supersymmetric effective theory in the type of $U(1)$ embedding into $E_8$ used in this paper.

As discussed in the text and Appendix A, the formalism in this paper is based on using the linear approximation  to solve the five-dimensional BPS vacuum equations -- first described in \cite{Lukas:1998tt}. Although we find that the linear expansion is not sufficiently accurate in the case that we are interested in, heterotic M-theory has never been solved to include terms of orders higher than  $\kappa_{11}^{2/3}$ and $\kappa_{11}^{4/3}$. In the absence of an exact solution, accurate to all orders, we have no choice but to use the existing results in the literature, which are expansions to linear order. However, in the following, we will try to get an idea of how our equations and results might change if we were to work to one order higher than in this paper.

Our analysis follows the solution presented in \cite{Lukas:1998tt}, but we will extend it to one order higher. Note that the notation will be slightly changed. In \cite{Lukas:1998tt}, the BPS state equation 
\begin{equation}
d_{ijk}f^jf^k=H_i(z)
\label{AppD1}
\end{equation}
is solved to linear order only, with the ansatz
\begin{equation}
f^i(z)=A^i+B^iz \ ,
\label{AppD20}
\end{equation}
where $A^i$ and $B^i$ are constants which depend on the K\"ahler moduli averaged over the fifth dimension and  $z$ is the space coordinate across the fifth dimension, defined as $z=\frac{y}{\pi \rho} \in[0,1]$. Ideally, $A^i\ll B^i$. We will assume the five-brane is very close to the hidden wall, that is, $\lambda=z_{1}-\frac{1}{2} \approx \frac{1}{2}$. In this limit, $H_i(z)$ is given by the expression
\begin{equation}
H_i(z)=-4\epsilon_S^\prime k\beta_iz+k_i , \quad\text{with } k_i,k>0\text{ and }\beta_i =(2/3,-1/3,4)_i\ .
\label{AppD3}
\end{equation}
We will now extend this analysis to second order and, hence, write
\begin{equation}
f^i(z)=A^i+B^iz+C^iz^2\ .
\label{AppD2}
\end{equation}
Since eleven-dimensional Hořava--Witten theory, as well as its reduction to a five-dimensional effective theory, are both valid only to order $\kappa_{11}^{2/3}$, it is unclear what going to even higher order in the solution of the BPS vacuum equation would mean. Be that as it may, it does give an indication of how various physical quantities might behave at higher order. 
Matching the powers of $z$ in  equation \eqref{AppD1} we obtain
\begin{equation}
\begin{split}
d_{ijk}A^jA^k&=k_i\ ,\\
2d_{ijk}A^jB^k&=-4\epsilon_S^\prime k\beta_i\ ,\\
2d_{ijk}A^jC^k+d_{ijk}B^jB^k&=0\ .
\end{split}
\label{AppD4}
\end{equation}

Let us attempt to solve these equalities, making use of the fact, presented in \cite{Lukas:1998tt}, that 
\begin{equation}
\begin{split}
V(z)&=\left(\frac{d_{ijk}f^if^jf^k}{6} \right)^2\ ,\\
a(z)&=\tilde kV^{1/6}\ ,\\
b(z)&=kV^{2/3}\ ,\\
b^i(z)&=f^iV^{-1/6} \ .
\end{split}
\label{AppD5}
\end{equation}
To second order in the $z$ dependence, these quantities can be expressed as
\begin{equation}
\begin{split}
V(z)&=\bar V+V^{(1)}z+V^{(2)}z^2\ ,\\
a(z)&=\bar a+a^{(1)}z+a^{(2)}z^2\ ,\\
b(z)&=\bar b+b^{(1)}z+b^{(2)}z^2\ ,\\
b^i(z)&=\bar b^i+{b^i}^{(1)}z+{b^i}^{(2)}z^2 \ .
\end{split}
\label{AppD6}
\end{equation}
Matching the powers in $z$ at zeroth order, we can now proceed to evaluate $A^i$, $B^i$ and $C^i$. We find that
\begin{equation}
\ A^i=\bar b^i{\bar V}^{1/6} \ ,
\label{AppD7}
\end{equation}
while $B^j$ is given by
\begin{equation}
B^j=2\frac{\epsilon_S^\prime k }{\bar V^{1/6}}\left( \beta^{j}-\frac{1}{2}{\bar b^i}\beta_i{\bar b^j} \right)  \ ,
\label{AppD8}
\end{equation}
and finally
\begin{equation}
 C^j=(\bar g^{ij}-\frac{1}{2} \bar b^i\bar b^j)\frac{d_{imn}B^mB^n}{2\bar V}^{1/6} \ .
 \label{AppD9}
\end{equation}
Using $f^i=A^i+B^iz+C^iz^2$, one can then compute the values of $V$, $a$, $b$ and $b^i$ to second order.
For the volume $V(z)$ and the metric functions $a(z)$ and $b(z)$, we find  
\begin{align}
	V(z)&=\bar V\Bigg[1+\frac{d_{ijk}A^iA^jB^k}{\bar V^{1/2}}z+\frac{(d_{ijk}A^iA^jB^k)^2}{4\bar V} z^2+\frac{d_{ijk}A^iB^jB^k}{2\bar V^{1/2}}z^2\Bigg]\label{AppD10} \ , \\
	a(z)&={\tilde k}{\bar V^{1/6}}\left[ 1+\frac{d_{ijk}A^iA^jB^k}{6\bar V^{1/2}}z-\frac{(d_{ijk}A^iA^jB^k)^2}{36\bar V} z^2 +\frac{d_{ijk}A^iB^jB^k}{12\bar V^{1/2}}z^2\right]\label{AppD11} \ , \\
	b(z)&={k}{\bar V^{2/3}}\left[ 1+2\frac{d_{ijk}A^iA^jB^k}{3\bar V^{1/2}}z+\frac{(d_{ijk}A^iA^jB^k)^2}{9\bar V} z^2 +\frac{d_{ijk}A^iB^jB^k}{3\bar V^{1/2}}z^2\right]	\label{AppD12} \ ,
\end{align}
while for  the moduli $b^i$ we simply have 
\begin{equation}
\begin{split}
b^i(z)&=f^iV(z)^{-1/6}=(A^i+B^iz+C^iz^2)V(z)^{-1/6}\ .
\end{split}
\label{AppD13}
\end{equation}
One can further convert the $b^i$ shape moduli to the $a^i$ K\"ahler moduli via
\begin{equation}
\begin{split}
a^i=b^iV^{1/3}&=(A^i+B^iz+C^iz^2)V(z)^{1/6}\\
&=\Bigg[
A^i+\left( B^i+A^i\frac{d_{ijk}A^iA^jB^k}{6\bar V^{1/2}}\right)z\\
&\eqspace+\left( C^i+B^i\frac{d_{ijk}A^iA^jB^k}{6\bar V^{1/2}}
-A^i\frac{(d_{ijk}A^iB^jB^k)^2}{36\bar V^{1/2}}+A^i\frac{d_{ijk}A^iB^jB^k}{12\bar V^{1/2}}
\right)z^2
\Bigg] \ .
\end{split}
\label{AppD14}
\end{equation}

The only thing left to do before one can compute how the volume $V(z)$, the metric moduli  $a(z)$, $b(z)$ and the K\"ahler moduli $a^{i}(z), i=1,2,3$ vary across the fifth dimension is to fix the constant $k$. It follows from \eqref{AppD12} that
\begin{equation}
 k=\frac{\bar b}{\bar V^{2/3}}\ .
 \label{AppD15}
 \end{equation}
 In the $D=4$ theory, we can identify the averaged value of $b(z)$ over the fifth dimension with $\hat R/2$. Furthermore, in the $D=4$ effective theory we know that we can go to unity gauge and set $\epsilon_S^\prime \hat R/V^{1/3}=1$, where $V$ is the average of the volume $V(z)$ over the fifth dimension.
 In our second order expansion, the tree level values of our moduli do not equal the average of these moduli over the fifth dimension.  However, we will show that for $b(z)$ and $V(z)$, one can approximate the tree level values in the Taylor expansions, $\bar b$ and $\bar V$, to be equal to the $D=4$ effective theory values $b=\hat R/2$ and $V$.
  Hence, we set
 \begin{equation}
 k=\frac{\hat{R}}{2V^{2/3}}
 \label{AppD16}
 \end{equation}
 and  ${\epsilon_S^\prime  \hat{R}}/{\bar V^{1/3}}=1$. In this limit, we find
 \begin{equation}
B^j=\frac{\epsilon_S^\prime \hat R }{V^{5/6}}\left( \beta^{j}-\tfrac{1}{2}{\bar b^i}\beta_i{\bar b^j} \right) =\frac{1 }{V^{1/2}}\left( \beta^{j}-\tfrac{1}{2}{\bar b^i}\beta_i{\bar b^j} \right) \ ,
\end{equation}
\label{AppD17}
which allows us to also determine $C^i$ exactly.

\begin{figure}[t]
\centering
\begin{subfigure}[b]{0.7\textwidth}
\includegraphics[width=1\textwidth]{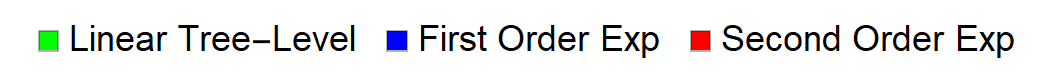}
\end{subfigure}\\
\begin{subfigure}[b]{0.32\textwidth}
\includegraphics[width=1\textwidth]{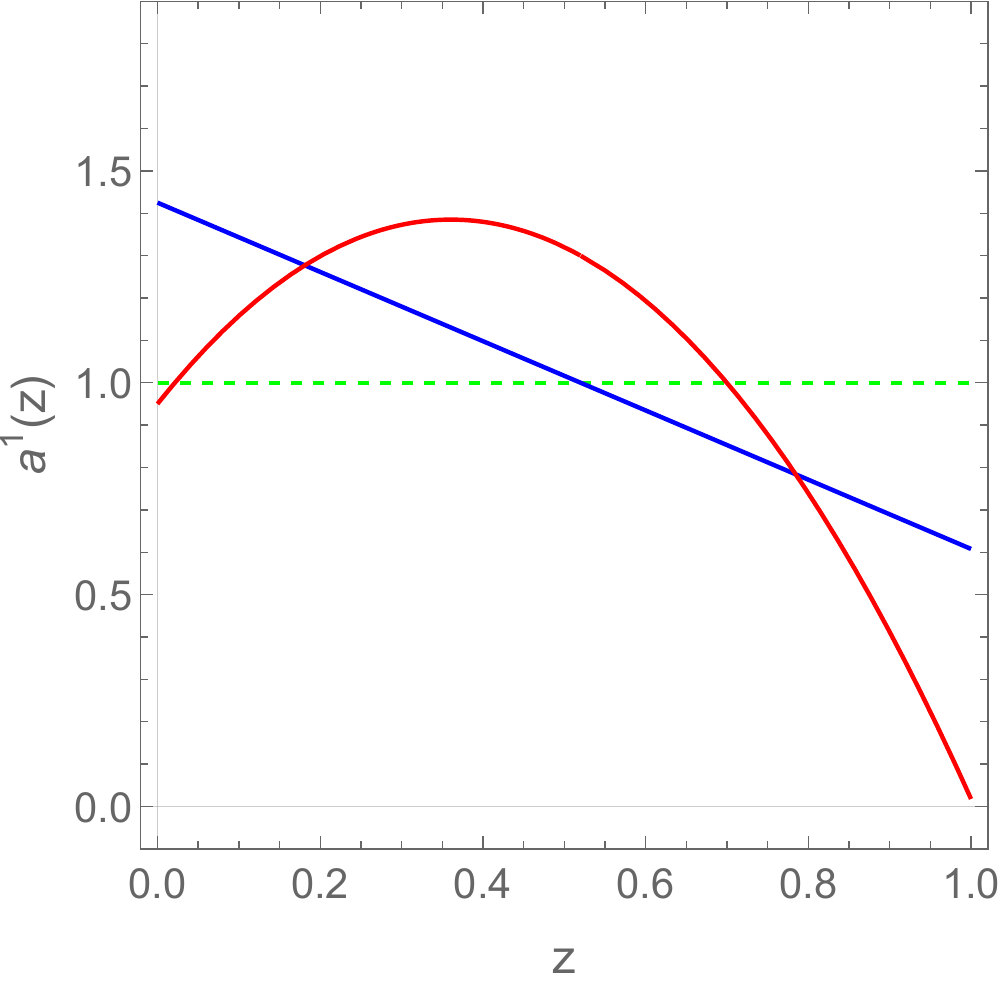}
\end{subfigure}
\begin{subfigure}[b]{0.32\textwidth}
\includegraphics[width=1\textwidth]{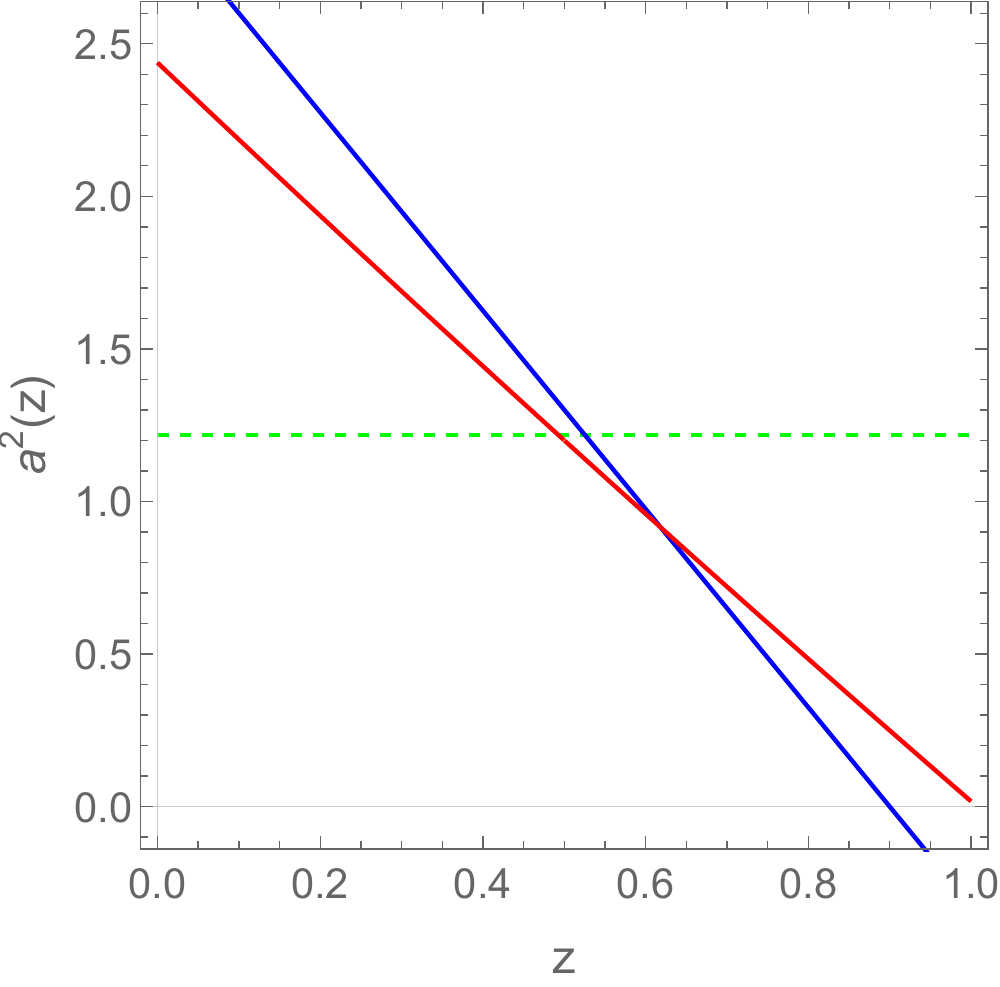}
\end{subfigure}
\begin{subfigure}[b]{0.32\textwidth}
\includegraphics[width=1\textwidth]{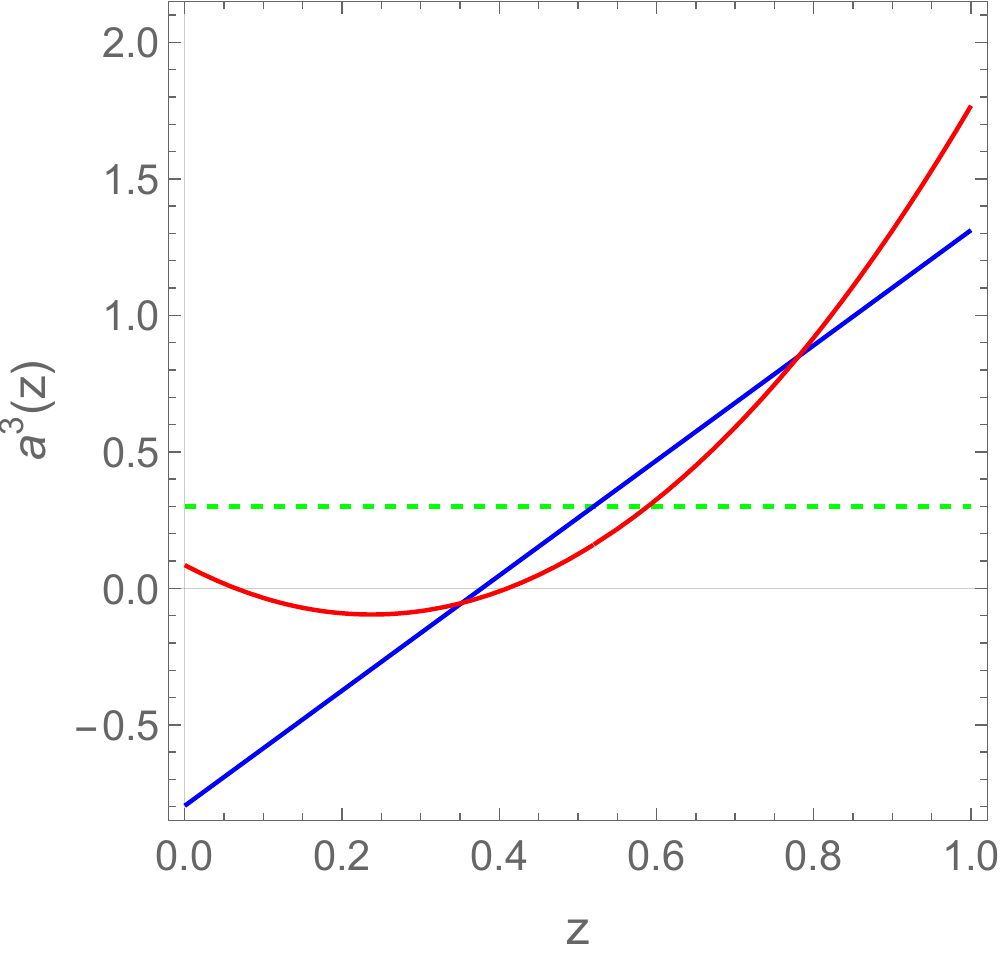}
\end{subfigure}
\caption{Plots of the K\"ahler moduli $a^1(z),a^2(z),a^3(z)$ respectively for $z \in [0,1]$. The red curves are these functions plotted to order $z^{2}$ whose zeroth-order coefficients are chosen to be $(\bar a^1, \bar a^2, \bar a^3)=(1.3,1.2,0.16)$. The orbifold averages over these functions are displayed as the dashed green lines and take the values $({a^1},{a^2},{a^3})=(1.0,1.2,0.3)$. The blue curves then represent the plots of these K\"ahler moduli in the linearized approximation using $({a^1},{a^2},{a^3})=(1.0,1.2,0.3)$ as their zeroth-order coefficients. That is, these graphs show how the K\"ahler moduli change when one goes from the first order linearized approximation (blue) to the second order $z^{2}$ expansion (red).}.
\label{fig:moduliAppD}
\end{figure}

Having determined $A^i, B^i, C^i$ exactly, one can now evaluate how the volume $V(z)$, the metric  functions $a(z), b(z)$ and the K\"ahler moduli  $ a^i(z)$  change as functions of the fifth dimensional coordinate $z$. We are mainly interested in the region of K\"ahler moduli space, analyzed in the text, where, in addition to all necessary constraints being satisfied, the genus-one corrected FI-term also vanishes; that is, the magenta region of Figure 10. In order to compare our results to the linear case studied in \cite{Lukas:1998tt}, we will write the Taylor expansions around the center of the interval $z\in [0,1]$, which is equivalent to shifting 
\begin{equation}
z\rightarrow z-\frac{1}{2}
\end{equation}
in \eqref{AppD2}.
To determine how $V(z)$, $a(z)$, $b(z)$ and $a^i(z)$ behave for points within the solution space on which the FI-term vanishes--the magenta region of Figure 10--naively, one would choose a point within this solution space and compute the behavior of these functions. However, the solution space we found was in terms of the $D=4$ effective moduli $a^i=\langle a^i(z) \rangle$.  Here, where we work to order $z^{2}$, one can only compute the functions $V(z)$, $a(z)$, $b(z)$ and $a^i(z)$ once we specify the zeroth-order values $\bar a^i$, which, in general, are not equal to the average of $a^i(z)$ over the interval $z\in[0,1]$ (unlike in the linear case). Therefore, our problem is that we can find if we are within the solution space or not only after we sample the values for  $\bar a^i$, compute $a^i(z)$ and do the average
\begin{equation}
a^i=\int_{0}^1 \dd z\> a^i(z)\ .
\label{AppD19}
\end{equation}
For example, let us choose $(\bar a^1, \bar a^2, \bar a^3)=(1.3,1.2,0.16)$. From this, one can compute all coefficients $A^{i}, B^{i}, C^{i}$ specified above and, hence, $\bar{V}, V^{(1)}, V^{(2)}$ as well as $\bar{a}, a^{(1)}, a^{(2)}$ and  $\bar{b}, b^{(1)}, b^{(2)}$ and the coefficients  ${\bar{a}}^{i}, a^{i(1)}, a^{i(2)}$ for $i=1,2,3$. We begin by plotting the the $z$-dependent functions $a^{1}(z), a^{2}(z), a^{3}(z)$. They appear as the {\it red} lines in Figure \ref{fig:moduliAppD} from left to right respectively. Taking the average \eqref{AppD19} over each of these three functions, we find that $( a^1, a^2, a^3)=(1.0,1.2,0.3)$. These values are plotted as the dashed {\it green} lines in Figure \ref{fig:moduliAppD}.
We find that $(a^1,a^2,a^3)=(1.0, 1.2, 0.3)$ indeed sits on the surface of vanishing slope, that is
\begin{equation}
FI(a^1=1.0,a^2=1.2,a^3=0.3)=0 \ ,
\end{equation}
and is a point in the interior of the physical magenta region shown in Figure 10. It is of interest to compare the order $z^{2}$ solutions to the linear order results.  To do so, we truncate the expansions in (D.12) at order $z$ and now, using the averaged values $(a^1,a^2,a^3)=(1.0, 1.2, 0.3)$, compute all coefficients to linear order, that is, $A^{i}, B^{i}$ specified above and, hence, $V, V^{(1)} $ as well as $a, a^{(1)}$ and  $b, b^{(1)}$ and the coefficients  $a^{i}, a^{i(1)}$ for $i=1,2,3$.
The linear order results are shown as the {\it blue} curves in Figure \ref{fig:moduliAppD}. The first thing to note is that the linearized approximations to $a^{2}(z)$ and, more significantly, $a^{3}(z)$ both become unacceptably negative in the interval $z \in [0,1]$.  In particular,  $a^{3}(z)$ is deeply negative over much of the interval $[0,\sim.4]$. However, we see that by going to order $z^{2}$, $a^{2}(z)$ is positive over the entire interval, while $a^{3}(z)$ -- although still slightly negative in the interval $[0.5,\sim.4]$ -- is much less negative than in the linearized case and is clearly correcting the behavior of $a^{3}(z)$ significantly.

\begin{figure}[t]
\centering
\begin{subfigure}[b]{0.7\textwidth}
\includegraphics[width=1\textwidth]{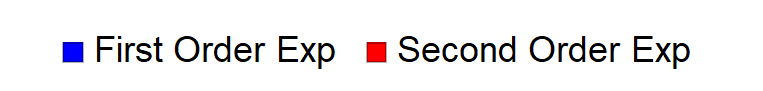}
\end{subfigure}\\
\begin{subfigure}[b]{0.32\textwidth}
\includegraphics[width=1\textwidth]{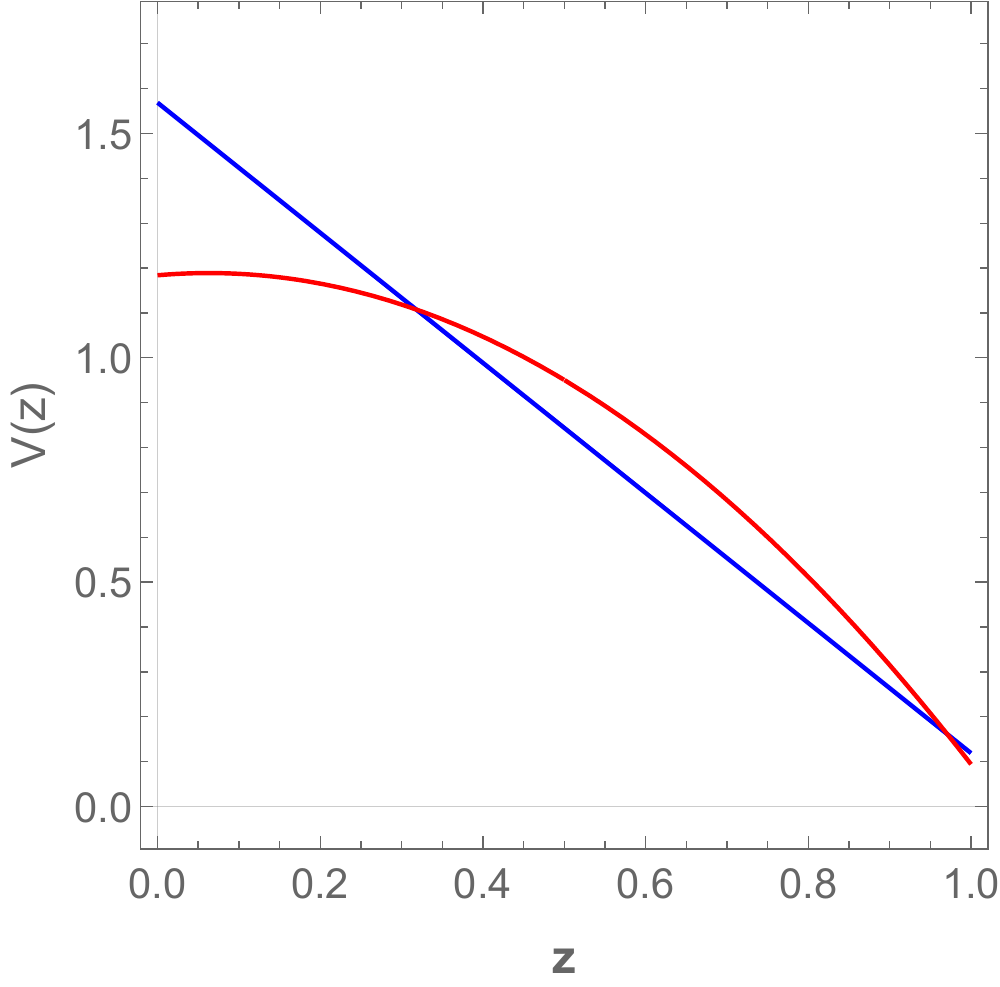}
\end{subfigure}
\begin{subfigure}[b]{0.32\textwidth}
\includegraphics[width=1\textwidth]{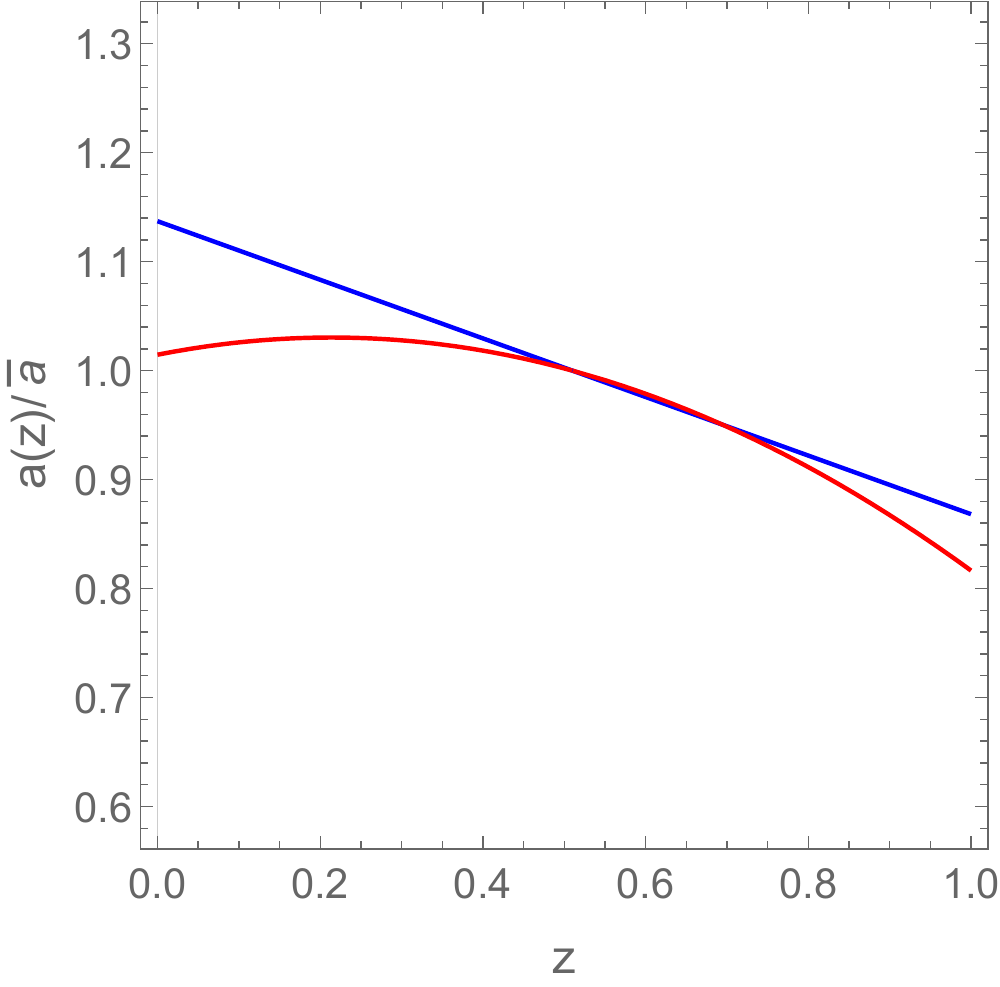}
\end{subfigure}
\begin{subfigure}[b]{0.32\textwidth}
\includegraphics[width=1\textwidth]{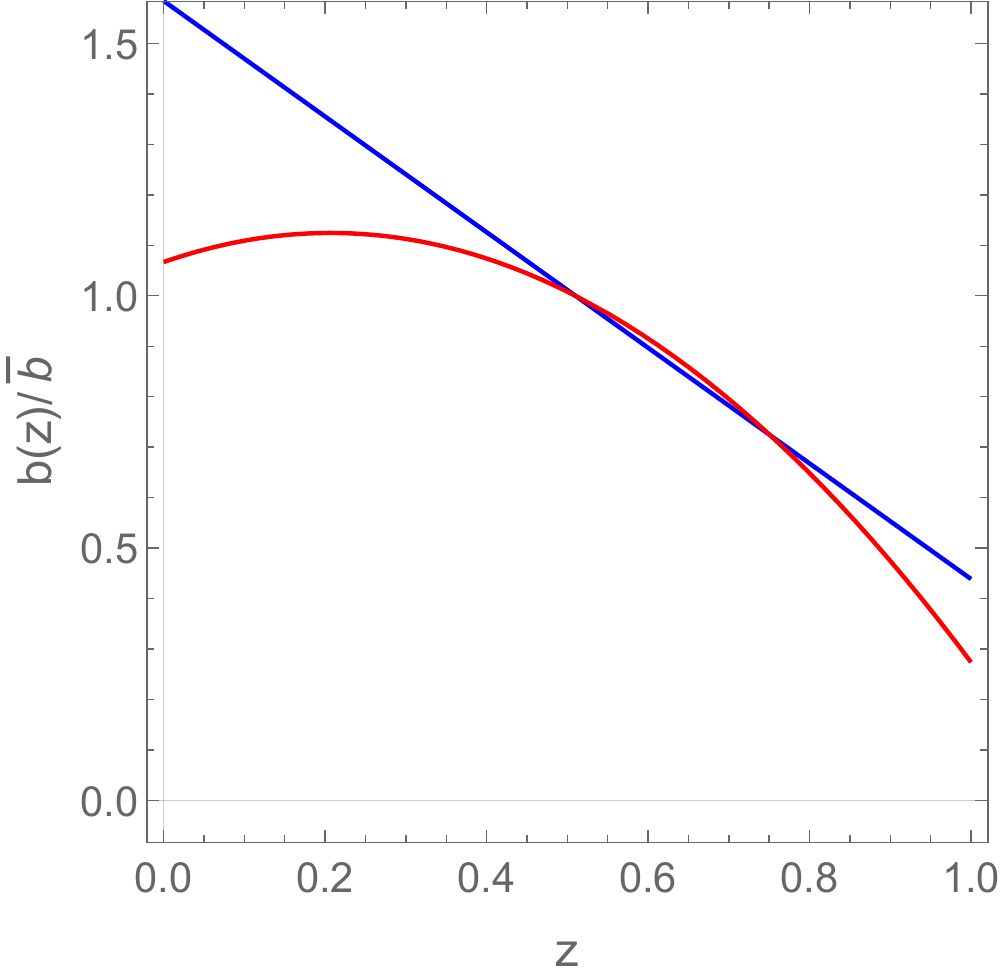}
\end{subfigure}
\caption{The quadratic $z^{2}$ order expansion of $V(z)$, $a(z)/\bar a$ and $b(z)/\bar b$ are shown as the red curves in each graph, where we have chosen the zeroth order values in each expansion to be $(\bar a^1, \bar a^2, \bar a^3)=(1.3,1.2,0.16)$. Similarly, using the average K\"ahler moduli $({a^1},{a^2},{a^3})=(1.0,1.2,0.3)$ shown as the dashed green lines in Figure \ref{fig:moduliAppD}, the plots of $V(z)$, $a(z)/\bar a$ and $b(z)/\bar b$ expanded to linear order in $z$ are shown in blue. This clearly demonstrates how each of the solutions change when one goes from a first order (blue) to a second order (red) expansion. }
\label{fig:VolumeAppD}
\end{figure}

As a second check on the relationship between the order $z^{2}$ and the linearized approximations, we now plot $V(z)$, $a(z)/\bar a$ and $b(z)/\bar b$ in both cases -- using, as above, $(\bar a^1, \bar a^2, \bar a^3)=(1.3,1.2,0.16)$ as the order-zero $z^{2}$ coefficients and $(a^1,a^2,a^3)=(1.0, 1.2, 0.3)$ as the order-zero linear approximation coefficients. These are plotted as the {\it red} and {\it blue} lines from left to right respectively in Figure \ref{fig:VolumeAppD}. Among the conclusions we can draw from this second-order extended analysis is that the moduli $V(z), b(z)$ and $a(z)$ do not change much when including the second order corrections, even in the strongly coupled regime we have to work in. The behavior of $V(z)$ is particularly encouraging, because it is directly related to the kinetic functions $f_1$ and $f_2$, and hence, to the gauge couplings  $(g^1)^{2}$ and $(g^2)^{2}$ on the observable and hidden sector, respectively. Our finding implies that the regions of validity for $(g^1)^{2}>0$ and $(g^2)^{2}>0$ do not change significantly.

Finally, we note that these results have been obtained under the assumption that  $\bar{ b}=k{\bar V^{2/3}}$ can be approximated by ${\hat R}/{2}=kV^{2/3}$. So was that approximation justified in our case?
We find for $(\bar a^1, \bar a^2, \bar a^3)=(1.3,1.2,0.16)$ and $(a^1,a^2,a^3)=(1.0, 1.2, 0.3)$ that
\begin{equation}
\bar V^{2/3}=0.93 \ ,
\end{equation}
while\begin{equation}
V^{2/3}=0.87 \ ,
\end{equation}
respectively. Hence, the approximation is well justified.

\bibliographystyle{utphys}
\bibliography{citations}

\end{document}